\documentclass{jfm}

\usepackage{graphicx,color}

\newcount\ndots
\def\drawline#1#2{\raise 2.5pt\vbox{\hrule width #1pt height #2pt}}
\def\spacce#1{\hskip #1pt}
\def\solid{\drawline{24}{.5}\nobreak}
\def\bsolid{\drawline{24}{0.8}\nobreak}
\def\bdash{\hbox{\drawline{4}{.5}\spacce{2}}}
\def\dashed{\bdash\bdash\bdash\bdash\nobreak}
\def\chndot{\hbox{\drawline{5}{.5}\spacce{1.5}\drawline{1.5}{.5}\spacce{1.5}
\drawline{5}{.5}\spacce{1.5}\drawline{1.5}{.5}\spacce{1.5}\drawline{5}{.5}} \nobreak}
\def\trian{\raise 1.25pt\hbox{$\scriptstyle\triangle$}\nobreak}
\def\circle{$\circ$\nobreak}
\def\solidcircle{$\bullet$\nobreak}
\newcommand{\GK} [1] {{\color{black}#1}}
\newcommand{\GGK} [1] {{\color{black}#1}}
\newcommand{\GKK} [1] {{\color{black}#1}}
\newcommand{\SM} [1] {{\color{black}#1}}
\newcommand{\SMM} [1] {{\color{black}#1}}

\shorttitle{Optimal heat transfer enhancement in plane Couette flow} 
\shortauthor{S. Motoki et al} 

\title{Optimal heat transfer enhancement in plane Couette flow}

\author
 {
  \GK{Shingo}  Motoki\aff{1},
  \GK{Genta} Kawahara\aff{1}
  \corresp{\email{kawahara@me.es.osaka-u.ac.jp}} 
  \and
  \GK{Masaki} Shimizu\aff{1}
  }

\affiliation
{
\aff{1}
Graduate School of Engineering Science, Osaka University, 
1-3 Machikaneyama, Toyonaka, Osaka 560--8531, Japan
}

\begin{document}

\maketitle

\begin{abstract} 
\GK{Optimal} heat transfer enhancement
\GK{has been explored}
theoretically in plane Couette flow.
The velocity field 
\GK{to be optimised is}
time-independent and incompressible, and
\GKK{temperature} is determined
\GK{in terms of the velocity}
as a solution to an advection-diffusion equation.
\GK{The Prandtl number is set to
unity, and
consistent boundary conditions are imposed on the velocity
and the temperature fields.}
The excess of
\GK{a} wall heat flux (\GKK{or \SM{equivalently}} \GK{total} scalar dissipation)
over
total energy dissipation is taken as an objective functional, and
\GK{by using a variational method}
the Euler--Lagrange equations are derived,
\GK{which are solved numerically
to obtain the optimal states in the sense of
maximisation of the functional.}
\GK{The laminar conductive field is an optimal state at low Reynolds number
$Re\sim \GKK{10^0}$.}
\GK{At higher Reynolds number $Re\sim \GKK{10^1}$, however,
the optimal state exhibits a streamwise-independent two-dimensional
velocity field.}
The \GK{two-dimensional}
field consists of large-scale circulation rolls
that play a role in heat transfer
\GKK{enhancement} \GK{with respect to the conductive state as in
thermal convection.}
\GKK{A further} \GK{increase of the Reynolds number leads
to a three-dimensional optimal state at $Re\gtrsim \GKK{10^2}$.}
\GK{In the three-dimensional velocity field
there appear smaller-scale \SMM{hierarchical} quasi-streamwise vortex tubes
near the walls in addition to
the large-scale rolls.}
The \GK{streamwise}
vortices are tilted in the spanwise direction
so that they may produce the
\GK{anticyclonic} vorticity
\GK{antiparallel to the mean-shear vorticity,
bringing about} significant three-dimensionality.
\GK{The isotherms wrapped around the tilted anticyclonic vortices
undergo the cross-axial shear of the mean flow,
so that the spacing of the wrapped isotherms is narrower
and so the temperature gradient is steeper than those around
a purely streamwise (two-dimensional) vortex tube,
intensifying scalar dissipation and so a wall heat flux.}
\GK{Moreover, the tilted anticyclonic vortices induce
the flow towards the wall to push low- (or high-) temperature
fluids on the hot (or cold) wall,
enhancing a wall heat flux.}
\GK{The optimised three-dimensional velocity fields
achieve a much higher wall heat flux and much lower energy dissipation
than those of plane Couette turbulence.}
\end{abstract}


\section{Introduction}\label{sec1} 
\GK{The improvement of}
the performance of heat exchangers has always been
\GK{crucial to efficient energy use.}
What is a flow field with the ability of the most efficient heat transfer?
This may be a naive question in the development of flow control technique aimed at heat transfer enhancement.
If we can find an optimal state in which heat transfer will be enhanced while suppressing energy loss, it will serve as
\GK{the} target of flow control.

For buoyancy-driven convection, \cite{Malkus1954}
\GK{raised} a question of an upper limit to heat transfer and an optimal state
\GK{realising}
it, and \cite{Howard1963} formulated a variational problem to derive
a theoretical upper bound on
\GK{a}
heat flux.
\GK{In the case of velocity fields satisfying}
the equation of continuity,
Howard obtained the upper bound under the assumption that
\GK{an}
optimal state has a single horizontal 
\GK{wavenumber},
and subsequently \cite{Busse1969}
\GK{considered the same}
problem by using a multiple-boundary-layer approach
\GK{to derive} an asymptotic upper bound
\GK{in a large-Rayleigh-number} limit.
Moreover, by using the same method, \cite{Busse1970}
\GK{found}
an upper bound on momentum transfer
\GK{(or equivalently energy dissipation and so enstrophy)},
and
\GK{conjectured an optimal velocity field consisting of hierarchical}
streamwise vortices
in plane Couette flow
\GK{with the aid of}
the so-called attached-eddy hypothesis \citep{Townsend1976}.
Later Doering and Constantin developed a new variational approach called 
\GK{`the background method'} and
\GK{obtained} a rigorous upper bound on energy dissipation
in plane Couette flow \citep{Doering1992,Doering1994} and a heat flux
in Rayleigh--B\GKK{\'e}nard convection \citep{Doering1996}.
The development of
\GK{this method has
\SMM{triggered} remarkable advancements 
in
identification} of optimal momentum transfer in shear flows \citep{Nicodemus1997,Nicodemus1998a,Nicodemus1998b,Plasting2003} and optimal heat transfer in thermal
\GK{convection} \citep{Kerswell2001,Otero2002,Doering2006}.
Recently, \cite{Hassanzadeh2014} have reported optimal velocity fields
\GK{for}
two-dimensional thermal convection without
\GK{taking account of} the Navier--Stokes equation as a constraint
\GK{on} fluid motion.
They
\GK{considered}
\GK{two-dimensional} divergence-free velocity fields
\GK{under} one of the two types of constraints
\GK{on} velocity fields (fixed
\GK{kinetic}
energy
\GK{or} fixed enstrophy).
On the other hand, \cite{Sondak2015} have
\GK{taken another approach to optimal heat
transfer in the two-dimensional thermal convection,
in which they optimised the aspect ratio
of the two-dimensional domain for their
steady solutions
to the Navier--Stokes equation
to achieve
maximal heat transfer.}
\GK{To our knowledge},
however,
\GK{the velocity field for}
optimal heat transfer
has \GK{not} been
\GK{determined in shear flows} as yet.

\GK{Shear flow}
turbulence has an ability of significantly high heat transfer in comparison with laminar flow, but it results in friction loss as a consequence of simultaneous promotion of momentum transfer.
This is well known as the similarity between heat and momentum transfer in engineering \citep{Reynolds1874,Chilton1934}.
Especially, when the Prandtl number, the ratio between thermal and momentum diffusivity, is close to unity, the heat and momentum transfer exhibit strong similarity \citep{Dipprey1962}.
Although one of the ultimate goals in
\GK{turbulence control} is
\GK{more heat transfer and less friction},
\GK{their}
achievement has been extremely difficult due to the similarity.
Recently, however, control strategies to break the similarity have been proposed.
Momentum transfer by
\GK{viscous} flow obeys the Navier--Stokes equation,
whereas heat transfer is governed by the
\GK{energy equation, i.e.} an advection-diffusion equation for temperature.
By applying the feedback control using the adjoint equations \citep{Bewley2001,Kasagi2012} to heat transfer in turbulent channel flow with blowing and suction on the wall, \cite{Hasegawa2011} and \cite{Yamamoto2013} have numerically
achieved the dissimilarity even when the Prandtl number is equal to unity.
Experimentally, several practicable passive or active control techniques have been developed using riblet \citep{Sasamori2014}, permeable wall \citep{Suga2011}, micro actuator and sensor \citep{Kasagi2009}, and so on.
Hence, the
accomplishment of dissimilar heat transfer enhancement by flow control
\GK{might} be expected
\GKK{in the near future}.

The aim of this study is to find an optimal velocity field which will provide new insight into such flow control.
We focus on optimal heat transfer in plane Couette flow.
In this flow, the temperature and velocity field  exhibit complete similarity in a laminar
\GK{and conductive}
state, when a constant temperature difference is imposed
\GK{between the two parallel} walls.
Furthermore, when the Prandtl number is set to unity, a turbulent state shows strong similarity between heat and momentum transfer.
By taking the excess of the scalar dissipation (or equivalently,
\GKK{a} \GK{total}
wall heat flux) over the energy dissipation as an objective functional,
an optimal velocity field is determined by maximising the functional
under the constraint of the continuity equation for the velocity
and the advection-diffusion equation for the temperature.

This paper is organized as follows.
In \S\ref{sec2} we describe the flow configuration and the mathematical relation between
scalar (or energy) dissipation and
\GKK{a}
wall heat (or momentum) flux, and then present the formulation of a variational problem.
In \S\ref{sec3}, the details of numerical procedures and the parameter dependence of optimal velocity fields are explained.
The spatial structures and the statistics of the optimal states are presented in \S\ref{sec4}, and the hierarchical structure of the optimal fields is discussed in \S\ref{sec5}.
In \S\ref{sec6} we discuss the significant effects of three-dimensional vortical structures on
heat transfer enhancement \SMM{in the optimal state}.
Section \ref{sec7} is
\GK{devoted} to \GKK{summary and discussion}.
\GK{Properties of the external body force to be}
\GKK{added}
\GK{for the achievement of the optimal velocity field are
described in Appendix~\ref{appA}.
An analytical interpretation of
heat transfer
intensification mechanisms through the spanwise inclination
of a quasi-streamwise tubular vortex
is given in Appendix~\ref{appB}.}

\begin{figure} 
\centering
	\begin{minipage}{.55\linewidth}
	\includegraphics[clip,width=\linewidth]{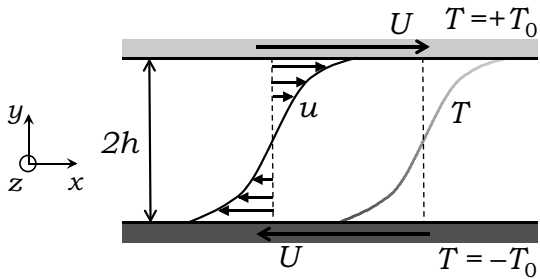}
	\end{minipage}
	
\caption{Configuration of the velocity and temperature fields.\label{fig1}}
\end{figure}

\section{Mathematical formulation}\label{sec2} 
\subsection{Flow configuration} 
Figure \ref{fig1} shows the configuration of the velocity and temperature fields.
The
\GK{flow} is driven by the \GK{two} parallel plates
moving in the opposite directions at a constant speed
\GK{$U$}.
The upper (or lower) wall surface is held at higher (or lower) constant temperature
\GK{$+T_0$ (or $-T_0$)}.  
The coordinates, $x$, $y$ and $z$ (or $x_{1}$, $x_{2}$ and $x_{3}$), are used for the representation of the streamwise, the wall-normal and the spanwise directions, respectively,
and their origin is on the midplane of the
\GK{two plates separated by a distance $2h$}. 
The corresponding components of the velocity $\mbox{\boldmath$u$}(x,y,z)$ are given by $u$, $v$ and $w$ (or $u_{1}$, $u_{2}$ and $u_{3}$) in the streamwise, the wall-normal and the spanwise directions, respectively.

Let us consider heat transfer in an incompressible time-independent velocity field fulfilling the continuity equation
\begin{equation}
\label{eq2-1}
\frac{\partial u_{i}}{\partial x_{i}}=0.
\end{equation}
We suppose that the temperature field $T(x,y,z)$ is determined as a solution to an advection-diffusion equation
\begin{equation}
\label{eq2-2}
\frac{\partial}{\partial x_{j}}\left( u_{j}T \right)=\kappa\frac{\partial^{2}T}{\partial x_{j}^2},
\end{equation}
where $\kappa$ denotes
thermal diffusivity.
The velocity and temperature fields are supposed to be periodic in the $x$- and $z$-directions with the periods, $L_{x}$ and $L_{z}$.
The boundary conditions
\begin{eqnarray}
\label{eq2-3}
u(y=\pm h)=\pm U,\hspace{1em} v(y=\pm h)=w(y=\pm h)=0;
\quad T(y=\pm h)=\pm T_{0}
\end{eqnarray}
are imposed on the walls.

\subsection{Scalar dissipation and heat flux} 
We introduce the $xz$-average and the $xyz$-average
\GK{given respectively by}
\begin{eqnarray}
\label{eq2-4}
{\left< \cdot \right>}_{xz}&=&\frac{1}{L_{x}L_{z}}\int_{0}^{L_{x}}\int_{0}^{L_{z}}\  \cdot \  {\rm d}x{\rm d}z,\\
\label{eq2-5}
{\left< \cdot \right>}_{xyz}&=&\frac{1}
{\GK{2h L_x L_z}}
\int_{0}^{L_{x}}\int_{-h}^{h}\int_{0}^{L_{z}}\  \cdot \  {\rm d}x{\rm d}y{\rm d}z.
\end{eqnarray}
We
\GK{take the $xz$-average of}
equation (\ref{eq2-2})
to obtain
\begin{eqnarray}
\label{eq2-6}
\frac{\rm d}{{\rm d}y}{\left< vT \right>}_{xz}=\kappa\frac{\rm d^2}{{\rm d}y^2}{\left< T \right>}_{xz},
\end{eqnarray}
the $y$-integration of which \SMM{gives us} the total heat flux
\begin{eqnarray}
\label{eq2-7}
{\left< vT \right>}_{xz}-\kappa\frac{\rm d}{{\rm d}y}{\left< T \right>}_{xz}
                       =-\kappa\frac{\rm d}{{\rm d}y}{\left< T \right>}_{xz}\left( y=-h \right)={\rm const.}
\end{eqnarray}
\GK{at any wall-normal position $y$.}
By further taking the average of equation (\ref{eq2-7}) in the $y$-direction we have
\begin{eqnarray}
\label{eq2-8}
{\left< vT \right>}_{xyz}-\kappa\frac{T_{0}}{h}
                       =-\kappa\frac{\rm d}{{\rm d}y}{\left< T \right>}_{xz}\left( y=-h \right).
\end{eqnarray}
\indent
Now we decompose the temperature into a \GKK{conductive
state} and a fluctuation about it as 
\begin{eqnarray}
\label{eq2-9}
T=T_{0}\frac{y}{h}+\theta.
\end{eqnarray}
The boundary conditions of the temperature fluctuation $\theta$ are
\begin{eqnarray}
\label{eq2-10}
\theta (y=\pm h)=0.
\end{eqnarray}
The substitution of decomposition (\ref{eq2-9}) in (\ref{eq2-8}) yields
\begin{eqnarray}
\label{eq2-11}
{\left< v\theta \right>}_{xyz}=-\kappa\frac{\rm d}{{\rm d}y}{\left< \theta \right>}_{xz}\left( y=-h \right).
\end{eqnarray}

We substitute decomposition (\ref{eq2-9}) for equation (\ref{eq2-2}) to have 
\begin{eqnarray}
\label{eq2-12}
\frac{\partial}{\partial x_{j}}\left( u_{j}\theta \right)+v\frac{T_{0}}{h}
=\kappa\frac{\partial^{2}\theta}{\partial x_{j}^2}.
\end{eqnarray}
The product of equation (\ref{eq2-12}) with $\theta$ \SMM{gives us}
\begin{eqnarray}
\label{eq2-13}
\frac{\partial}{\partial x_{j}}\left( u_{j}\frac{1}{2}\theta^2 \right)+v\theta\frac{T_{0}}{h}
= \kappa\frac{\partial}{\partial x_{j}}\left( \theta\frac{\partial \theta}{\partial x_{j}} \right)
 -\kappa{\left( \frac{\partial \theta}{\partial x_{j}} \right)}^2.
\end{eqnarray}
By further taking the $xyz$-average of equation (\ref{eq2-13}) and taking into account
boundary conditions (\ref{eq2-10}), we obtain
\begin{eqnarray}
\label{eq2-14}
\frac{T_{0}}{h}{\left< v\theta \right>}_{xyz}=-\kappa{\left< {\left| \nabla \theta \right|}^2 \right>}_{xyz},
\end{eqnarray}
where the right-hand side represents the dissipation of variance
\GK{$\theta^2$,
hereafter referred to as} \GKK{`scalar dissipation'.}
As a consequence, we have the expression of the wall heat flux
\begin{eqnarray}
\label{eq2-15}
-\kappa\frac{\rm d}{{\rm d}y}{\left< T \right>}_{xz}\left( y=+h \right)
=-\kappa\frac{\rm d}{{\rm d}y}{\left< T \right>}_{xz}\left( y=-h \right)
=-\frac{\kappa h}{T_{0}}{\left< {\left| \nabla \theta \right|}^2 \right>}_{xyz}
 -\kappa\frac{T_{0}}{h}.
\end{eqnarray}

\subsection{Energy dissipation and momentum flux} 
The steady motion of a viscous fluid is described by the Navier--Stokes equation
\begin{eqnarray}
\label{eq2-16}
\frac{\partial}{\partial x_{j}}\left( u_{i}u_{j} \right)
=-\frac{1}{\rho}\frac{\partial p}{\partial x_{i}}
+\nu\frac{\partial^2 u_{i}}{\partial x_{j}^2}+f_{i},
\end{eqnarray}
where \GK{$p$, $\rho$, $\nu$ and $f_{i}$
are pressure, mass density, kinematic viscosity,
and the $i$-th component of
external body force per unit mass, respectively.}
The inner product of equation (\ref{eq2-16}) with the velocity yields the local energy budget equation
\begin{eqnarray}
\label{eq2-17}
\frac{\partial}{\partial x_{j}}\left( u_{j}\frac{1}{2}{\left| \mbox{\boldmath $u$} \right|}^2 \right)
=-\frac{1}{\rho}\frac{\partial}{\partial x_{i}}\left( p u_{i} \right)
 +\nu\frac{\partial}{\partial x_{j}}\left( u_{i}\frac{\partial u_{i}}{\partial x_{j}} \right)
 -\nu{\left( \frac{\partial u_{i}}{\partial x_{j}} \right)}^2+u_{i}f_{i}.
\end{eqnarray}
By taking the $xyz$-average of equation (\ref{eq2-17}) and taking into account
boundary \GKK{conditions} (\ref{eq2-3}), we obtain the total energy budget equation
\begin{eqnarray}
\label{eq2-18}
\frac{\nu}{2}\left[ \frac{\rm d}{{\rm d}y}{\left< u \right>}_{xz}(y=+h)+\frac{\rm d}{{\rm d}y}{\left< u \right>}_{xz}(y=-h) \right]
=\frac{\nu h}{U}{\left< {\left| \nabla \mbox{\boldmath $u$} \right|}^2 \right>}_{xyz}
-\frac{h}{U}{\left< u_{i}f_{i} \right>}_{xyz},
\end{eqnarray}
where we have supposed a null mean pressure gradient in the wall-parallel ($x$- and $z$-) directions.
\GKK{The left-hand side of equation (\ref{eq2-18})
is mean wall shear stress,
while the} first and second terms in the right-hand side
\GKK{originate from}
energy dissipation and an energy input by the external force, respectively.

\subsection{Objective functional and Euler-Lagrange equation} 
Let us now consider the \GK{optimisation} problem of a functional
to find an optimal state for the dissimilarity between momentum and heat transfer.
As can be seen from equation (\ref{eq2-15}) and (\ref{eq2-18}), the wall heat flux corresponds to the
\GK{total}
scalar dissipation, and the wall momentum flux (i.e. the skin friction) is related with the
\GK{total} energy dissipation.
Therefore, we shall
\GK{search for an}
optimal state
\GK{which maximises}
the functional $J\{\theta,\mbox{\boldmath$u$},p^{*},\theta^{*}\}$
\GK{given by}
\begin{eqnarray}
\label{eq2-19}
&\displaystyle
\hspace{-15em}
J= \frac{1}{2}\epsilon\frac{c_{p}\kappa}{T_{0}}{\left< {\left| \nabla \theta \right|}^2 \right>}_{xyz}
  -\frac{1}{2}\nu{\left< {\left| \nabla \mbox{\boldmath $u$} \right|}^2 \right>}_{xyz}& \nonumber \\
&\hspace{7em}
\displaystyle
+\frac{1}{\rho}{\left< p^{*}\frac{\partial u_{i}}{\partial x_{i}} \right>}_{xyz}
  +\epsilon\frac{c_{p}}{T_{0}}
  {\left< \theta^{*}\left( u_{j}\frac{\partial \theta}{\partial x_{j}}
                          +v\frac{T_{0}}{h}
                          -\kappa\frac{\partial^2 \theta}{\partial x_{j}^2} \right)
   \right>}_{xyz}, &
\end{eqnarray}
where $\epsilon$ \GK{($>0$)}
is the dimensionless weight of the contribution from the heat transfer against the momentum transfer,
\GK{$c_{p}$} is specific heat at constant pressure,
\GK{and}
$p^{*}(x,y,z)$ and $\theta^{*}(x,y,z)$ represent Lagrangian multipliers.
We suppose that $\theta^{*}$ satisfies boundary conditions on the walls,
\begin{eqnarray}
\label{eq2-20}
\theta^{*} (y=\pm 1)=0.
\end{eqnarray}
By
\GK{introducing} the following dimensionless quantities
\begin{eqnarray}
\label{eq2-21}
\GK{x_i'=\frac{x_i}{h}},\hspace{1em} J'=\frac{J}{U^3 / h},\hspace{1em} \theta'=\frac{\theta}{T_{0}},\hspace{1em}
\GK{u_i'=\frac{u_i}{U}},\hspace{1em} {p^{*}}'=\frac{p^{*}}{\rho U^2},\hspace{1em}
\GK{{\theta^{*}}'=\frac{\theta^{*}}{T_{0}}},
  \nonumber \\
\end{eqnarray}
we rewrite the objective functional (\ref{eq2-19}) in
\GK{a} dimensionless form as
\GK{
\begin{eqnarray}
\label{eq2-22}
&\displaystyle
 J= \frac{1}{2}\lambda\frac{1}{RePr}{\left< {\left| \nabla \theta \right|}^2 \right>}_{xyz}
  -\frac{1}{2}\frac{1}{Re}{\left< {\left| \nabla {\mbox{\boldmath $u$}} \right|}^2 \right>}_{xyz}
  +{\left< p^{*}\frac{\partial u_{j}}{\partial x_{j}} \right>}_{xyz}& \nonumber \\
&\displaystyle
 +\lambda
  {\left< {\theta^{*}}\left( u_{j}\frac{\partial \theta}{\partial x_{j}}
                          +v
                          -\frac{1}{RePr}\frac{\partial^2 \theta}{\partial x_{j}^2} \right) 
   \right>}_{xyz}, &
\end{eqnarray}
}where
\GK{we have dropped the primes attached to the dimensionless quantities,}
\begin{eqnarray}
\label{eq2-23}
\lambda
= \frac{\epsilon}{Ec}
\end{eqnarray}
\GK{is
new dimensionless weight, and}
\begin{eqnarray}
\label{eq2-24}
Re=\frac{Uh}{\nu},\hspace{1em} Pr=\frac{\nu}{\kappa},\hspace{1em} Ec=\frac{U^2}{c_{p}T_{0}}
\end{eqnarray}
are the Reynolds number, the Prandtl number, and the Eckert number, respectively.
\GK{Usually $Ec \ll 1$, and thus we suppose that $\epsilon \ll 1$
so that we may take $\lambda \sim \GKK{10^0}$.}
We decompose $\mbox{\boldmath$u$},\theta,p^{*}$ and $\theta^{*}$ into
references and
infinitesimal perturbations as
\begin{eqnarray}
\label{eq2-25}
\theta=\overline{\theta} +\delta \theta,\hspace{1em}
\mbox{\boldmath $u$}=\overline{\mbox{\boldmath $u$}} +\delta \mbox{\boldmath $u$},\hspace{1em} 
p^{*}=\overline{p^{*}} +\delta p^{*},\hspace{1em} 
\theta^{*}=\overline{\theta^{*}} +\delta \theta^{*}.
\end{eqnarray}
Substituting (\ref{eq2-25}) into (\ref{eq2-22}), using integration by parts and recalling
boundary conditions (\ref{eq2-3}), (\ref{eq2-10}) and (\ref{eq2-20}), we obtain the first variation of $J$ as
\begin{eqnarray}
\label{eq2-26}
&\displaystyle
 \delta J= {\left< \left( -\frac{\partial \overline{p^{*}}}{\partial x_{i}}
                          +\frac{1}{Re}\frac{\partial^2 \overline{u_{i}}}{\partial x_{j}^2}
                          +\lambda\overline{\theta^{*}}\frac{\partial \overline{\theta}}{\partial x_{i}}
                          +\lambda\overline{\theta^{*}}\delta_{i2}
                   \right) \delta u_{i} \right>}_{xyz}& \nonumber \\
&\displaystyle
  -{\left< \lambda \left(  \overline{u_{j}}\frac{\partial \overline{\theta^{*}}}{\partial x_{j}}
                          +\frac{1}{RePr}\frac{\partial^2 \overline{\theta^{*}}}{\partial x_{j}^2}
                          +\frac{1}{RePr}\frac{\partial^2 \overline{\theta}}{\partial x_{j}^2}
                   \right) \delta \theta \right>}_{xyz}& \nonumber \\
&\displaystyle
  +{\left< \lambda \left(  \overline{u_{j}}\frac{\partial \overline{\theta}}{\partial x_{j}}
                          +\overline{v}
                          -\frac{1}{RePr}\frac{\partial^2 \overline{\theta}}{\partial x_{j}^2}
                   \right) \delta \theta^{*} \right>}_{xyz}& \nonumber \\
&\displaystyle
          +{\left< \left(  \frac{\partial \overline{u_{i}}}{\partial x_{i}}
                   \right) \delta p^{*} \right>}_{xyz}. &
\end{eqnarray}
A stationary point, at which the first variation of $J$ vanishes, i.e. $\delta J=0$, for any perturbation $\delta\mbox{\boldmath$u$},\delta\theta,\delta\theta^{*}$ and $\delta p^{*}$, is determined by
the Euler--Lagrange equations
\begin{eqnarray}
\label{eq2-27}
\displaystyle
\frac{\delta J}{\delta u_{i}}&\equiv&
-\frac{\partial {p^{*}}}{\partial x_{i}}
+\frac{1}{Re}\frac{\partial^2 {u_{i}}}{\partial x_{j}^2}
+\lambda{\theta^{*}}\frac{\partial {\theta}}{\partial x_{i}}
+\lambda{\theta^{*}}\delta_{i2}=0, \\
\label{eq2-28}
\displaystyle
\frac{\delta J}{\delta \theta}&\equiv&
 {u_{j}}\frac{\partial {\theta^{*}}}{\partial x_{j}}
+\frac{1}{RePr}\frac{\partial^2 {\theta^{*}}}{\partial x_{j}^2}
+\frac{1}{RePr}\frac{\partial^2 {\theta}}{\partial x_{j}^2}=0, \\
\label{eq2-29}
\displaystyle
\frac{\delta J}{\delta \theta^{*}}&\equiv&
 {u_{j}}\frac{\partial {\theta}}{\partial x_{j}}
+{v}
-\frac{1}{RePr}\frac{\partial^2 {\theta}}{\partial x_{j}^2}=0, \\
\label{eq2-30}
\displaystyle
\frac{\delta J}{\delta p^{*}}&\equiv&
\frac{\partial {u_{i}}}{\partial x_{i}}=0.
\end{eqnarray}

\section{Numerical procedures}\label{sec3} 
\subsection{Finding initial guesses} 
We find velocity fields which give a maximal
\GK{value} of the objective functional $J$ with
a combination of the steepest ascent method and the Newton--Krylov method.
To globally seek a maximal point of $J$, we start with finding
initial guesses in the following procedure.

\begin{description}
\setlength{\parskip}{1em}
\item[Step. 1] {\ }Start from an arbitrary
velocity field $\mbox{\boldmath$u$}$ which satisfies the continuity equation (\ref{eq2-1}) and 
boundary condition (\ref{eq2-3}).
\item[Step. 2] {\ }For given $\mbox{\boldmath$u$}$, solve equation (\ref{eq2-29}) to obtain $\theta$.
\item[Step. 3] {\ }For given $\mbox{\boldmath$u$}$ and obtained $\theta$ from Step. 2, solve equation (\ref{eq2-28}) to find $\theta^{*}$.
\item[Step. 4] {\ }Since $\mbox{\boldmath$u$}$, $\theta$, and $\theta^{*}$ respectively satisfy the Euler--Lagrange equations (\ref{eq2-28})--(\ref{eq2-30}), the first variation $\delta J$ can be expressed, in terms of
\GK{\GKK{the only} functional derivative
$\widehat{u_{i}}\equiv \delta J/\delta u_{i}$,}
as
\begin{equation}
\label{eq3-1}
\delta J={\left< \widehat{u_{i}} \delta u_{i} \right>}_{xyz}.
\end{equation}
Using the
\GK{functional derivative $\widehat{\mbox{\boldmath$u$}}$,
renew the velocity field} as 
\begin{equation}
\label{eq3-2}
\mbox{\boldmath$u$}^{\rm new}=\mbox{\boldmath$u$}+s\widehat{\mbox{\boldmath$u$}},
\end{equation}
where $s$ is a small positive constant to be taken as $s\sim 10^{-2}$.
\item[Step. 5] {\ }Repeat Step. 2 \GKK{--} Step. 4.
\vspace{1em}
\end{description}

Since
\GK{(\ref{eq3-2}) implies that}
\begin{equation}
\label{eq3-3}
\frac{\partial u_{i}^{\rm new}}{\partial x_{i}}=\frac{\partial u_{i}}{\partial x_{i}}+s\frac{\partial \widehat{u}_{i}}{\partial x_{i}}=s\frac{\partial \widehat{u}_{i}}{\partial x_{i}},
\end{equation}
\GK{in order that $\mbox{\boldmath$u$}^{\rm new}$ is incompressible
we invoke the continuity for $\widehat{\mbox{\boldmath$u$}}$,}
\begin{equation}
\label{eq3-4}
\frac{\partial \widehat{u}_{i}}{\partial x_{i}}=0.
\end{equation}
\GK{Taking the double curl, the single curl and the $xz$-average of
equation (\ref{eq2-27}), we have the equations
for the wall-normal velocity $v$, the wall-normal vorticity
$\omega_y$, and the averaged velocities,}
\begin{eqnarray}
\label{eq3-5}
\displaystyle
\frac{\partial^{2} \widehat{v}}{\partial x_{j}^{2}}&=&\frac{1}{Re}\frac{\partial^{2}}{\partial x_{j}^{2}}\frac{\partial^{2}v}{\partial x_{j}^{2}}+\lambda H_{v}+\lambda\left( \frac{\partial^{2}}{\partial x^{2}}+\frac{\partial^{2}}{\partial z^{2}} \right)\theta^{*}, \\
\label{eq3-6}
\displaystyle
\widehat{\omega}_{y}&=&\frac{1}{Re}\frac{\partial^{2}\GK{\omega_{y}}}{\partial x_{j}^{2}}+\lambda H_{\omega}, \\
\label{eq3-7}
\displaystyle
{\left< \widehat{u} \right>}_{xz}&=&
\GK{\frac{1}{Re}\frac{{\rm d}^{2} {\left< u \right>}_{xz}}{{\rm d} y^{2}}+\lambda{\left< \theta^{*}\frac{\partial\theta}{\partial x} \right>}_{xz},} \\
\label{eq3-8}
\displaystyle
{\left< \widehat{w} \right>}_{xz}&=&
\GK{\frac{1}{Re}\frac{{\rm d}^{2} {\left< w \right>}_{xz}}{{\rm d} y^{2}}+\lambda{\left< \theta^{*}\frac{\partial\theta}{\partial z} \right>}_{xz}}
\end{eqnarray}
\GK{with}
\begin{eqnarray}
\label{eq3-9}
\displaystyle
H_{v}
&=&-\frac{\partial}{\partial y}\left[ \frac{\partial}{\partial x}\left( \theta^{*}\frac{\partial\theta}{\partial x} \right)+\frac{\partial}{\partial z}\left( \theta^{*}\frac{\partial\theta}{\partial z} \right) \right]+\left( \frac{\partial^{2}}{\partial x^{2}}+\frac{\partial^{2}}{\partial z^{2}} \right)\left( \theta^{*}\frac{\partial\theta}{\partial y} \right), \\
\label{eq3-10}
\displaystyle
H_{\omega}
&=&\frac{\partial}{\partial z}\left( \theta^{*}\frac{\partial\theta}{\partial x} \right)-\frac{\partial}{\partial x}\left( \theta^{*}\frac{\partial\theta}{\partial z} \right),
\end{eqnarray}
from which $p^{*}$ has been eliminated.
The \GK{corresponding}
boundary conditions
\GK{are}
\begin{subequations}\label{eq3-11}
\begin{eqnarray}
v(y=\pm1)=\frac{\partial v}{\partial y}(y=\pm1)&=&0, \\
\omega_{y}(y=\pm1)&=&0, \\
{\left< u \right>}_{xz}(y=\pm1)&=&\pm1, \\
{\left< w \right>}_{xz}(y=\pm1)&=&0, \\
\theta(y=\pm1)&=&0, \\
\theta^{*}(y=\pm1)&=&0.
\end{eqnarray}
\end{subequations}

In the periodic streamwise and spanwise directions, $\mbox{\boldmath$u$}$,
\GK{vorticity $\mbox{\boldmath$\omega$}=\nabla\times \mbox{\boldmath$u$}$,}
$\theta$, and $\theta^{*}$
are approximated
\GK{by truncated Fourier series}
\begin{eqnarray}
\label{eq3-12}
\left(\mbox{\boldmath$u$},\GK{\mbox{\boldmath$\omega$},} \theta,\theta^{*}
\right)
&=&\sum_{k=-K}^{K}\sum_{m=-M}^{M}\left(\widetilde{\mbox{\boldmath$u$}}^{km},
\GK{\widetilde{\mbox{\boldmath$\omega$}}^{km},}
\widetilde{\theta}^{km},\widetilde{\theta^{*}}^{km}
\right){\rm e}^{{\rm i}2\pi k x/L_{x}}{\rm e}^{{\rm i}2\pi m z/L_{z}},
\end{eqnarray}
where $K$ and $M$ are positive integers.
\GK{Note that
the Fourier coefficients of $u$ and $w$ for $(k,m)\ne(0,0)$
are given by $\widetilde{v}^{km}$ and $\widetilde{{\omega}_{y}}^{km}$
through}
\begin{eqnarray}
\label{eq3-13}
\displaystyle
\GK{\widetilde{u}^{km}}
&=&
\GK{\mbox{i}\frac{
(2\pi k/L_x)\mbox{d}\widetilde{v}^{km}/\mbox{d}y
+(2\pi m/L_z)\widetilde{\omega_{y}}^{km}
}
{4\pi^2(k^2/L_{x}^2+m^2/L_{z}^2)},} \\
\label{eq3-14}
\displaystyle
\GK{\widetilde{w}^{km}}
&=&
\GK{\mbox{i}\frac{
(2\pi m/L_z)\mbox{d}\widetilde{v}^{km}/\mbox{d}y
+(2\pi k/L_x)\widetilde{\omega_{y}}^{km}
}
{4\pi^2(k^2/L_{x}^2+m^2/L_{z}^2)}.}
\end{eqnarray}
In the wall-normal direction,
\GK{$\widetilde{v}^{km}$, $\widetilde{\omega_y}^{km}$,
$\left<u \right>_{xz}-y$, $\left< w \right>_{xz}$,
$\widetilde{\theta}^{km}$ and $\widetilde{\theta^{*}}^{km}$
are
expanded, in terms of the polynomials}
which satisfy
boundary conditions (\ref{eq3-11}), as
\begin{subequations}\label{eq3-15}
\begin{eqnarray}
\widetilde{v}^{km}&=&\sum_{n=0}^{N-4}c_{v}^{kmn}\left( T_{n}-\GK{2
\frac{n+2}{n+3}}T_{n+2}+\frac{n+1}{n+3}T_{n+4} \right), \nonumber \\
\\
\left( \widetilde{\omega_{y}}^{km},
\widetilde{\theta}^{km},\widetilde{\theta^{*}}^{km} \right)&=&\sum_{n=0}^{N-2}\left( c_{\omega}^{kmn},
c_{\theta}^{kmn},c_{\theta^{*}}^{kmn} \right)\left( T_{n}-T_{n+2} \right), \nonumber \\
\\
\SM{\left( \left<u \right>_{xz}-y,\left<w \right>_{xz} \right)}&=&\SM{\frac{3}{2}\left( \left<u \right>_{xyz},\left<w \right>_{xyz} \right)\left(1-y^2\right)} \nonumber \\
&{\ }&\SM{+\frac{\rm d}{{\rm d}y}\left\{ \sum_{n=0}^{N-4}\left( c_{u}^{n},c_{w}^{n} \right)\left( T_{n}-2\frac{n+2}{n+3}T_{n+2}+\frac{n+1}{n+3}T_{n+4} \right) \right\}}, \nonumber \\
\end{eqnarray}
\end{subequations}
where $T_{n}(y)$ represents the \GK{$n$-th order} Chebyshev polynomial.
The nonlinear terms are evaluated using a spectral collocation method.
Aliasing errors are removed
\GK{with the aid of}
the $2/3$ rule for the Fourier transform and the $1/2$ rule for the Chebyshev transform.

\subsection{Newton--Krylov \GK{iteration}
} 
In nonlinear dynamical systems stemming from the Navier--Stokes equation, a combination of the Newton--Raphson method and the Krylov subspace method
\GK{has been}
used to efficiently obtain nonlinear equilibrium and periodic solutions \citep{Viswanath2007,Viswanath2009}.
In this study,
\GK{employing} the initial guesses obtained \GK{as mentioned}
in \S 3.1, we perform the Newton--Krylov iteration to find nonlinear solutions to the Euler--Lagrange equations (\ref{eq2-27})--(\ref{eq2-30}).

Introducing
pseudo-time derivative terms $\partial(\cdot)/\partial \tau$ into equations (\ref{eq3-5})--(\ref{eq3-8}), \SM{the $xyz$-average of equation (\ref{eq2-27}), equations} (\ref{eq2-28}) and (\ref{eq2-29}), we have the following nonlinear evolution equations
\begin{eqnarray}
\label{eq3-16}
\displaystyle
\frac{\partial}{\partial \tau}\frac{\partial^{2} v}{\partial x_{j}^{2}}&=&\frac{1}{Re}\frac{\partial^{2}}{\partial x_{j}^{2}}\frac{\partial^{2}v}{\partial x_{j}^{2}}+\lambda H_{v}+\lambda\left( \frac{\partial^{2}}{\partial x^{2}}+\frac{\partial^{2}}{\partial z^{2}} \right)\theta^{*}, \\
\label{eq3-17}
\displaystyle
\frac{\partial}{\partial \tau}\omega_{y}&=&\frac{1}{Re}
\GK{\frac{\partial^{2}\omega_{y}}{\partial x_{j}^{2}}}+\lambda H_{\omega}, \\
\label{eq3-18}
\displaystyle
\frac{\partial}{\partial \tau}{\left< u \right>}_{xz}&=&\frac{1}{Re}\frac{{\rm d}^{2} {\left< u \right>}_{xz}}{{\rm d} y^{2}}+\lambda{\left< \theta^{*}\frac{\partial\theta}{\partial x} \right>}_{xz}, \\
\label{eq3-19}
\displaystyle
\frac{\partial}{\partial \tau}{\left< w \right>}_{xz}&=&\frac{1}{Re}\frac{{\rm d}^{2} {\left< w \right>}_{xz}}{{\rm d} y^{2}}+\lambda{\left< \theta^{*}\frac{\partial\theta}{\partial z} \right>}_{xz}, \\
\label{eq3-20}
\displaystyle
\SM{\frac{\partial}{\partial \tau}{\left<u \right>}_{xyz}}&=&\SM{\frac{1}{2Re}
\GKK{\left[ \frac{{\rm d}{\left<u \right>}_{xz}}{{\rm d}y} \right]_{y=-1}^{y=+1}}+\lambda{\left< \theta^{*}\frac{\partial\theta}{\partial x} \right>}_{xyz}}, \\
\label{eq3-21}
\displaystyle
\SM{\frac{\partial}{\partial \tau}{\left<w \right>}_{xyz}}&=&\SM{\frac{1}{2Re}
\GKK{\left[ \frac{{\rm d}{\left<w \right>}_{xz}}{{\rm d}y} \right]_{y=-1}^{y=+1}}+\lambda{\left< \theta^{*}\frac{\partial\theta}{\partial z} \right>}_{xyz}}, \\
\label{eq3-22}
\displaystyle
\frac{\partial}{\partial \tau}\GK{\theta^{*}}&=&{u_{j}}\frac{\partial {\theta^{*}}}{\partial x_{j}}+\frac{1}{RePr}\frac{\partial^2 {\theta^{*}}}{\partial x_{j}^2}+\frac{1}{RePr}\frac{\partial^2 {\theta}}{\partial x_{j}^2}, \\
\label{eq3-23}
\displaystyle
\frac{\partial}{\partial \tau}\GK{\theta}&=&{u_{j}}\frac{\partial {\theta}}{\partial x_{j}}+{v}-\frac{1}{RePr}\frac{\partial^2 {\theta}}{\partial x_{j}^2},
\end{eqnarray}
\GK{`steady'} solutions to which satisfy the Euler--Lagrange equations
\GK{(\ref{eq2-27})--(\ref{eq2-30})}.
By using a spectral Galerkin method, we can convert equations (\ref{eq3-16})--(\ref{eq3-23}) into \GK{a nonlinear} autonomous dynamical system
\begin{eqnarray}
\label{eq3-24}
\frac{{\rm d}\mbox{\boldmath$X$}
}{{\rm d}\tau}=\mbox{\boldmath$F$}(\mbox{\boldmath$X$}
; Re,Pr,\lambda),
\end{eqnarray}
where $\mbox{\boldmath$X$}(\tau)$ represents a state point in the
phase space spanned by the Fourier--Chebyshev coefficients of $v,\omega_{y},{\left< u \right>}_{xz},{\left< w \right>}_{xz},\theta$ and $\theta^{*}$.
The time advancement is carried out by the Crank--Nicholson scheme for the diffusion terms and the 2nd-order Adams--Bashforth scheme for \GK{the others}.

Let $\overline{\mbox{\boldmath$X$}}$ be an initial guess to find a maximal point of $J$.
Integrating
equation (\ref{eq3-24}), we define the flow map as
\begin{eqnarray}
\label{eq3-25}
\mbox{\boldmath$X$}=\mbox{\boldmath$G$}(\overline{\mbox{\boldmath$X$}},\tau)
\end{eqnarray}
for any \GKK{`time'} $\tau$, where $\mbox{\boldmath$G$}(\overline{\mbox{\boldmath$X$}},0)=\overline{\mbox{\boldmath$X$}}$.
\GK{Supposing}
that $\overline{\mbox{\boldmath$X$}}$ is sufficiently close to a stationary point and
\GK{thus $\delta\mbox{\boldmath$X$}$ is a small correction vector},
\GK{we linearise equation (\ref{eq3-25})
at $\mbox{\boldmath$X$}=\overline{\mbox{\boldmath$X$}}$ to obtain}
\begin{eqnarray}
\label{eq3-26}
\mbox{\boldmath$A$}\delta\mbox{\boldmath$X$}=\mbox{\boldmath$B$},
\end{eqnarray}
where
\begin{eqnarray}
\label{eq3-27}
\mbox{\boldmath$A$}=\frac{\partial \mbox{\boldmath$G$}}{\partial \mbox{\boldmath$X$}}(\overline{\mbox{\boldmath$X$}})-\mbox{\boldmath$I$},
\hspace{3em} \mbox{\boldmath$B$}=\overline{\mbox{\boldmath$X$}}-\mbox{\boldmath$G$}(\overline{\mbox{\boldmath$X$}}),
\end{eqnarray}
\GK{$\mbox{\boldmath$I$}$ being}
an identity matrix with the same dimension as 
\GKK{the Jacobian matrix
$\partial \mbox{\boldmath$G$}/\partial \mbox{\boldmath$X$}$}.
By solving the linear equation (\ref{eq3-26}) the correction vector $\delta\mbox{\boldmath$X$}$ is obtained to
\GK{renew $\overline{\mbox{\boldmath$X$}}$} as
\begin{eqnarray}
\label{eq3-28}
\overline{\mbox{\boldmath$X$}}^{\rm new}=\overline{\mbox{\boldmath$X$}}+\delta\mbox{\boldmath$X$}.
\end{eqnarray}
We iterate the procedure until a termination criterion
\begin{eqnarray}
\label{eq3-29}
\frac{{|| \mbox{\boldmath$G$}(\overline{\mbox{\boldmath$X$}})-\overline{\mbox{\boldmath$X$}} ||}_{2}}{{|| \overline{\mbox{\boldmath$X$}} ||}_{2}}<10^{-12}
\end{eqnarray}
is satisfied, where ${||\cdot||}_{2}$ denotes 
\GK{an}
$L^{2}$-norm.
To solve equation (\ref{eq3-26}) we use the GMRES method \citep{Saad1986} which is one of the Krylov subspace methods.
In \GK{this} method, 
\GK{an approximation of a correction vector $\delta\mbox{\boldmath$X$}$
can be got without calculating matrix $A$, by minimising
the norm of a residual vector in a Krylov subspace.}

\subsection{Direct numerical simulation of turbulent plane Couette flow} 
For comparison with optimal states obtained by the \GK{maximisation}
of the functional $J$, 
\GK{direct numerical simulations are also performed
for turbulent momentum and heat transfer in plane Couette flow.
In the simulations we numerically integrate
the incompressible Navier--Stokes equation and the advection-diffusion equation for a passive scalar, i.e. temperature,
supplemented by boundary conditions (\ref{eq2-3}).}
\GK{The evolution equations are discretised
employing a spectral Galerkin method based on
the expansions (\ref{eq3-12}) and (\ref{eq3-15}).
Time marching of the equations is done with a
combination of the Crank--Nicholson and 2nd-order Adams--Bashforth scheme.}
The Prandtl number $Pr$ is set to unity.
\GK{Turbulence data are
accumulated in the range of 
the Reynolds number,
$Re=500$--$10000$,
\GKK{in a computational domain of}
$(L_{x},L_{y},L_{z})=(2\pi,2,\pi)$.}

\subsection{Global convergence and dependence on domain size} 
\begin{figure} 
\centering
	\begin{minipage}{.75\linewidth}
	\includegraphics[clip,width=\linewidth]{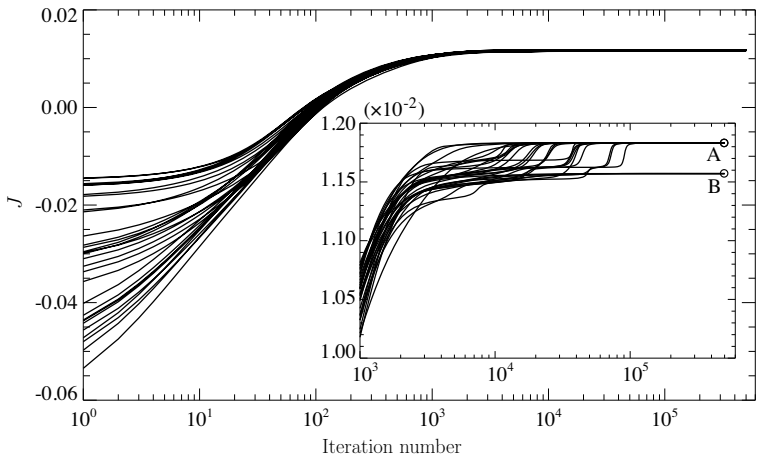}
	\end{minipage}

\caption{Objective functional $J$ against \GK{the number of the
iteration described in \S~3.1} for various initial guesses at $Re=100$ \GK{for} $Pr=1$ and $\lambda=1$.
\GK{30 initial guesses (10 each) are
taken from turbulent flow fields at $Re=1000$, $Re=2000$ and $Re=3000$.
26 of them converge to a local maximum A, and the others
tend to a local maximum B.}
\label{fig2}}
\end{figure}

\begin{table} 
  \begin{center}
  \def~{\hphantom{0}}
  \begin{tabular}{cccc}
      Label & $J$ & $\varepsilon_{S}$ & $\varepsilon_{E}$ \\
\\
       A  & $1.18\times10^{-2}$  & $5.17\times10^{-2}$ & $2.81\times10^{-2}$\\
       B  & $1.16\times10^{-2}$  & $5.27\times10^{-2}$ & $2.96\times10^{-2}$\\
  \end{tabular}
  \caption{Two local maxima of
  objective functional $J$ at $Re=100$ \GK{for} $Pr=1$ and $\lambda=1$.}
  \label{table1}
  \end{center}
\end{table}

We perform the
\GK{maximisation of the functional $J$
starting with 30 different initial guesses
extracted from turbulent flow fields in order to examine the 
dependence of
a computed local maximum on the initial guesses.}
Figure \ref{fig2} shows the functional $J$
against \GK{the number of the iteration
mentioned in \S~3.1}
for the distinct 30 initial guesses
at \GK{$Re=100$ for $Pr=1$ and $\lambda=1$}.
For all the initial guesses, \GK{the value of} $J$
rapidly approaches the vicinity of the two local maxima
to be mentioned just below.
We have eventually obtained the two different local maxima (the larger local maximum is labeled as A, and the smaller one as B).
Table \ref{table1} compares \GK{the values of}
$J$
\GKK{together with}
the total scalar dissipation $\varepsilon_{S}$ and the total energy dissipation \GK{$\varepsilon_{E}$},
\begin{eqnarray}
\label{eq3-30}
\varepsilon_{S}\GK{=}
\frac{1}{RePr}{\left< {|\nabla T|}^{2} \right>}_{xyz},\hspace{1em}
\varepsilon_{E}\GK{=}
\frac{1}{Re}{\left< {|\nabla\mbox{\boldmath$u$}|}^{2} \right>}_{xyz},
\end{eqnarray}
\GK{between the local maxima A and B}.
\GKK{The} present objective functional $J$ has multimodality, but there appear only two local maxima
\GKK{for many} initial guesses.

The optimal velocity fields exhibit three-dimensional structures which consist of large-scale streamwise circulation rolls and smaller-scale quasi-streamwise vortex tubes near the wall \GK{(see \SMM{figures}~\ref{fig5} and \ref{fig6})}.
The size of the large rolls corresponds to
\GK{half
the spanwise period $L_{z}/2$}.
Table \ref{table2} \GK{summarises} each maximal state obtained
in the various domains \GK{at $Re=1000$ for $Pr=1$} and $\lambda=0.1$.
It is found that $J$ takes the largest value for $(L_{x},L_{z})=(2\pi,\pi)$ among $(L_{x},L_{z})=(\pi,\pi)$, $(2\pi,\pi)$, $(3\pi,\pi)$\GKK{;}
$(2\pi,0.5\pi)$, $(2\pi,\pi)$, $(2\pi,1.5\pi)$, $(2\pi,2\pi)$, suggesting that there exist finite values of the streamwise and spanwise periods which lead to the global optimal.
However, our goal in this paper is not to identify the optimal domain for all
\GK{the} parameters including $Re$.
As will be described below, the velocity fields, which are \GK{realised} as a solution to the Euler--Lagrange equations (\ref{eq2-27})-(\ref{eq2-30}), have extremely high performance for heat transfer and exhibit qualitatively similar flow structures being roughly independent of $(L_{x},L_{z})$.
In the following, therefore, we \GKK{shall}
choose the domain size $(L_{x},L_{y},L_{z})=(2\pi,2,\pi)$
\GKK{to}
discuss
characteristic \GKK{flow} structures and
the mechanism of heat transfer enhancement.
\GKK{The effects of the domain size on the scalar dissipation and the energy dissipation
can be seen to be minor}.

\begin{table} 
  \begin{center}
  \def~{\hphantom{0}}
  \begin{tabular}{ccccccccc}
      $Re$ & $Pr$ & $\lambda$ & $L_{x}$ & $L_{z}$ & $(N_{x},N_{y,}N_{z})$ & $J$ & $\varepsilon_{S}$ & $\varepsilon_{E}$ \\
\\
      1000  & 1.0 & 0.1 & $\pi$   & $\pi$      & (32,\SM{128},256) & $0.65\times10^{-5}$ & $1.70\times10^{-2}$ & $1.69\times10^{-3}$\\
      1000  & 1.0 & 0.1 & $2\pi$ & $\pi$      & (32,\SM{128},256) & $4.42\times10^{-5}$ & $1.62\times10^{-2}$ & $1.53\times10^{-3}$\\
      1000  & 1.0 & 0.1 & $3\pi$ & $\pi$      & (64,\SM{128},256) & $3.97\times10^{-5}$ & $1.58\times10^{-2}$ & $1.50\times10^{-3}$\\
\\
      1000  & 1.0 & 0.1 & $2\pi$ & $0.5\pi$ & (32,\SM{128},256) & $3.39\times10^{-5}$ & $1.63\times10^{-2}$ & $1.56\times10^{-3}$\\
      1000  & 1.0 & 0.1 & $2\pi$ & $1.5\pi$ & (64,\SM{128},256) & $2.65\times10^{-5}$ & $1.60\times10^{-2}$ & $1.55\times10^{-3}$\\
      1000  & 1.0 & 0.1 & $2\pi$ & $2\pi$    & (64,\SM{128},512) & $1.24\times10^{-5}$ & $1.60\times10^{-2}$ & $1.58\times10^{-3}$\\
  \end{tabular}
  \caption{Dependence on
  computational domain size}
  \label{table2}
  \end{center}
\end{table}

\subsection{\GK{Effects} of weight $\lambda$ on optimal state} 
\begin{figure} 
\centering
	\begin{minipage}{.75\linewidth}
	\includegraphics[clip,width=\linewidth]{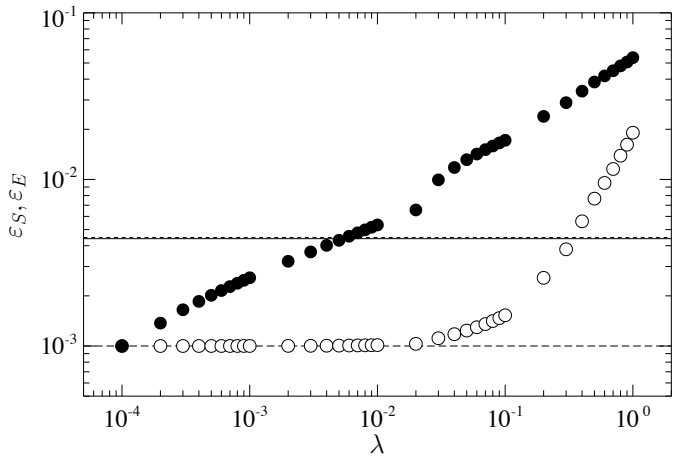}
	\end{minipage}

\caption[]{Total scalar and energy dissipation as a function of the weight $\lambda$ for the optimal state at $Re=1000$
\GKK{for $Pr=1$}:
\solidcircle, the total scalar dissipation $\varepsilon_{S}$; \circle, the energy dissipation $\varepsilon_{E}$.
\SM{The long-dashed line indicates $\varepsilon_{S}$ or $\varepsilon_{E}$
\GKK{in} a laminar state, $1/Re=10^{-3}$.}
The solid and \SM{short-}dashed line respectively represent the time-averaged $\varepsilon_{S}$ and $\varepsilon_{E}$ 
\GKK{in}
\GK{turbulent flow at $Re=1000$
for $Pr=1$.}
\label{fig3}}
\end{figure}

We next consider the effects on the optimal state of the weight
$\lambda$ of the scalar dissipation
in the objective functional $J$.

\SM{
\GKK{Let us first note that
in the special case
of a streamwise-independent two-dimensional
velocity field
the weight $\lambda$ can be removed
from the Euler--Lagrange equations
(\ref{eq2-27})--(\ref{eq2-30})
as follows.}
Taking the curl of equation (\ref{eq2-27}), we have
\begin{eqnarray}
\label{eq3-31}
\displaystyle
\frac{1}{\lambda Re}\frac{\partial^{2}\omega_{x}}{\partial x_{j}^{2}}+\frac{\partial(\theta^{*},\theta)}{\partial(y,z)}+\frac{\partial\theta^{*}}{\partial z}=0,
\end{eqnarray}
where 
\begin{eqnarray}
\label{eq3-32}
\displaystyle
\frac{\partial(\beta,\gamma)}{\partial(y,z)}=\frac{\partial\beta}{\partial y}\frac{\partial\gamma}{\partial z}-\frac{\partial\beta}{\partial z}\frac{\partial\gamma}{\partial y}
\end{eqnarray}
is the Jacobian determinant.
\GKK{The streamwise-independent velocity field can be
expressed, in terms of the streamfunction $\psi(y,z)$, as}
\begin{eqnarray}
\label{eq3-33}
\displaystyle
(u,v,w)=\left( y,\frac{\partial \psi}{\partial z},-\frac{\partial \psi}{\partial y} \right),
\end{eqnarray}
where the streamwise velocity $u$
\GKK{has been}
uniquely determined
from the $x$-component of equation (\ref{eq2-27}) and
\GKK{its independence of} the $x$-direction ($\partial\GKK{u}/\partial x=0$).
\GKK{The streamfunction is related to the streamwise vorticity via the}
\GKK{Poisson} equation
\begin{eqnarray}
\label{eq3-34}
\displaystyle
\frac{\partial^{2}\psi}{\partial x_{j}^{2}}=-\omega_{x}.
\end{eqnarray}
Using $\psi$, equations (\ref{eq2-28}) and (\ref{eq2-29}) are respectively rewritten as
\begin{eqnarray}
\label{eq3-35}
&\displaystyle
RePr\left[ -\frac{\partial(\psi,\theta^{*})}{\partial(y,z)} \right]+\frac{\partial^{2}}{\partial x_{j}^{2}}(\theta^{*}+\theta)=0,& \\
\label{eq3-36}
&\displaystyle
RePr\left[-\frac{\partial(\psi,\theta)}{\partial(y,z)}+\frac{\partial \psi}{\partial z} \right]-\frac{\partial^{2}\theta}{\partial x_{j}^{2}}=0.&
\end{eqnarray}
The boundary conditions of $\psi$ are
\begin{eqnarray}
\label{eq3-37}
\displaystyle
\GKK{\psi(y=\pm1)}=
\frac{\partial \psi}{\partial y}(y=\pm1)=0.
\end{eqnarray}
\GKK{Introducing the new variable 
and the new parameter} 
as
\begin{eqnarray}
\label{eq3-38}
\displaystyle
\psi^{*}=\frac{\psi}{\sqrt{\lambda}},\hspace{1em}Re^{*}=\sqrt{\lambda}Re
\end{eqnarray}
and using equation (\ref{eq3-34}), the Euler--Lagrange equations
\GKK{can be rewritten in the streamwise-independent case as}
\begin{eqnarray}
\label{eq3-39}
&\displaystyle
-\frac{1}{Re^{*}}\frac{\partial^{2}}{\partial x_{j}^{2}}\frac{\partial^{2} \psi^{*}}{\partial x_{j}^{2}}+\frac{\partial(\theta^{*},\theta)}{\partial(y,z)}+\frac{\partial\theta^{*}}{\partial z}=0,& \\
\label{eq3-40}
&\displaystyle
Re^{*}Pr\left[ -\frac{\partial(\psi^{*},\theta^{*})}{\partial(y,z)} \right]+\frac{\partial^{2}}{\partial x_{j}^{2}}(\theta^{*}+\theta)=0,& \\
\label{eq3-41}
&\displaystyle
Re^{*}Pr\left[-\frac{\partial(\psi^{*},\theta)}{\partial(y,z)}+\frac{\partial \psi^{*}}{\partial z} \right]-\frac{\partial^{2}\theta}{\partial x_{j}^{2}}=0,&
\end{eqnarray}
\GKK{which are \SM{supplemented} by
the boundary conditions of $\psi^{*}$,}
\begin{eqnarray}
\label{eq3-42}
\displaystyle
\frac{\partial \psi^{*}}{\partial z}(y=\pm1)=\frac{\partial \psi^{*}}{\partial y}(y=\pm1)=0.
\end{eqnarray}
}

Figure \ref{fig3} shows the total scalar dissipation $\varepsilon_{S}$ and the energy dissipation $\varepsilon_{E}$ as a function of $\lambda$ for the
optimal \GKK{state} at $Re=1000$, \GKK{which will turn out to
be three-dimensional \GKK{for $\lambda=0.1$} in \S~\ref{sec4-1}}.
Note that
\GKK{in the case of} $\lambda=0$, \GK{a laminar state
\GKK{indicated by the long-dashed line}
is optimal
since laminar
flow gives lower bound for the energy dissipation of an arbitrary velocity field,
that is,}
\begin{eqnarray}
\varepsilon_{E}=\frac{1}{Re}{\left< {|\nabla\mbox{\boldmath$u$}|}^{2} \right>}_{xyz}\ge\frac{1}{Re}.
\end{eqnarray}
The weight $\lambda$ is a factor to control the contribution of the scalar dissipation
\GK{against} the energy dissipation
\GK{in} the \GK{optimisation} of $J$.
Decreasing $\lambda$ means that the weight of the energy dissipation is increased relative to the scalar dissipation.
As a result,
\GK{an} optimal state has the smaller energy dissipation (in other words, the smaller enstrophy).
\GK{For} large $\lambda$, the significantly high scalar dissipation (i.e., high heat flux) is achieved with the larger energy dissipation.
In light of
heat transfer enhancement, we should find an optimal state with a higher heat flux and
smaller energy dissipation than 
uncontrolled turbulent flow.
In
plane Couette flow at $Re=1000$, 
\GKK{a} sustained turbulent state \GKK{can be observed}.
In \SMM{figure} \ref{fig3}, the solid and \SM{short-}dashed line respectively represent the time-averaged scalar and energy dissipation of
turbulent flow at \GK{$Re=1000$
for $Pr=1$}.
Their consistency is attributed to the similarity between 
\GK{turbulent} heat and momentum transfer.
In the range of $\SM{0.02}<\lambda<0.4$, we can find the optimal states of higher heat transfer and lower energy dissipation than the turbulent state.
\GKK{In this range of the weight $\lambda$}
we
\GK{have confirmed} that the optimised velocity fields do not
\GK{exhibit} significantly different flow structures depending on $\lambda$.

In the following sections we shall present the optimal states in the domain
$(L_{x},L_{y},L_{z})=(2\pi,2,\pi)$ \GK{for} $Pr=1$ and $\lambda=0.1$.
The numerical solutions have been obtained on grid points
$(N_{x},N_{y},N_{z})=(32,\SM{128},256)$ \GK{at} $Re=30-900$ and 
$(N_{x},N_{y},N_{z})=(32,\SM{256},512)$ \GK{at} $Re=1000-10000$.

\section{Characteristics of optimal state\SMM{s}}\label{sec4} 
\subsection{Flow and temperature field}\label{sec4-1} 
\begin{figure} 
\centering
	\begin{minipage}{.75\linewidth}
	\includegraphics[clip,width=\linewidth]{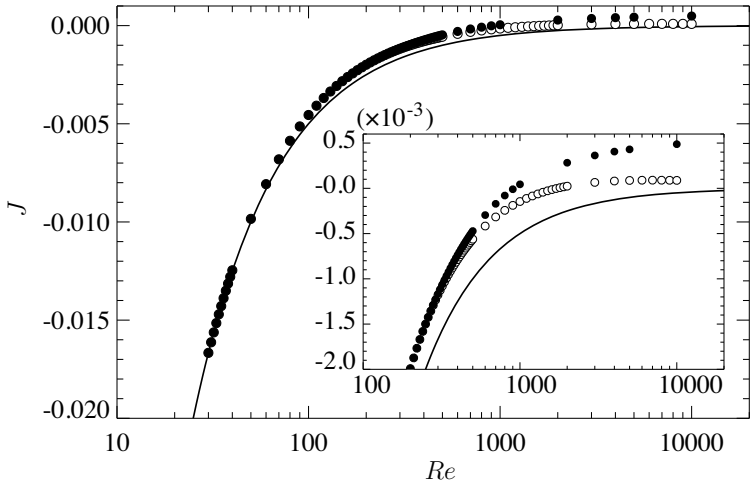}
	\end{minipage}

\caption{\GGK{The maximal value of} $J$ as a function of $Re$ for $\lambda=0.1$.
Filled circles represent \GGK{the value of} $J$ \GGK{for} the optimal states, and open circles denote \GGK{maximised} $J$ \GGK{only within} a streamwise-independent two-dimensional velocity field.
The solid curve represents \GGK{the value of} $J$ \GGK{for} the laminar solution, $J=-1/(2Re)$.
\label{fig4}}
\end{figure}

\begin{figure} 
\centering
        \begin{minipage}{.45\linewidth}
	(\textit{a})\\
	\includegraphics[clip,width=\linewidth]{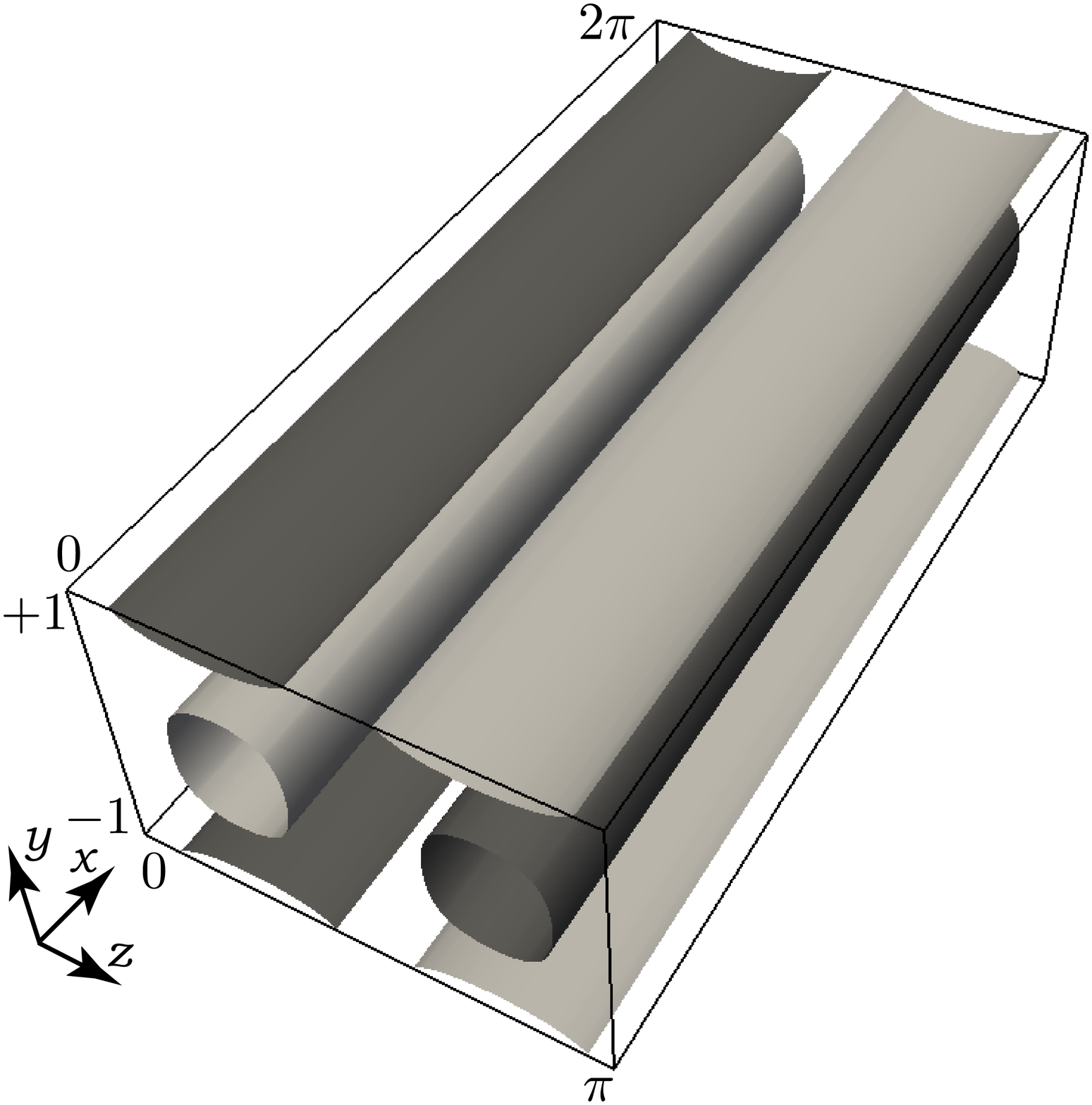}
	\end{minipage}
	\hspace{1em}
	\begin{minipage}{.45\linewidth}
	(\textit{b})\\
	\includegraphics[clip,width=\linewidth]{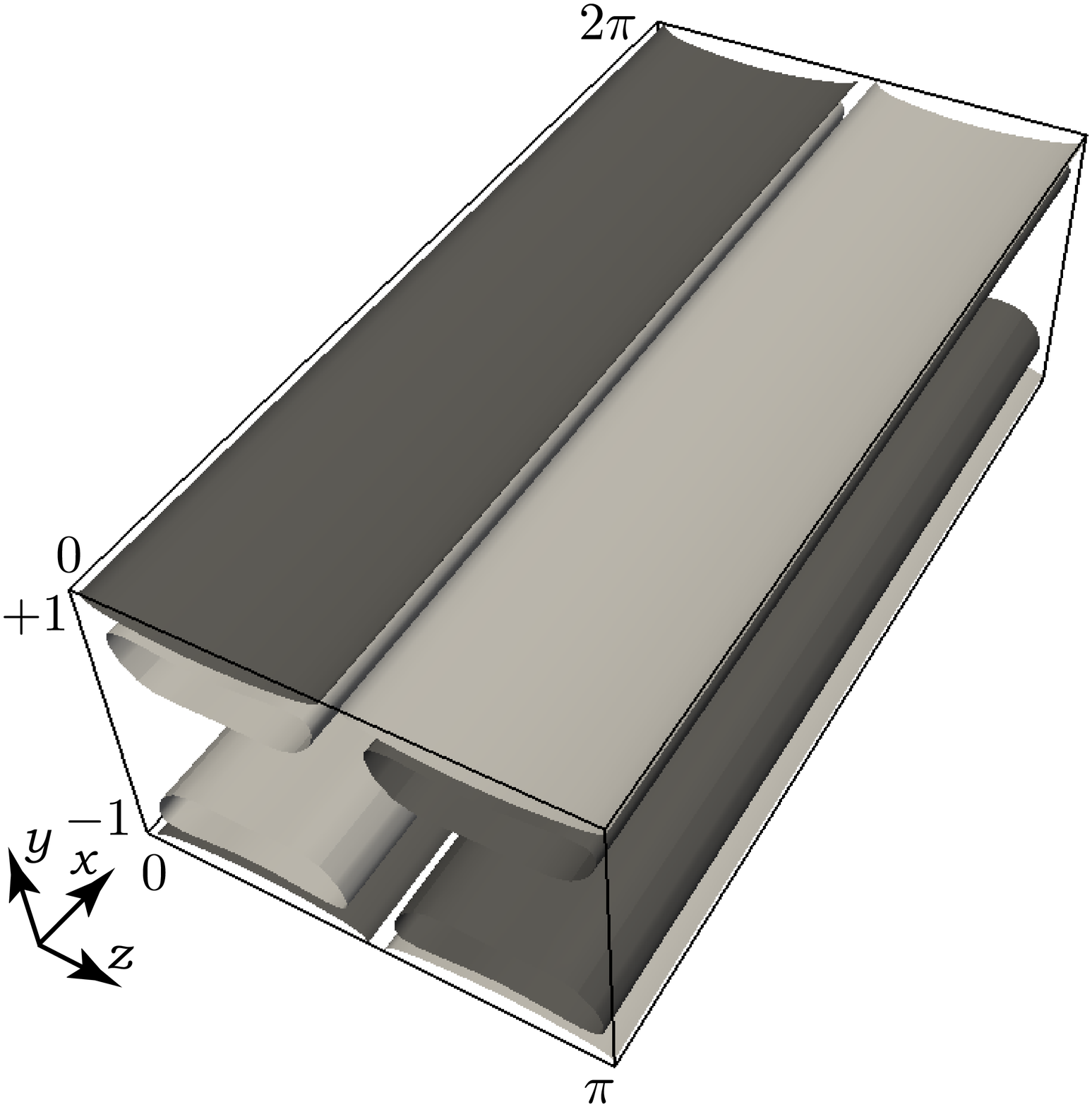}
	\end{minipage}

	\vspace{1em}

	\begin{minipage}{.45\linewidth}
	(\textit{c})\\
	\includegraphics[clip,width=\linewidth]{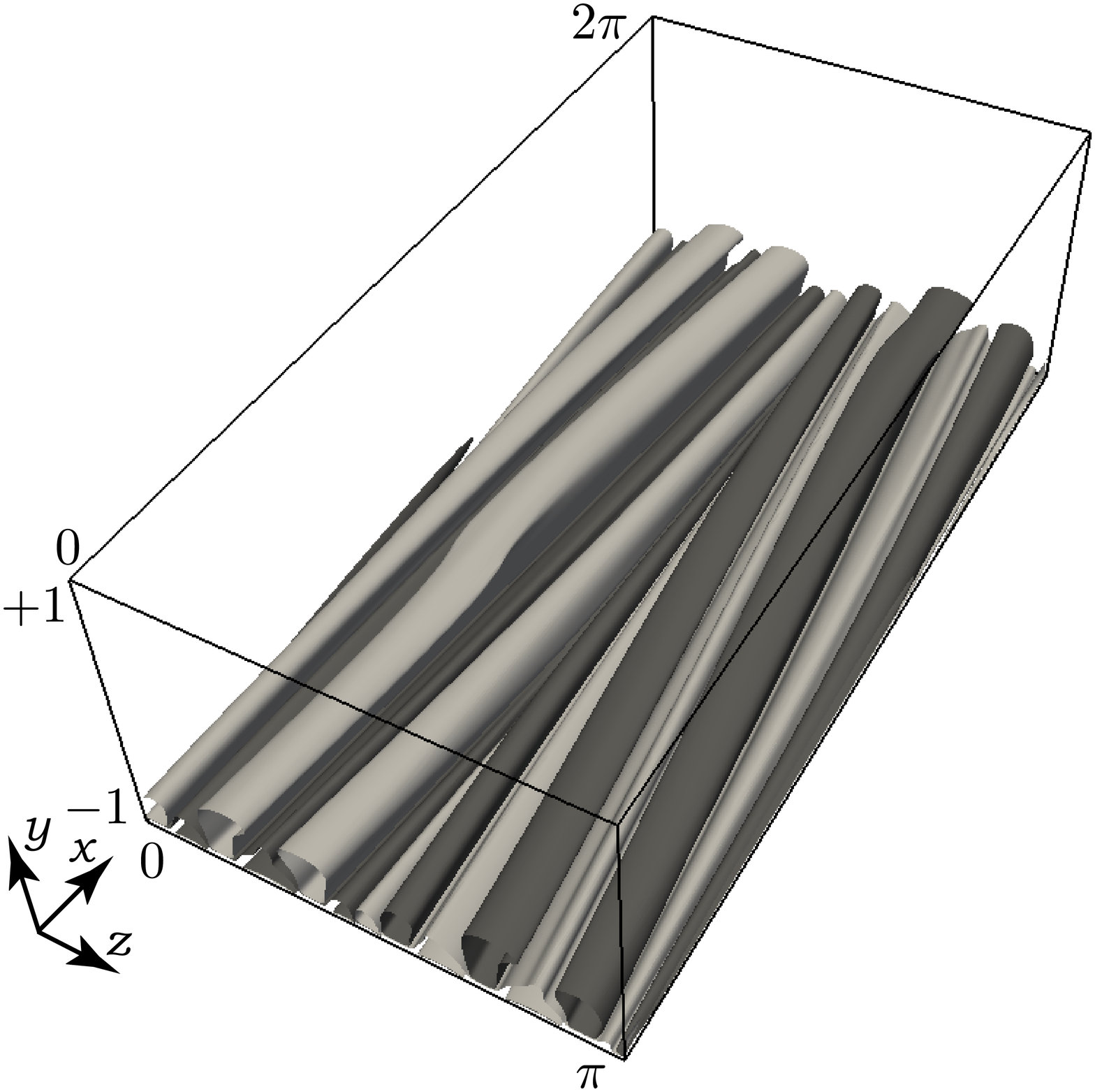}
	\end{minipage}
	\hspace{1em}
	\begin{minipage}{.45\linewidth}
	(\textit{d})\\
	\includegraphics[clip,width=\linewidth]{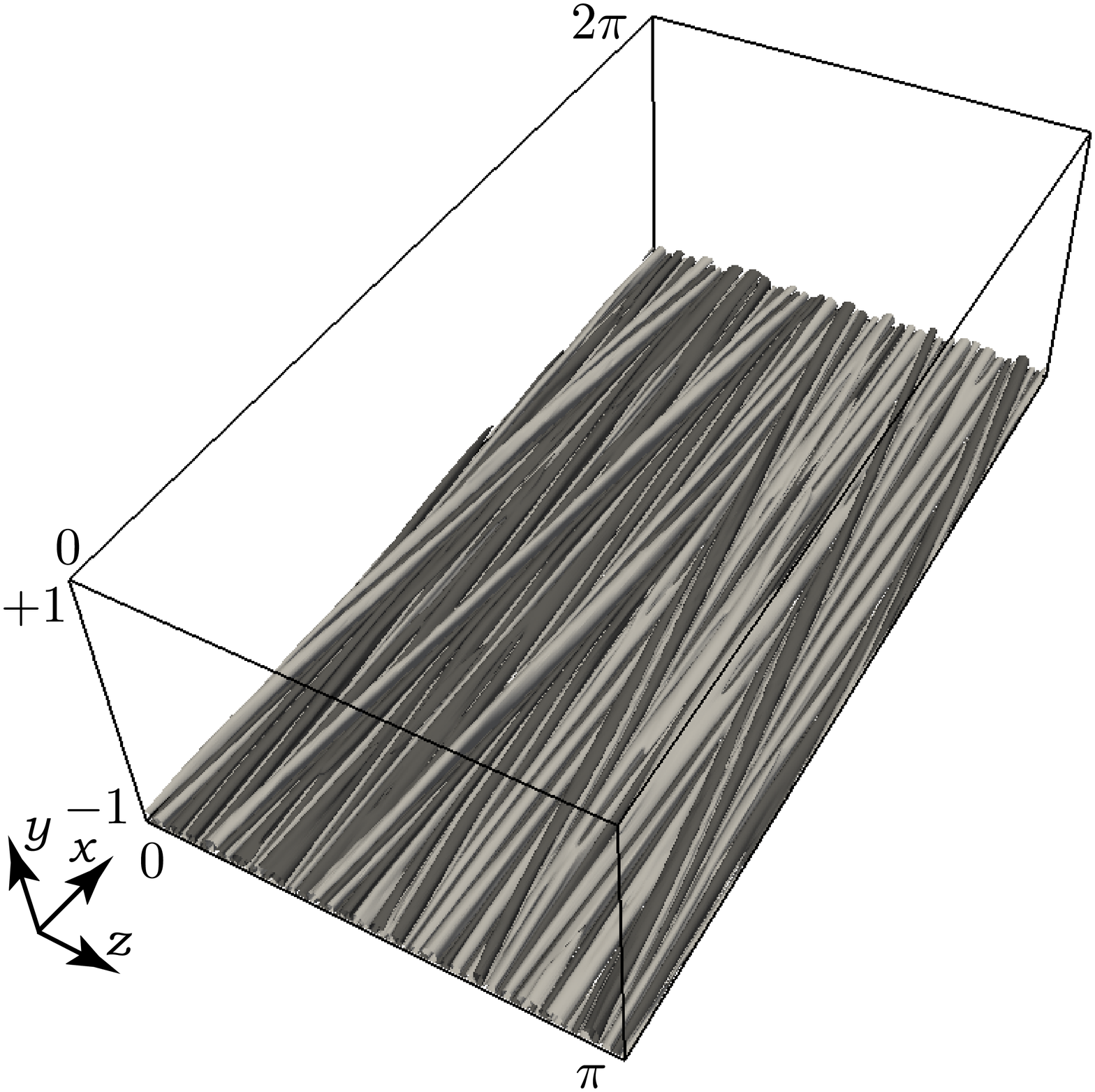}
	\end{minipage}

\caption{Isosurfaces of the streamwise vorticity in the optimal states at (\textit{a}) $Re=50$, (\textit{b}) $Re=200$, (\textit{c}) $Re=1000$ and (\textit{d}) $Re=5000$, for $\lambda=0.1$.
\GGK{In}
(\textit{c}) and (\textit{d}), only \GGK{the isosurfaces} in the lower half of the channel ($y<0$) are shown for visualization of the small-scale vortex structures near the lower wall.
Light \GKK{(or dark)} \GKK{grey} objects represent 
\GGK{isosurfaces} of (\textit{a}) $\omega_{x}=+0.25$ \GKK{(or $-0.25$)}, (\textit{b}) $\omega_{x}=+0.25$ \GKK{(or $-0.25$)}, (\textit{c}) $\omega_{x}=+0.5$ \GKK{(or $-0.5$)}, (\textit{d}) $\omega_{x}=+2$ \GKK{(or $-2$)}.
\label{fig5}}
\end{figure}

\begin{figure} 
\centering
        \begin{minipage}{.48\linewidth}
        	(\textit{a})\\
	\includegraphics[clip,width=\linewidth]{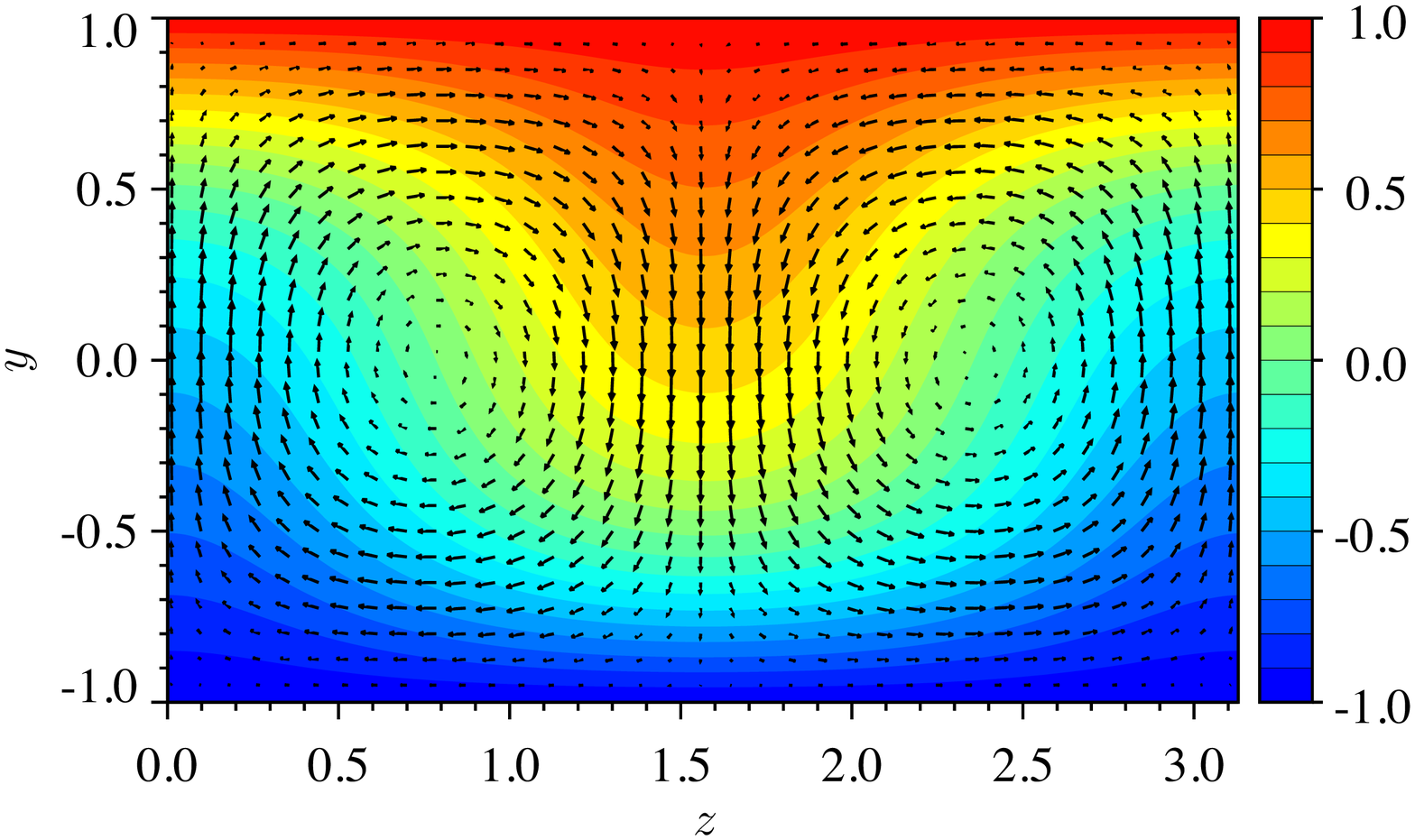}
	\end{minipage}
	\hspace{1em}
	\begin{minipage}{.48\linewidth}
	(\textit{b})\\
	\includegraphics[clip,width=\linewidth]{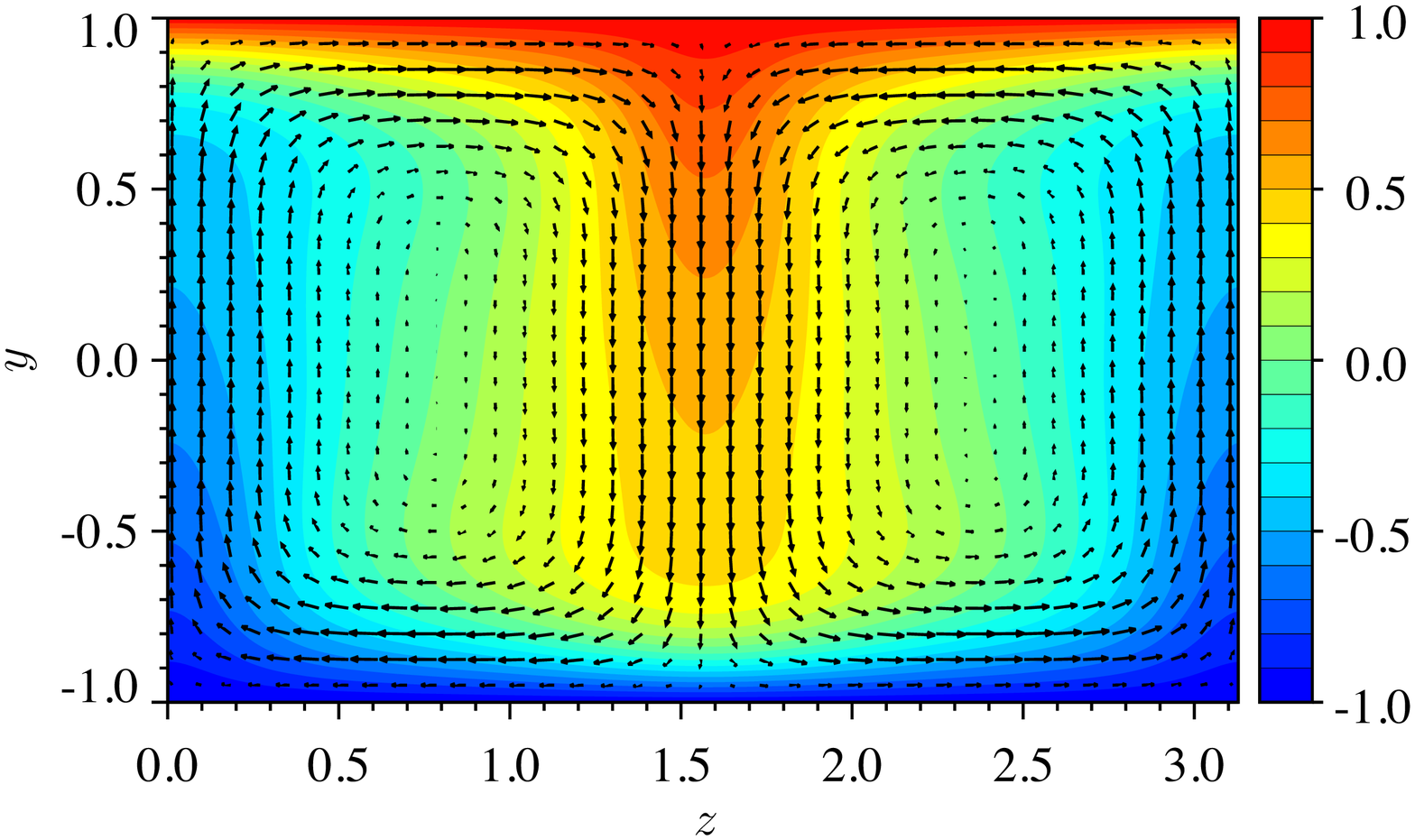}
	\end{minipage}

	\vspace{1em}
	
	\begin{minipage}{.48\linewidth}
	(\textit{c})\\
	\includegraphics[clip,width=\linewidth]{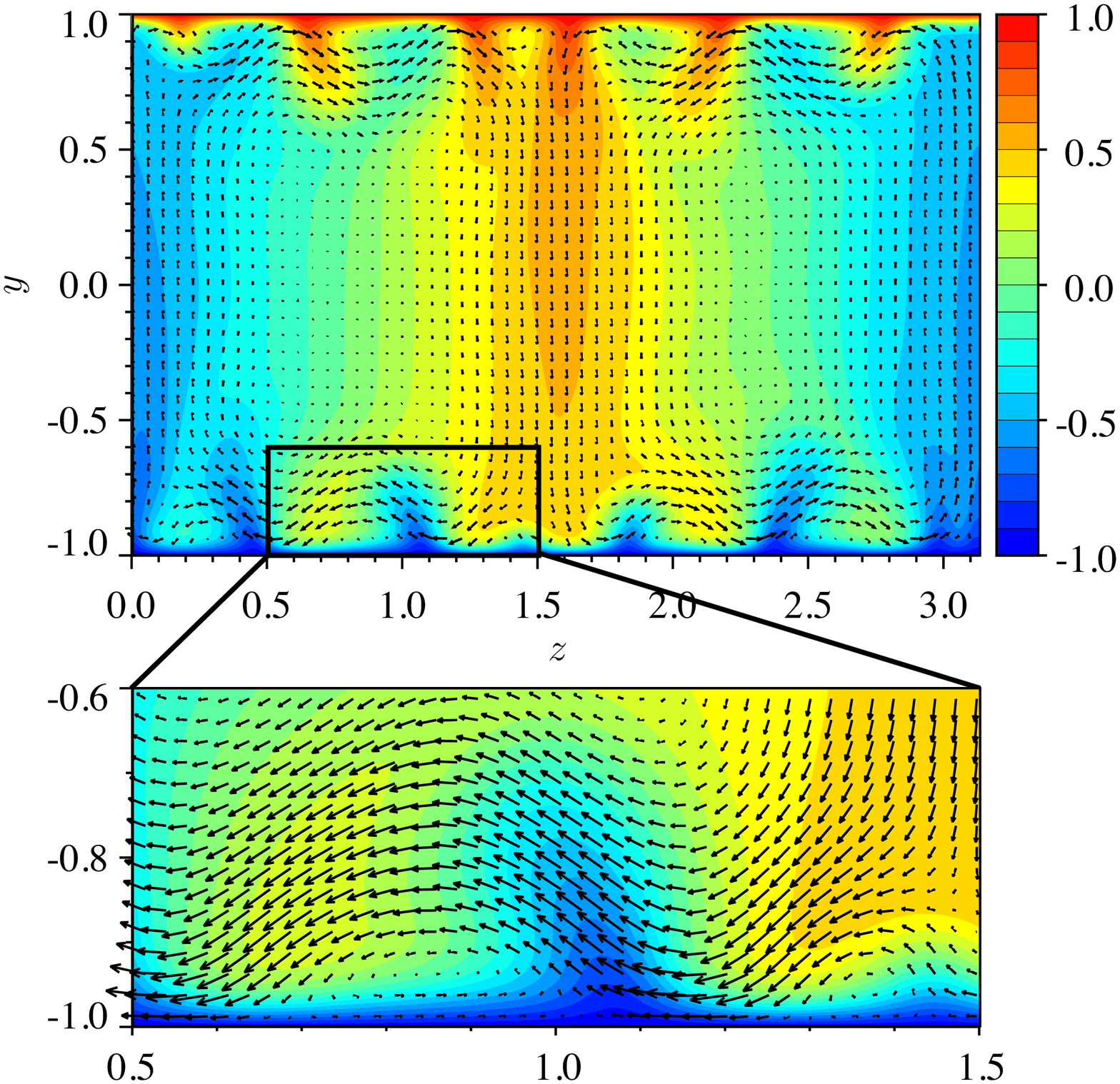}
	\end{minipage}
	\hspace{1em}
	\begin{minipage}{.48\linewidth}
	(\textit{d})\\
	\includegraphics[clip,width=\linewidth]{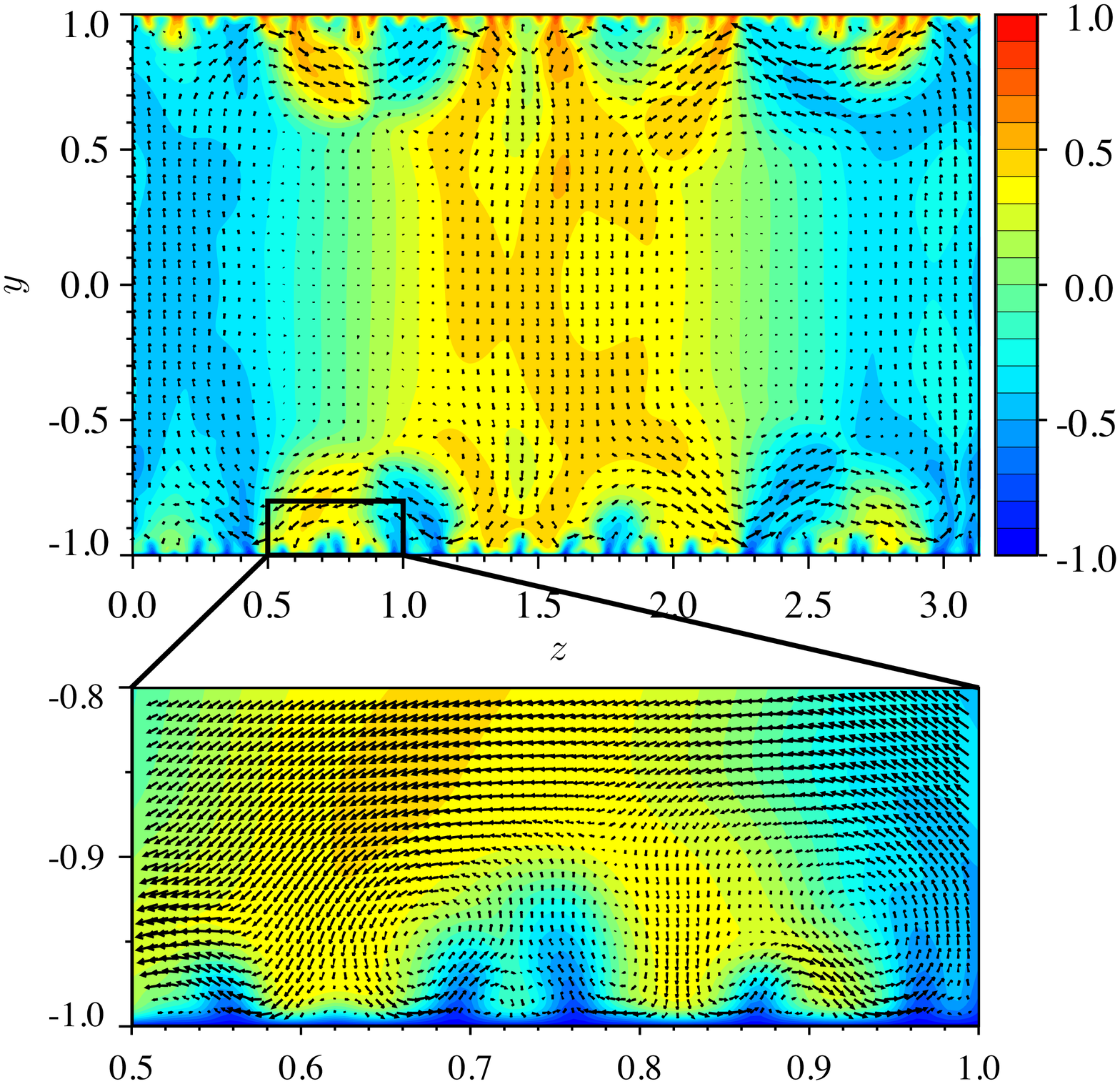}
	\end{minipage}

	\vspace{1em}

	\begin{minipage}{.48\linewidth}
	(\textit{e})\\
	\includegraphics[clip,width=\linewidth]{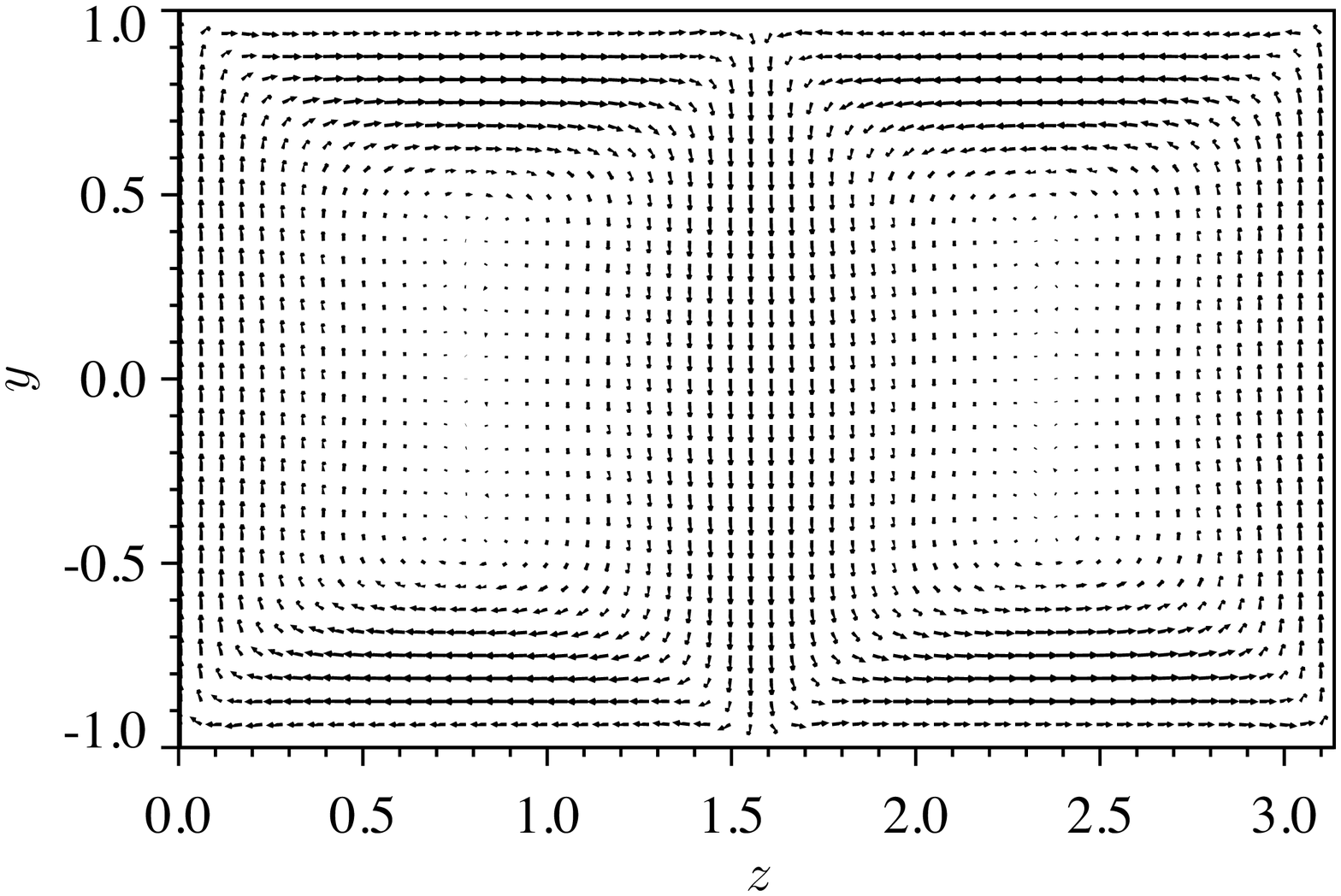}
	\end{minipage}
	\hspace{1em}
	\begin{minipage}{.48\linewidth}
	(\textit{f})\\
	\includegraphics[clip,width=\linewidth]{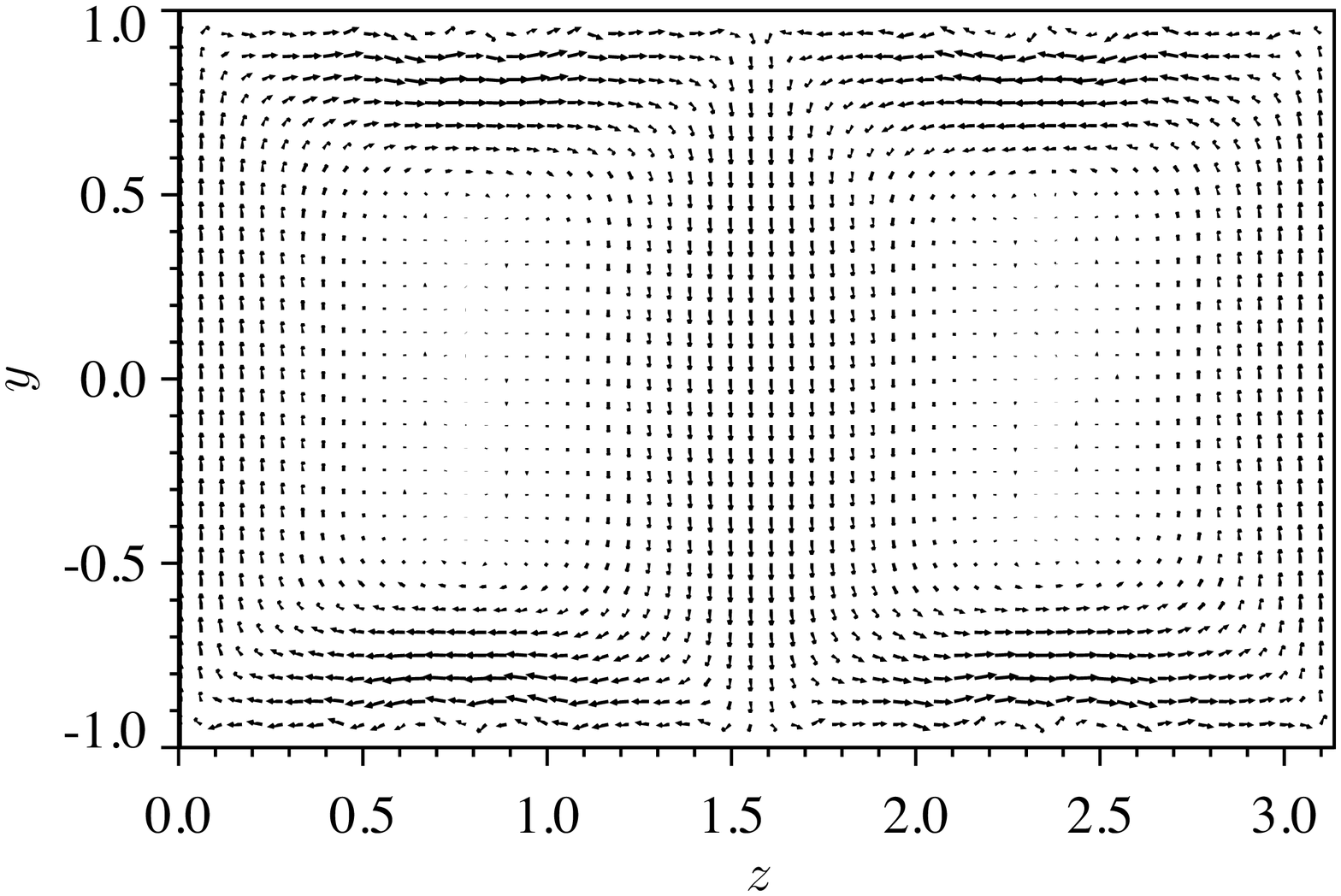}
	\end{minipage}

\caption{(Colour online) Velocity and temperature fields
\GGK{in} the optimal state at (\textit{a}) $Re=50$, (\textit{b}) $Re=200$, (\textit{c,e}) $Re=1000$ and (\textit{d,f}) $Re=5000$, for $\lambda=0.1$.
(\textit{a,b,c,d}) The cross-\GKK{stream} velocity vectors ($w,v$) and contours of the temperature field $T$ in the plane $x=0$.
In (\textit{c}) and (\textit{d}), the enlarged views of the near-wall region are \SMM{also} shown.
(\textit{e,f}) The streamwise-averaged cross-\GKK{stream} velocity vectors
\GGK{(${\left< w \right>}_x,{\left< v \right>}_x$)}.
\label{fig6}}
\end{figure}

\begin{figure} 
\centering
        \begin{minipage}{.65\linewidth}
       	(\textit{a})\\
	\includegraphics[clip,width=\linewidth]{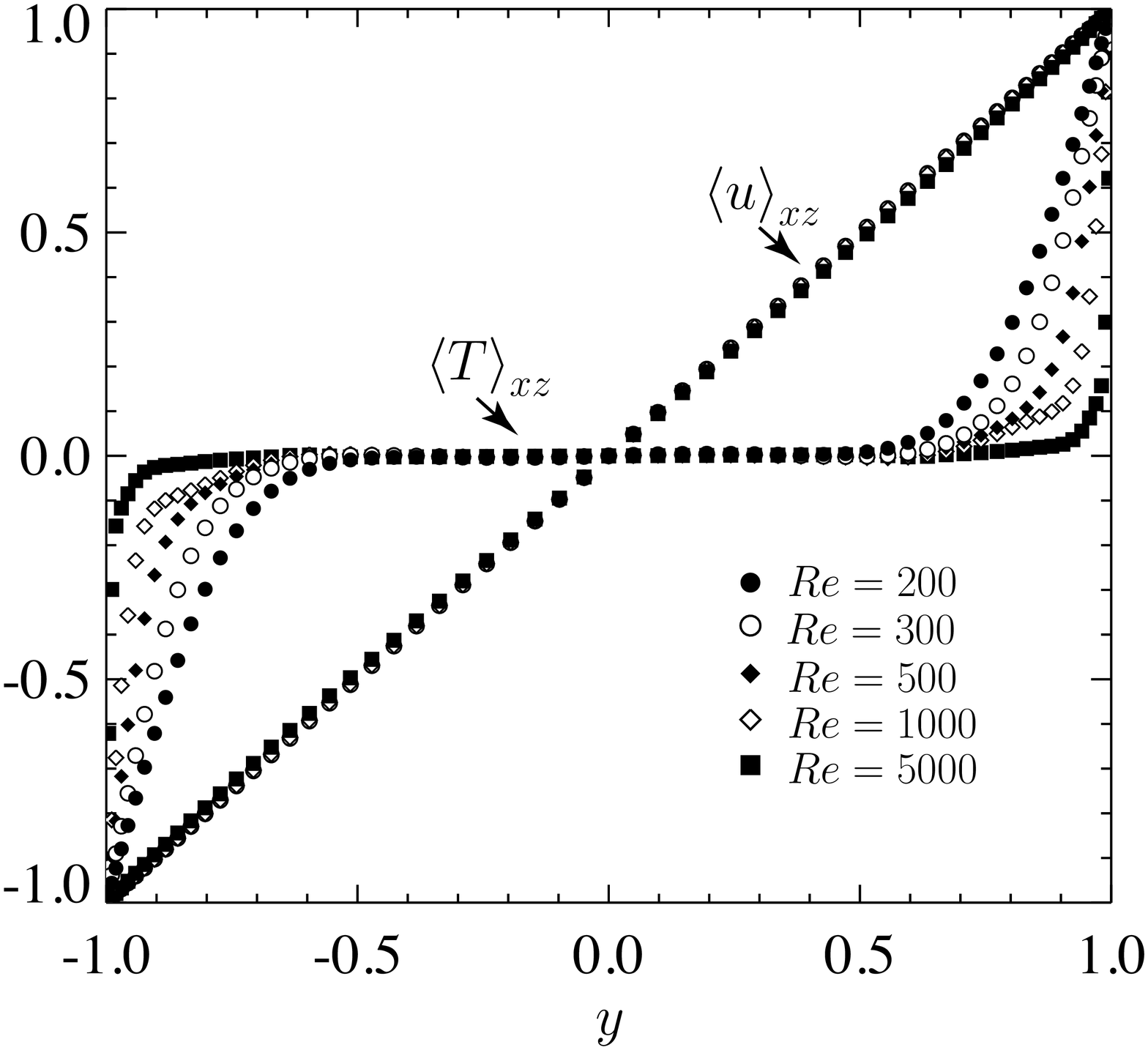}
	\end{minipage}

	\vspace{1em}

	\begin{minipage}{.65\linewidth}
	(\textit{b})\\
	\includegraphics[clip,width=\linewidth]{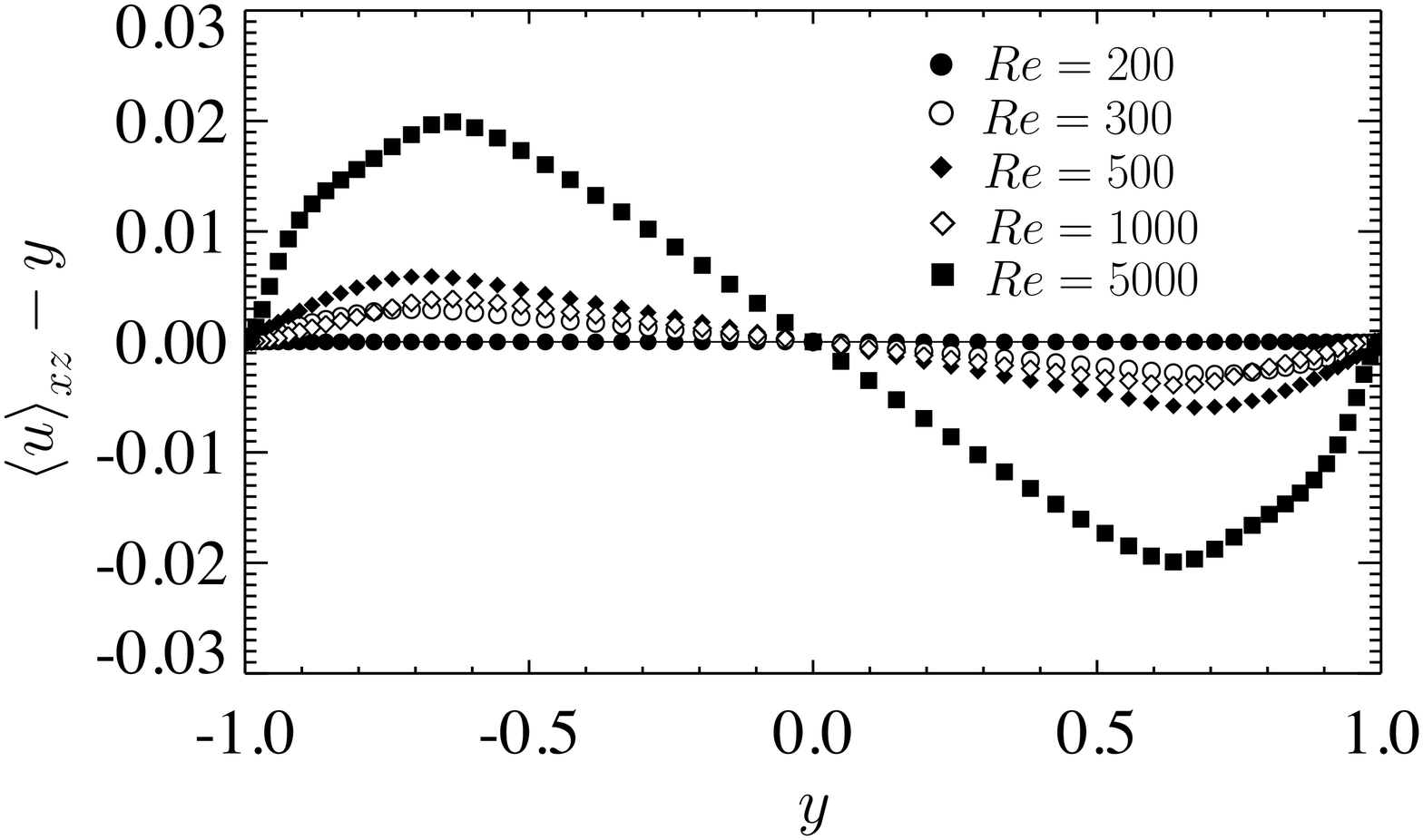}
	\end{minipage}

\caption{Mean temperature and velocity profiles
\GGK{in} the optimal state for $\lambda=0.1$.
(\textit{a}) The mean temperature ${\left< T \right>}_{xz}$ and the mean streamwise velocity ${\left< u \right>}_{xz}$.
(\textit{b}) The 
\GGK{deviation} of \SMM{the} mean streamwise velocity
\GGK{from} the laminar flow, ${\left< u \right>}_{xz}-y$.
\label{fig7}}
\end{figure}

Figure \ref{fig4} shows the maximal \GK{value of}
$J$ as a function of $Re$ for $\lambda=0.1$.
The
\GK{maximised}
$J$
\GK{only within}
\GK{a}
streamwise-independent two-dimensional velocity field is also plotted
\GK{for comparison purposes}.
The solid curve represents 
\GK{the value of}
$J$ for the laminar solution $\mbox{\boldmath$u$}=y\mbox{\boldmath$\rm e$}_{x}$, which is the lower bound of $J$.
At
\GK{low values of} $Re$, the optimal state exhibits \GK{a}
streamwise-independent two-dimensional velocity field
as
\GK{seen} in \SMM{figure} \ref{fig5},
\GK{where}
the isosurfaces of the positive and negative streamwise vorticity $\omega_{x}$
\GK{are shown for} the optimal states at $Re=50$, $200$, $1000$ and $5000$.
At \GK{the}
higher \GK{Reynolds numbers}
$Re \GK{=1000,\,5000}$,
\GK{streamwise-dependent three-dimensional structures,
\GK{i.e. spanwise-inclined vortex tubes},
with stronger $\omega_{x}$ appear near the
wall}.
\GK{We shall discuss the onset of the spanwise inclination
of the structures and its effects on heat transfer in \S~\ref{sec6}.}
In
\SMM{figure} \ref{fig5}(\textit{c,d}),
only the isosurfaces in the lower half of the channel ($y<0$)
\GK{are} shown for visualization of the \GK{near-wall}
vortices.
\GK{At} $Re \GK{=5000}$
a \GKK{large} number of
smaller vortex structures with \GK{much} stronger $\omega_{x}$
\GK{are observed in the close vicinity of
the wall}.
\GK{Note that such small-scale intense vortices do not
emerge near the wall
in the two-dimensional optimal velocity fields (\SMM{figure}~\ref{fig5}\textit{a,b}).}
Although the formation of the small-scale
\GK{structures} increases the energy dissipation, it also significant\SMM{ly} enhance\SMM{s} the scalar dissipation.
As a result of these counteractions, the three-dimensional 
\GK{fields could be an}
optimal state
at higher $Re$.
Even when the domain size or the weight $\lambda$ is changed, we have obtained the optimal states which represent qualitatively similar three-dimensional structures.

Figure \ref{fig6}(\textit{a-d}) shows the cross-stream velocity field $(w,v)$ and the temperature field $T$ in the optimal states at $Re=50$, $200$, $1000$ and $5000$.
The two-dimensional optimal states at $Re=50,200$
\GK{(\SMM{figure}~\ref{fig6}\textit{a,b})}
consist of the two large-scale circulation rolls which
\GK{are reminiscent of thermal convection and}
play a role in heat transfer from the upper wall to the lower wall.
Near the walls, the regions with the high temperature gradient are formed by the \GK{effects of the} circulation rolls, and
\GK{the gradient gets higher} as $Re$ increases.
Meanwhile, the three-dimensional optimal state at $Re=1000,5000$
\GK{(\SMM{figure}~\ref{fig6}\textit{c,d})} exhibit
the more complicated flow and temperature field.
Similar to the two-dimensional states, there are the large-scale circulation rolls with the
\GK{width} of
\GK{half} the spanwise period $L_{z}$ \GKK{($=\pi$)}.
\GK{The wall-normal component of the velocity is
dominant in between the circulation rolls.}
Small-scale vortices are observed only in the near-wall regions.
They produce the regions with high temperature gradient by pushing the hotter (or colder) fluid onto the \SM{lower (or upper)} wall.
\GK{In \SMM{figure} \ref{fig6}(\textit{e,f}) \SMM{are shown} the
streamwise-averaged cross-stream velocity fields at $Re=1000$ and $5000$.}
\GK{Even in the averaged fields, the effects of the small-scale vortices can be seen close to the wall at $Re=5000$.}

The mean temperature and the mean streamwise velocity profiles of the optimal states are shown in \SMM{figure} \ref{fig7}(\textit{a}).
As $Re$ increases, the mean temperature \GK{profile}
is more flattened in the central region of the channel,
while the temperature gradient at the walls is higher.
On the other hand, the mean velocity profile is
\GK{almost} consistent with a laminar profile.
For the streamwise-independent case
\GK{(i.e., $\partial(\cdot)/\partial x=0$)}
\GKK{seen in} the optimal state at $Re=200$,
the streamwise velocity $u$ is 
\GK{strictly consistent with the laminar profile $u=y$,
which is a consequence of the $x$-component of the Euler--Lagrange equation (\ref{eq2-27}).}
At large $Re$, the three-dimensional vortex structure \GK{has appeared to}
\SMM{bring about} the deviation from the laminar profile, shown in \SMM{figure} \ref{fig7}(\textit{b}),
\GK{although the \SM{deviation} is small even at $Re=5000$.}

\subsection{Wall heat flux and energy dissipation} 
\begin{figure} 
\centering
	\begin{minipage}{.7\linewidth}
	(\textit{a})\\
	\includegraphics[clip,width=\linewidth]{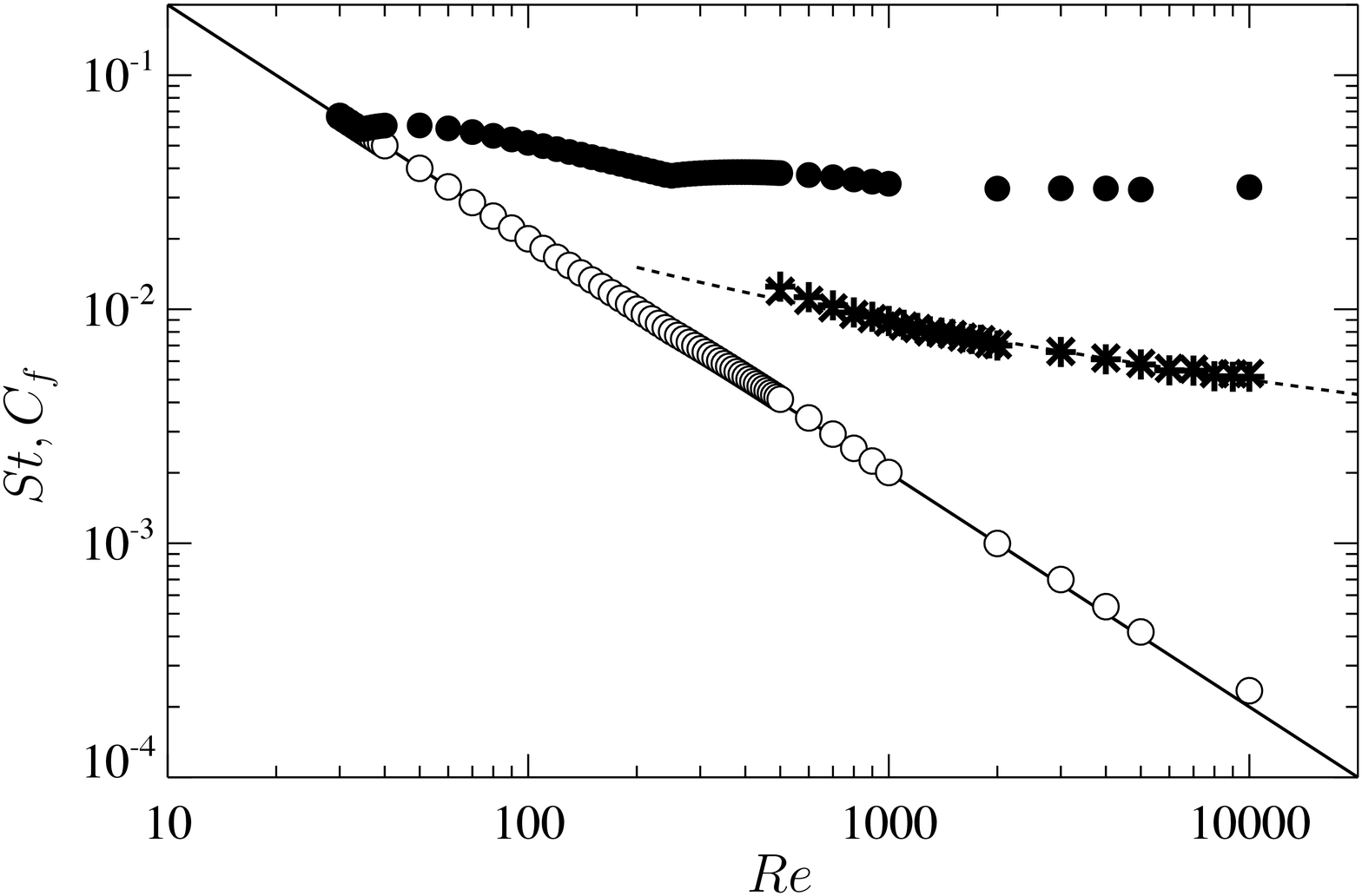}
	\end{minipage}

	\vspace{1em}

	\begin{minipage}{.7\linewidth}
	(\textit{b})\\
	\includegraphics[clip,width=\linewidth]{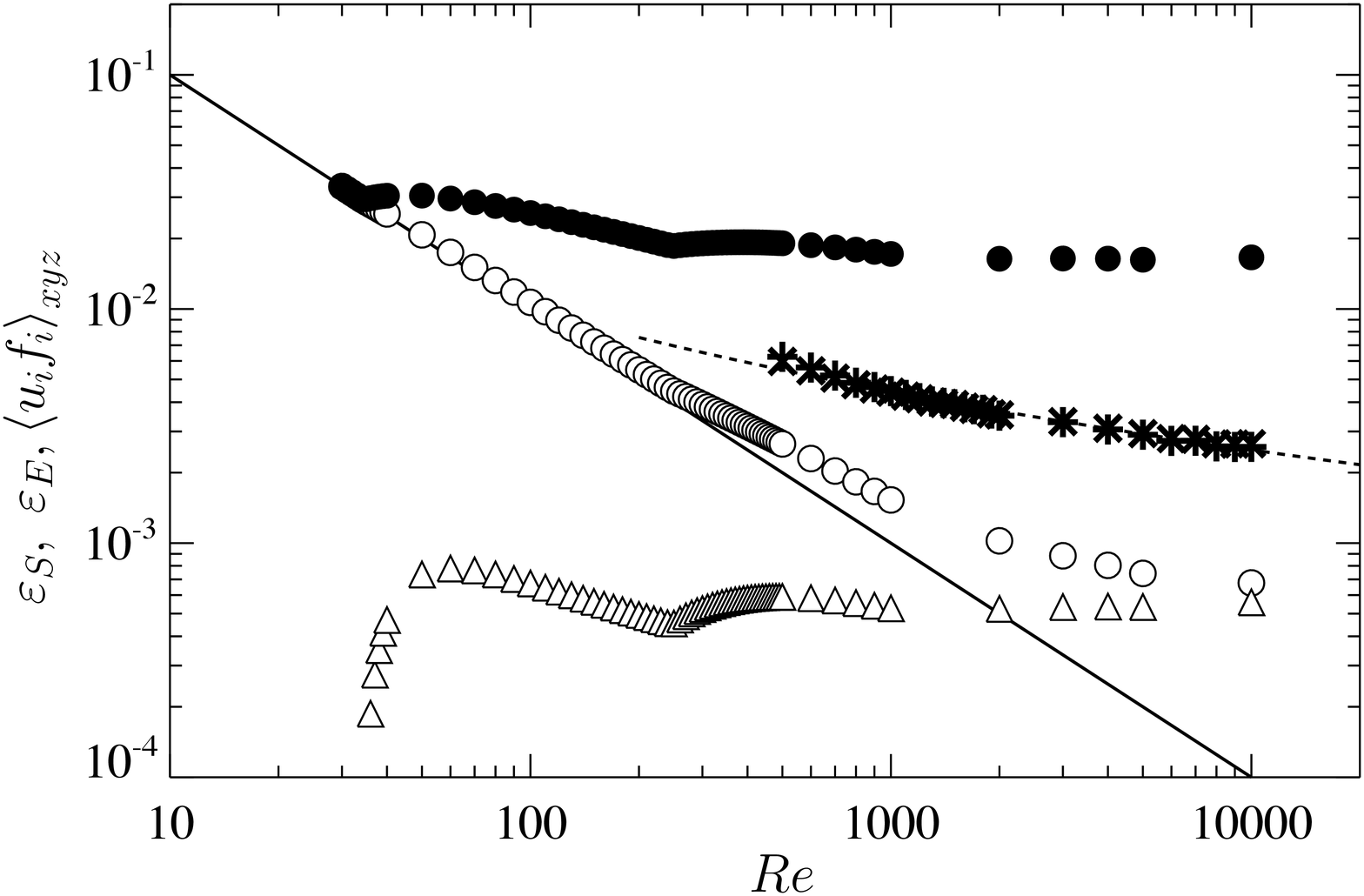}
	\end{minipage}

\caption[]{(\textit{a}) Stanton number $St$ and friction coefficient $C_{f}$
\GGK{as a function of $Re$}.
The optimal state for $\lambda=0.1$: \fullcirc, $St$; \opencirc, $C_{f}$.
The turbulent flow: $\times$, \GGK{$St$}; $+$, $C_{f}$.
The solid line and dashed curve represent $C_{f}=2/Re$ for the laminar flow and the empirical formula $C_{f}=2(0.2/\log(Re))^{2}$ for the turbulent flow, respectively.
(\textit{b}) Total scalar dissipation $\varepsilon_{S}$, energy dissipation $\varepsilon_{E}$ and energy input ${\left< u_{i}f_{i} \right>}_{xyz}$.
The optimal state for $\lambda=0.1$: \fullcirc, $\varepsilon_{S}$; \opencirc, $\varepsilon_{E}$; \opentriangle, ${\left< u_{i}f_{i} \right>}_{xyz}$.
The turbulent flow: $\times$, $\varepsilon_{S}$; $+$, $\varepsilon_{E}$.
The solid line and dashed curve represent $1/Re$ and $C_{f}/2$ for the laminar flow and the turbulent flow, respectively.
\label{fig8}}
\end{figure}

\GK{Let us now consider}
the Stanton number $St$ and the friction coefficient $C_{f}$ which are respectively defined as
\begin{eqnarray}
\label{eq4-1}
St&\GKK{=}&\frac{1}{RePr}\left[ \frac{\rm d}{{\rm d}y}{\left< T \right>}_{xz}\left( y=+1 \right)+\frac{\rm d}{{\rm d}y}{\left< T \right>}_{xz}\left( y=-1 \right) \right], \\
\label{eq4-2}
C_{f}&\GKK{=}&\frac{1}{Re}\left[ \frac{\rm d}{{\rm d}y}{\left< u \right>}_{xz}(y=+1)+\frac{\rm d}{{\rm d}y}{\left< u \right>}_{xz}(y=-1) \right].
\end{eqnarray}
Note that the similarity between heat and momentum transfer
\GK{implies} that $St\approx C_{f}$ for $Pr=1$.
Figure \ref{fig8}(\textit{a}) shows $St$ and $C_{f}$
\GK{in} the optimal state for $\lambda=0.1$.
The solid line and the dashed curve represent the theoretical formula for laminar flow and the empirical formula \citep{Robertson1970} for turbulent flow, respectively.
The time-averaged $St$ and $C_{f}$
\GK{in} turbulent flow are also plotted
\GK{as a reference}.
In turbulent flow, $St$ and $C_{f}$ are almost the same due to the similarity between momentum and heat transfer.
\GK{In contrast},
\GK{in} the optimal state extremely high $St$ is achieved
\GK{in comparison with turbulence}
\GK{whilst} $C_{f}$ \GK{is comparably low to the laminar \SMM{flow},}
\GK{decreasing} as $Re^{-1}$ \GK{with increasing} $Re$.

The dimensionless versions of budget equations (\ref{eq2-15}) and (\ref{eq2-18}) are given respectively by
\begin{eqnarray}
\label{eq4-3}
\varepsilon_{S}&=&\frac{1}{2}St, \\
\label{eq4-4}
\varepsilon_{E}&=&\frac{1}{2}C_{f}+{\langle u_{i}f_{i} \rangle}_{xyz}.
\end{eqnarray}
The
optimal states \GK{found in this work}
require the external force $f_{i}$ to satisfy the
Navier--Stokes equation, since we \GK{have imposed} only \GK{the}
divergence-free condition on the velocity field.
The external force leads to the
\GK{additional} energy input ${\langle u_{i}f_{i} \rangle}_{xyz}$
\GK{which has appeared in equation (\ref{eq4-4}).}
Figure \ref{fig8}(\textit{b}) shows the total scalar dissipation $\varepsilon_{S}$, the energy dissipation $\varepsilon_{E}$ and the energy input ${\left< u_{i}f_{i} \right>}_{xyz}$.
Note that in the case of no external force, the energy budget implies $\varepsilon_{E}=C_{f}/2$.
The solid line represents the energy dissipation
\GK{in} the laminar flow, and provides
\GK{the} lower bound of $\varepsilon_{E}$.
As $Re$ increases, $\varepsilon_{E}$ for the optimal state gradually deviates from that of the laminar flow, because of the relatively greater contribution of the energy input ${\left< u_{i}f_{i} \right>}_{xyz}$.
The energy input is positive (or zero for a laminar solution) at any $Re$, but it is quite small in comparison to $\varepsilon_{E}$
\GK{in} the turbulent flow.
\GK{The} significant heat transfer enhancement
 with small energy dissipation 
\GK{can be confirmed}
for the optimal state (e.g. $\varepsilon_{S}/\varepsilon_{E}
\GKK{\approx} 24.6$ at $Re=\GKK{10000}$).
The spatial distribution of the external force and the local energy input are
\GK{depicted} in Appendix \ref{appA}.

\section{Hierarchical structure
\GK{in} optimal state}\label{sec5} 
\begin{figure} 
\centering
	\begin{minipage}{.48\linewidth}
	(\textit{a})\\
	\includegraphics[clip,width=\linewidth]{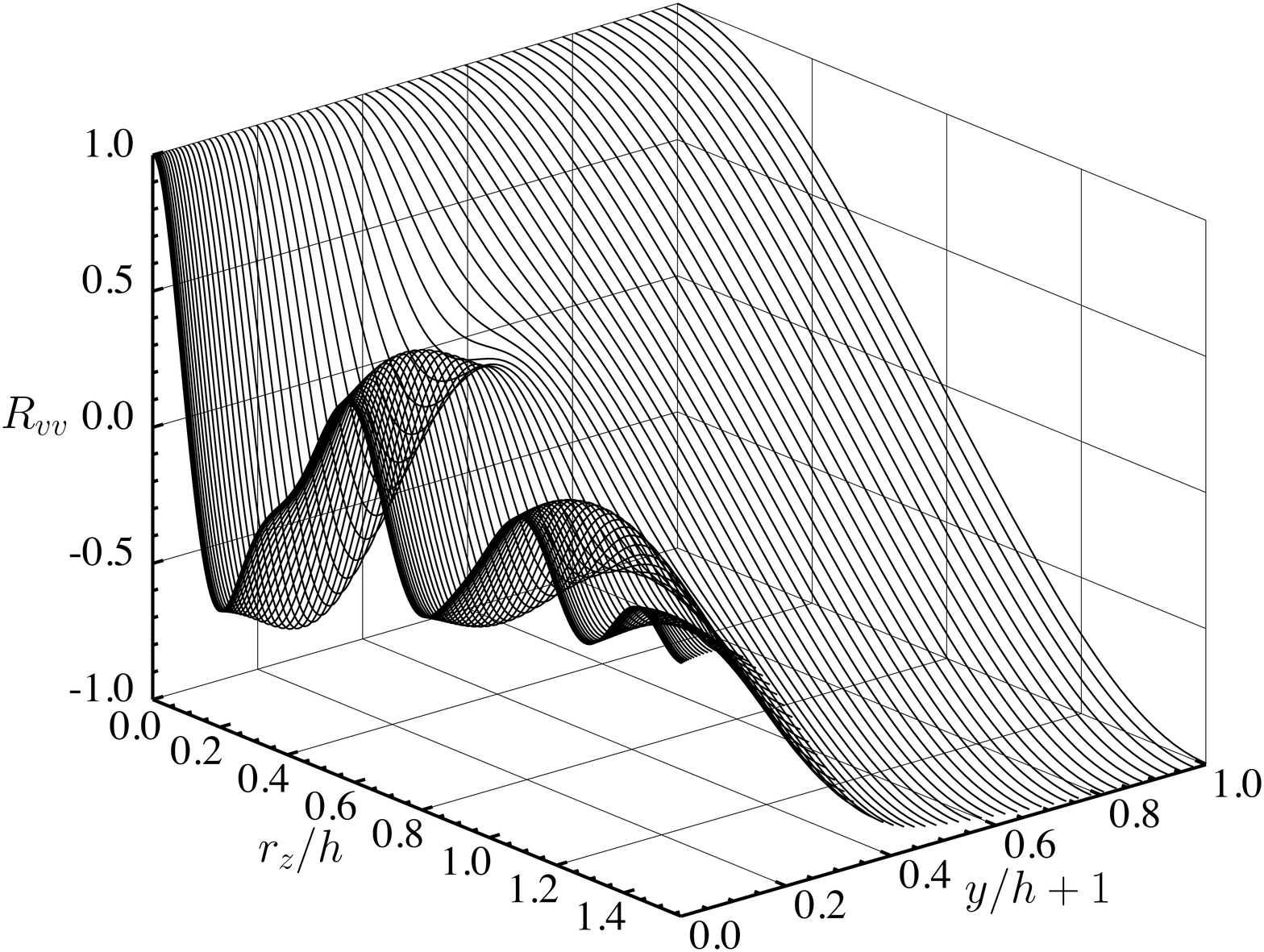}
	\end{minipage}
	\hspace{1em}
	\begin{minipage}{.48\linewidth}
	(\textit{b})\\
	\includegraphics[clip,width=\linewidth]{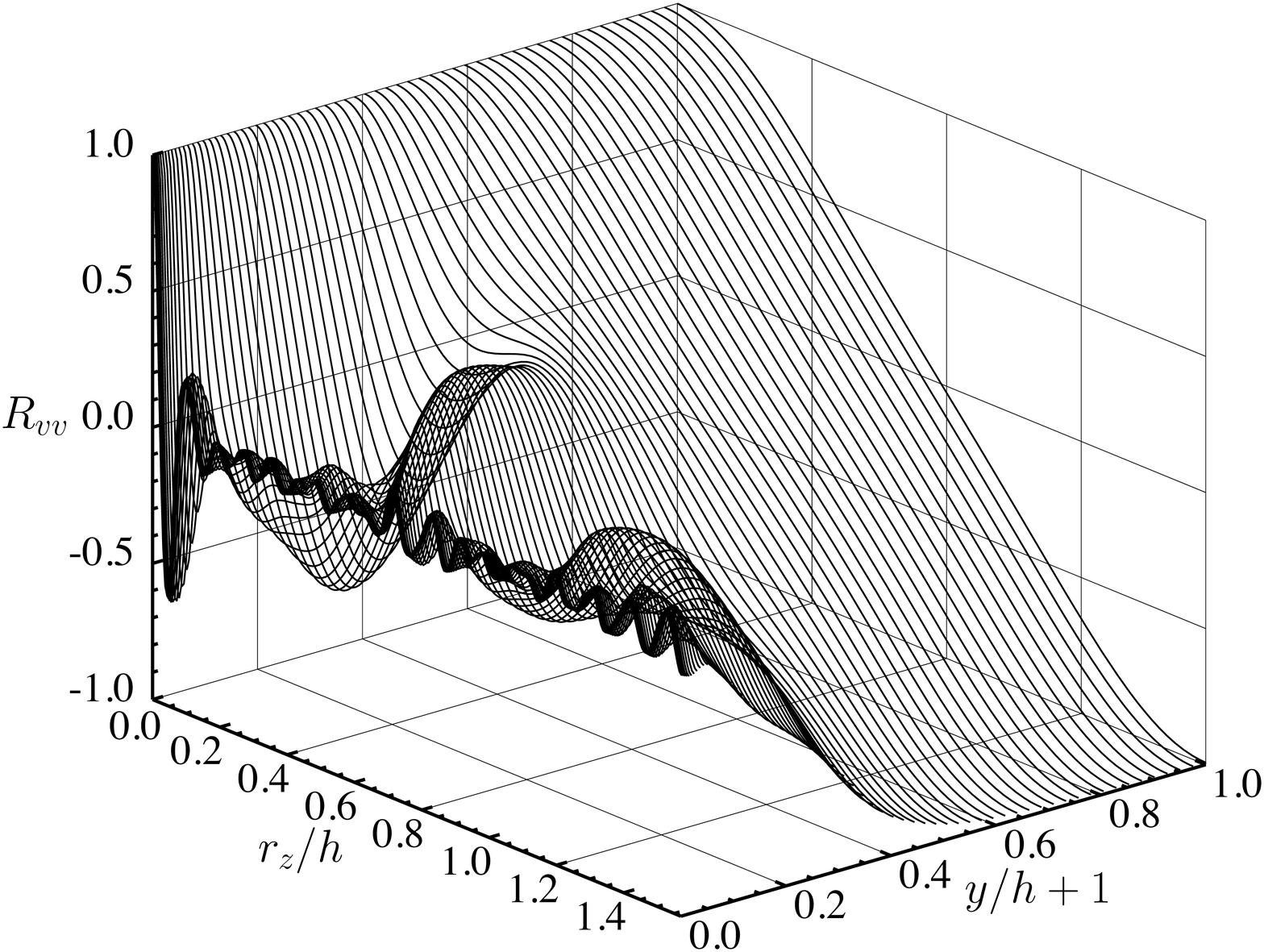}
	\end{minipage}

\caption{Two-point spanwise correlation of the wall-normal velocity $R_{vv}$ for $\lambda=0.1$.
(\textit{a}) $Re=1000$, (\textit{b}) $Re=5000$.
\label{fig9}}
\end{figure}

\begin{figure} 
\centering
	\begin{minipage}{.75\linewidth}
	(\textit{a})\\
	\includegraphics[clip,width=\linewidth]{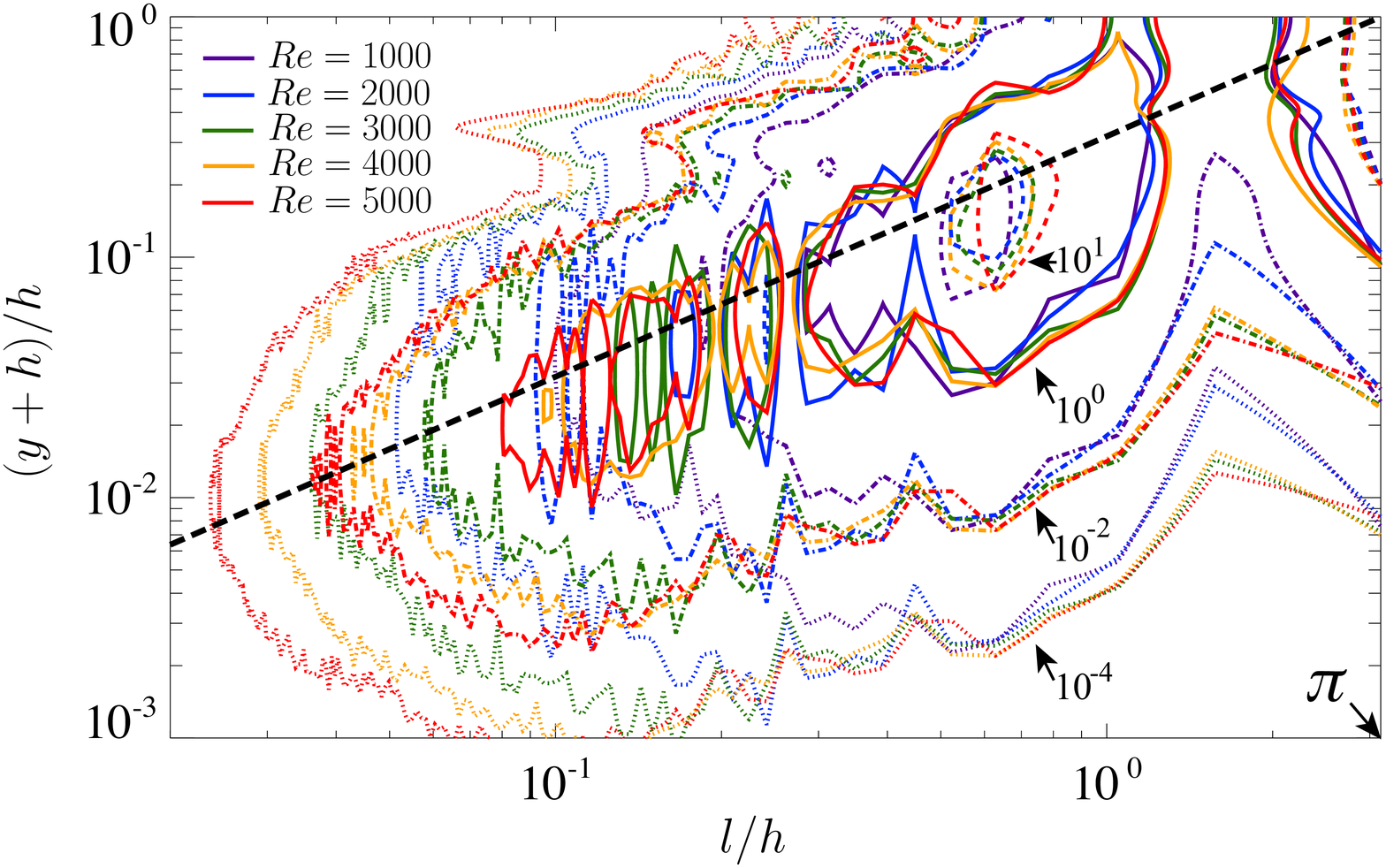}
	\end{minipage}

	\vspace{1em}

	\begin{minipage}{.75\linewidth}
	(\textit{b})\\
	\includegraphics[clip,width=\linewidth]{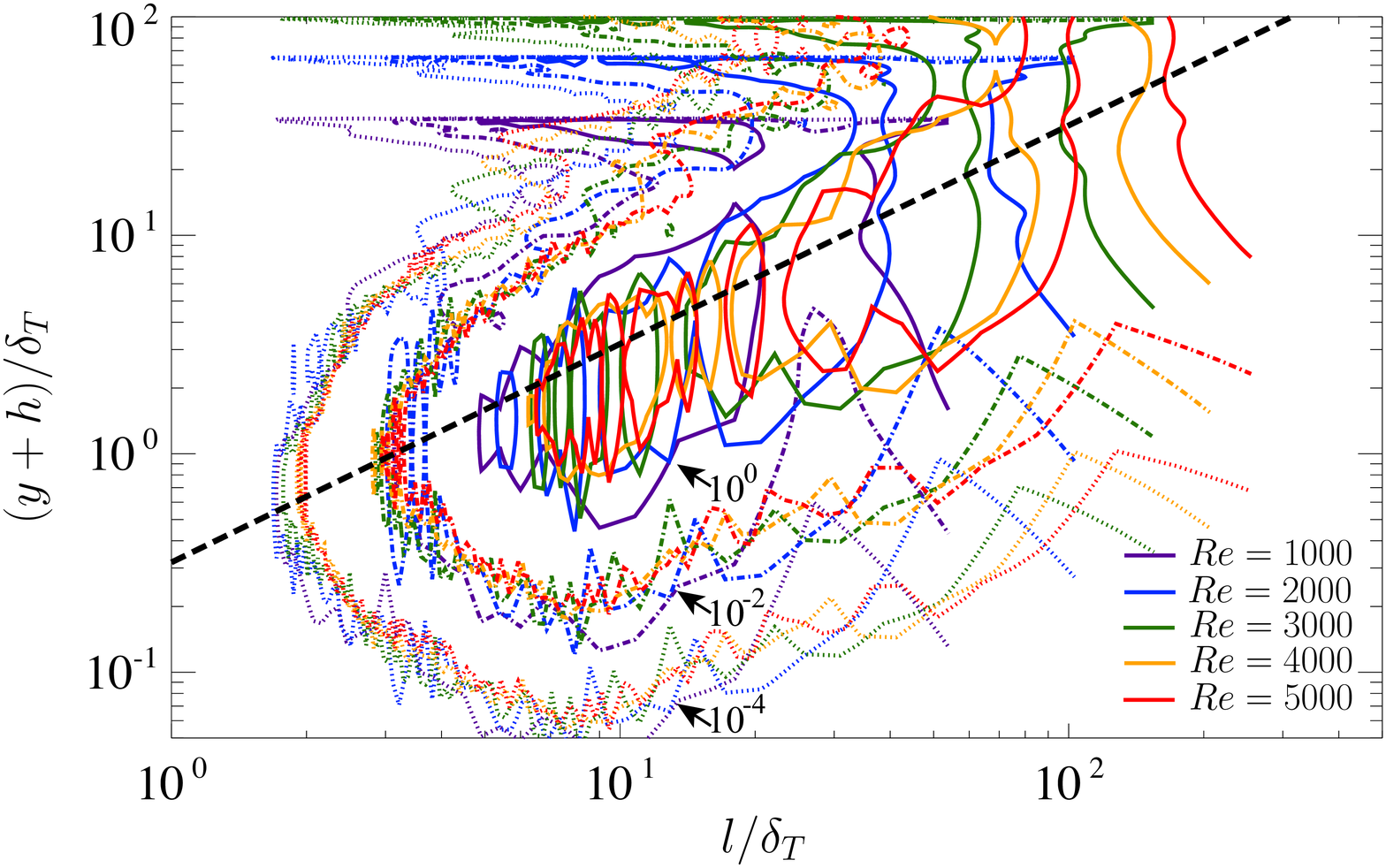}
	\end{minipage}

\caption{Spectral densities of the wall-normal velocity
\GGK{in} the optimal \SM{states} as a function of the spanwise wavelength $l$ for $\lambda=0.1$.
The 
\GGK{distance to the lower wall} $y+h$ and the wavelength $l$ are normalized by (\textit{a}) the channel half width $h$ and (\textit{b}) the inner length
$\SM{\delta_{T}=}T_{0}/({\rm d}{\left< T \right>}_{xz}/{\rm d}y)(y=-h)$,
respectively.
The thick dashed line \SMM{denotes} $l=\pi(y+h)$.
\label{fig10}}
\end{figure}

\begin{figure} 
\centering
	\begin{minipage}{.75\linewidth}
	\includegraphics[clip,width=\linewidth]{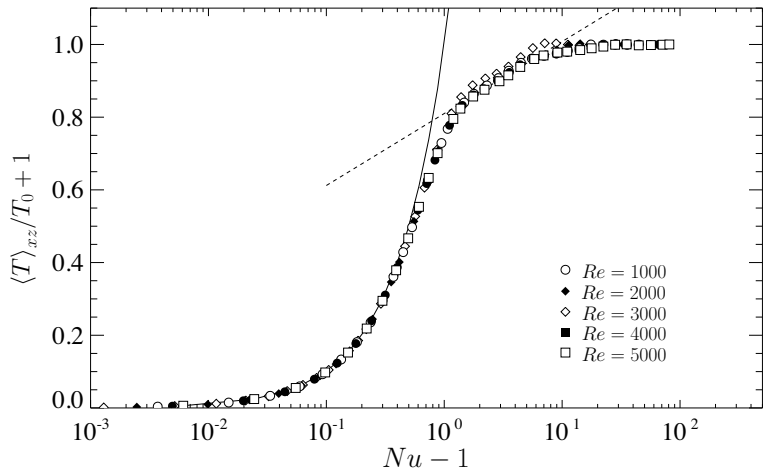}
	\end{minipage}

\caption{Mean temperature profile
\GGK{in} the optimal \GKK{states} for $\lambda=0.1$.
The distance to the wall $y+h$ and the temperature $T$ are normalized by the inner length and the wall temperature $T_{0}$, respectively.
The solid curve and dashed line
\GGK{denote}
${\left< T \right>}_{xz}/T_{0}+1=(y+h)/\SM{\delta_{T}}$
and the logarithmic fit
${\left< T \right>}_{xz}/T_{0}+1=0.086\ln{[(y+h)/\SM{\delta_{T}}]}+0.81$
determined in the range
$1<(y+h)/\SM{\delta_{T}}<10$,
respectively.
\label{fig11}}
\end{figure}

\begin{figure} 
\centering
	\begin{minipage}{.75\linewidth}
	\includegraphics[clip,width=\linewidth]{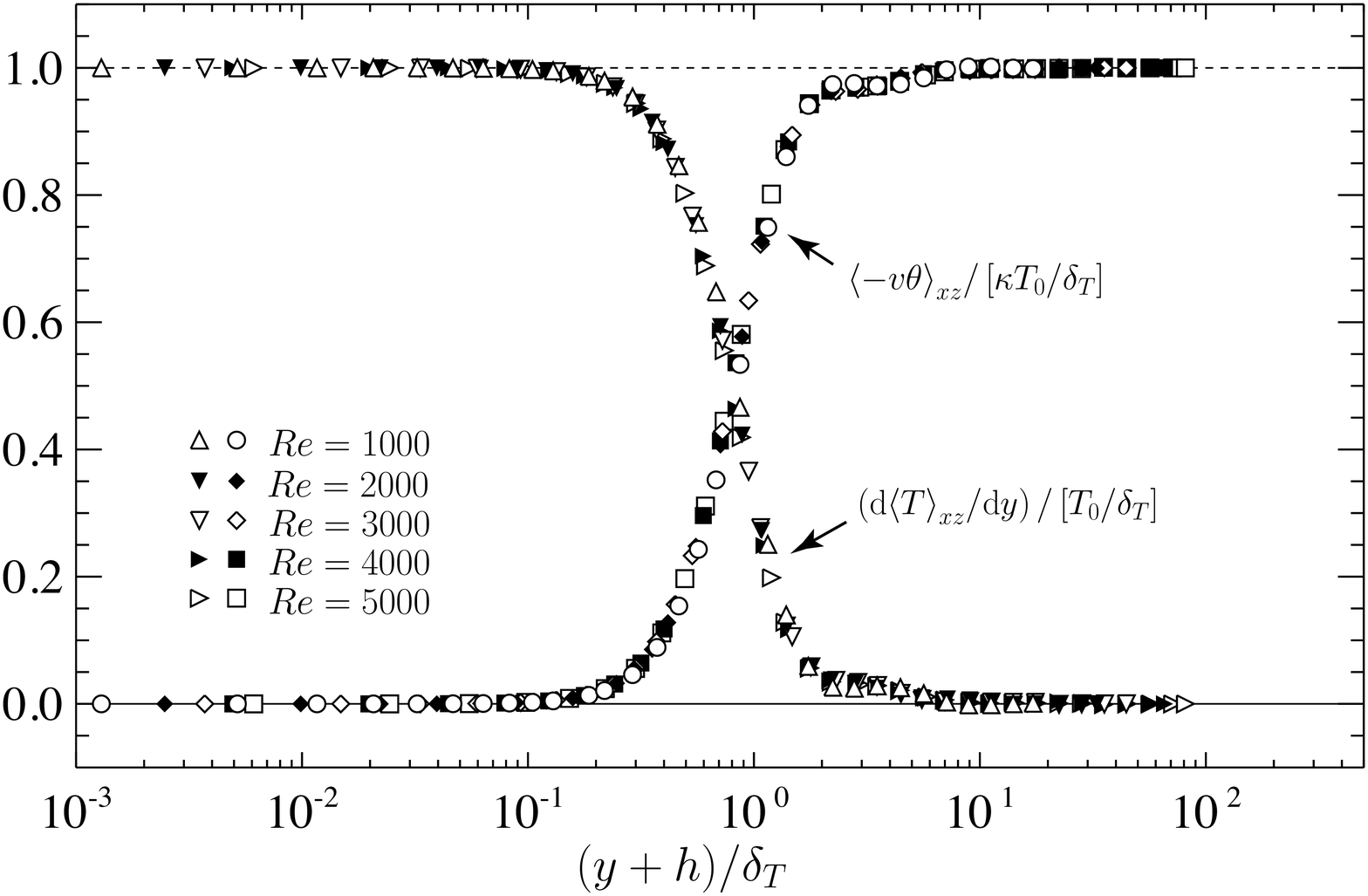}
	\end{minipage}
	
\caption{
\GGK{Convective and conductive heat fluxes in}
the optimal \GKK{states} \GGK{as a function of the distance to the wall}
for $\lambda=0.1$.
The heat \GGK{fluxes} and the
\GGK{distance} are normalized by the inner length. 
\label{fig12}}
\end{figure}

Let us discuss the hierarchical structure and its scaling for the optimal state. 
\GK{Just} in this section,
\GK{we restore the} dimensional variables,
\GK{such as} $\mbox{\boldmath$x$},\mbox{\boldmath$u$},T$.
As \SMM{mentioned} in \S\ref{sec4-1}, a number of quasi-streamwise vortex tubes
\GK{appear} near the \GKK{walls} at large $Re$.
In order to quantitatively evaluate their size at each distance to the wall, we introduce a two-point spanwise correlation function of the wall-normal velocity defined by
\begin{eqnarray}
\label{eq5-1}
R_{vv}(r_{z};y)=\frac{{\left< v(x,y,z)v(x,y,z+r_{z}) \right>}_{xz}}{{\left< v(x,y,z)^{2} \right>}_{xz}},
\end{eqnarray}
where $r_{z}$ indicates the
\GK{spacing} between two points \GK{separated in the $z$-direction}.
\GK{The correlation $R_{vv}$ is shown
for the optimal states at $Re=1000$ and $5000$
in \SMM{figure} \ref{fig9}, where}
\GK{$r_z$} and the distance from the 
\GK{lower} wall, $y+h$, are normalized by the channel half \GK{height} $h$.
At \GK{the midplane of the channel,}
$y/h+1=1$, $R_{vv}$ has only one local minimum, since the wall-normal velocity component 
\GK{is predominantly induced by the large-scale circulation rolls
in
the central region} of the channel.
At
\GK{the smaller values of $y/h+1$,
i.e. the closer positions to the wall},
however, oscillatory behaviour can be seen in $R_{vv}$.
The wavelength of the oscillation corresponds to the size of the spanwise-arranged quasi-streamwise vortex tubes.
The \GK{spanwise} wavelength is \GK{found to be}
smaller at smaller \GK{height} $y/h+1$.

Figure \ref{fig10} shows the one-dimensional spectral densities of the wall-normal velocity at $Re=1000$, $2000$, $3000$, $4000$, $5000$.
The lateral axis \GK{denotes}
the \GK{spanwise}
wavelength \SM{$l$} and the longitudinal one is the distance \GK{$y+h$}.
In \SMM{figure} \ref{fig10}(\textit{a}), the both axes are normalized by $h$.
The spectral peak at $y/h+1=1$
\GK{is relevant to} the large-scale circulation.
\GK{In the} near-wall region, \GK{$y/h+1\lesssim 2\times 10^{-1}$},
meanwhile, a lot of peaks are observed
\GK{along the `ridge' represented by the thick dashed line $l=\pi(y+h)$}.
That is \GK{to say, the near-wall} quasi-streamwise vortical structures
\GK{possess}
hierarchical self similarity.
Notice that the peaks are discretely located on the ridge unlike the 
\GK{continuously}
hierarchical structure \GK{statistically}
observed in wall turbulence, because in a
\GK{time-independent}
optimal field, there are no \GK{temporal}
fluctuations, i.e. jittering, of vortical structures.
The length scale of the largest vortices scales with the outer length $h$, and their height is around $y/h+1=0.2$ regardless of $Re$.
In \SMM{figure} \ref{fig10}(\textit{b}), $l$ and $y+h$ are normalized by the inner length
\SM{
\begin{eqnarray}
\label{eq5-2}
\delta_{T}=T_{0}{\left( \frac{{\rm d}{\left< T \right>}_{xz}}{{\rm d}y}(y=-h) \right)}^{-1}
\end{eqnarray}
}\GKK{which}
\GK{characterises the near-wall temperature profile}.
In the
close \GK{vicinity of} the wall,
the spectral density distribution can be seen to scale with the inner length.
The vortex structures exist in the range of $\SM{\delta_{T}}\lesssim y+h\lesssim0.2h$ for any $Re$. 

Figure \ref{fig11} shows the mean temperature profile as a function of
\GK{the distance to the wall,}
$(y+h)/\SM{\delta_{T}}$.
\GK{At} $(y+h)/\SM{\delta_{T}}\ll1$,
the mean temperature
\GK{is expressed by}
${\left< T \right>}_{xz}/T_{0}+1=(y+h)/\SM{\delta_{T}}$,
since
\GK{thermal conduction}
dominants over
\GK{convection} (see \SMM{figure} \ref{fig12}).
On the other hand, the hierarchical vortex structures lead to the logarithmic-like temperature profile in the intermediate region
$1\lesssim(y+h)/\SM{\delta_{T}}\lesssim10$.
The dashed line represents the logarithmic fit
${\left< T \right>}_{xz}/T_{0}+1=0.086\ln{[(y+h)/\SM{\delta_{T}}]}+0.81$
determined in the range
$1<(y+h)/\SM{\delta_{T}}<10$.
\GKK{It is well known that a logarithmic mean temperature
profile appears as a consequence of heat transfer
in near-wall turbulence
\SMM{where self-similar hierarchical vortical structures are commonly observed}.
Recently the logarithmic temperature distribution
has also been \SMM{found} \SMM{numerically and experimentally} in thermal convection turbulence
\SM{\citep{Ahlers2012,Ahlers2014}}.}
The \SMM{temperature} profiles are nearly uniform in the central region $y/h+1\approx1$ further away from the wall \SMM{because of} almost complete mixing of the scalar $T$ by the large-scale circulation.
The
\GK{convective} heat flux and the
\GK{conductive} heat flux are shown in \SMM{figure} \ref{fig12}.
Note that the total heat flux is strictly unity 
\GK{at any height} (see equation (\ref{eq2-7})).
It can be seen that around
$(y+h)/\SM{\delta_{T}}\approx1$
the two fluxes are comparable.
The
\GK{convective} heat flux is significant at
$(y+h)/\SM{\delta_{T}}\GKK{\gtrsim} 1$,
where the logarithmic-like mean temperature profile has been observed.
At
$(y+h)/\SM{\delta_{T}}\gtrsim10$
there is almost no contribution from the conduction,
i.e. no temperature gradient.

\section{Effects of \GK{flow three-dimensionality}
on local scalar dissipation}\label{sec6} 
\begin{figure} 
\centering
	\begin{minipage}{.75\linewidth}
	\includegraphics[clip,width=\linewidth]{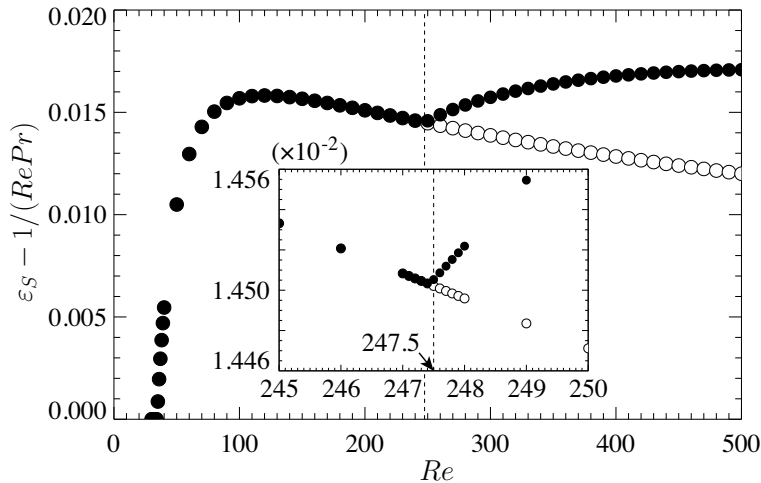}
	\end{minipage}

\caption{Scalar dissipation
as a function of $Re$ for $\lambda=0.1$.
Filled circles denote
\GGK{the scalar dissipation for} the optimal state and open circles stand for that
\GGK{obtained} from \GGK{the optimisation only within}
a streamwise-independent two-dimensional velocity field.
\label{fig13}}
\end{figure}

\begin{figure} 
\centering
	\begin{minipage}{.32\linewidth}
	(\textit{a})\\
	\includegraphics[clip,width=\linewidth]{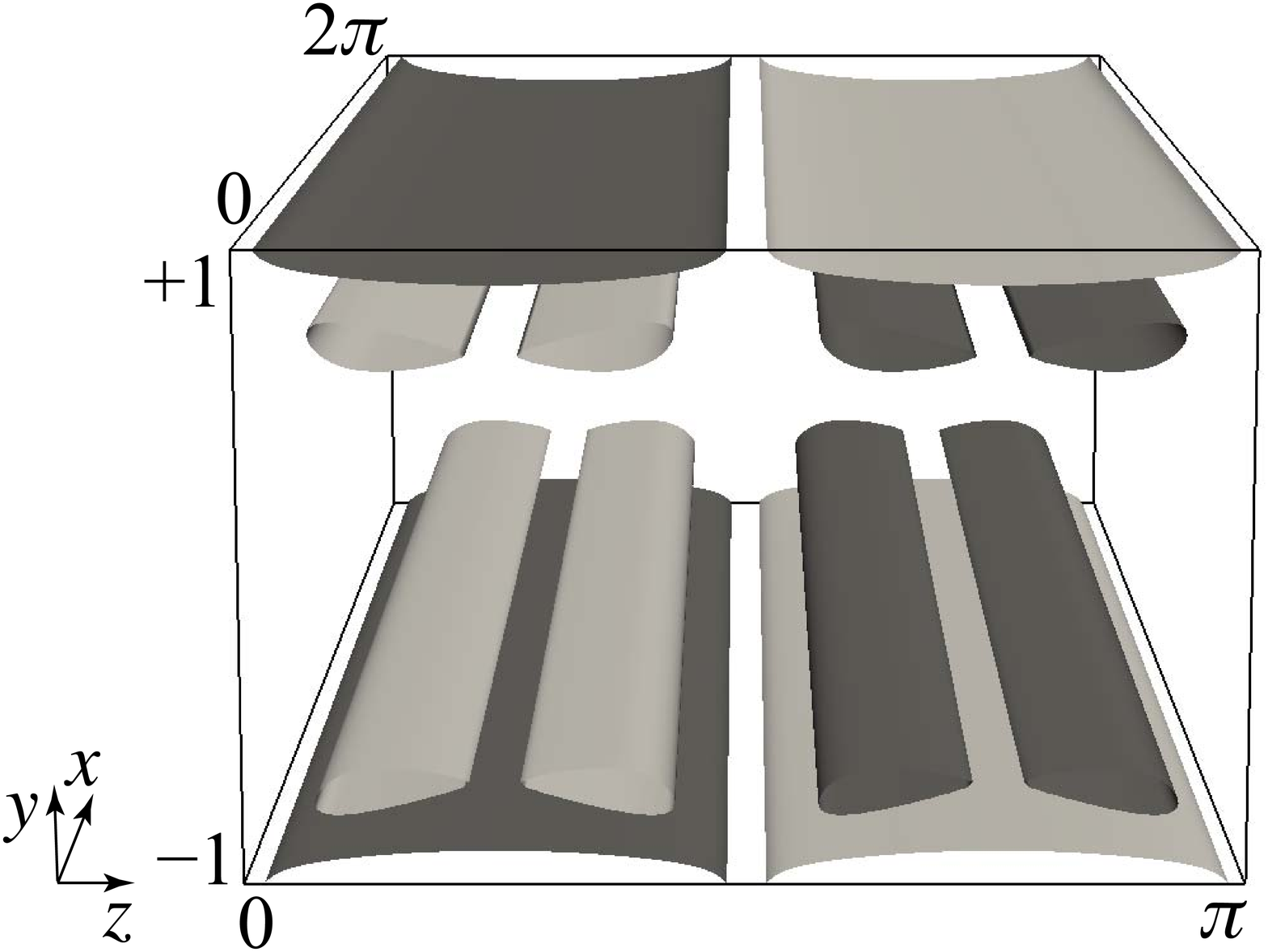}
	\end{minipage}
	\hspace{0.1em}
	\begin{minipage}{.32\linewidth}
	(\textit{b})\\
	\includegraphics[clip,width=\linewidth]{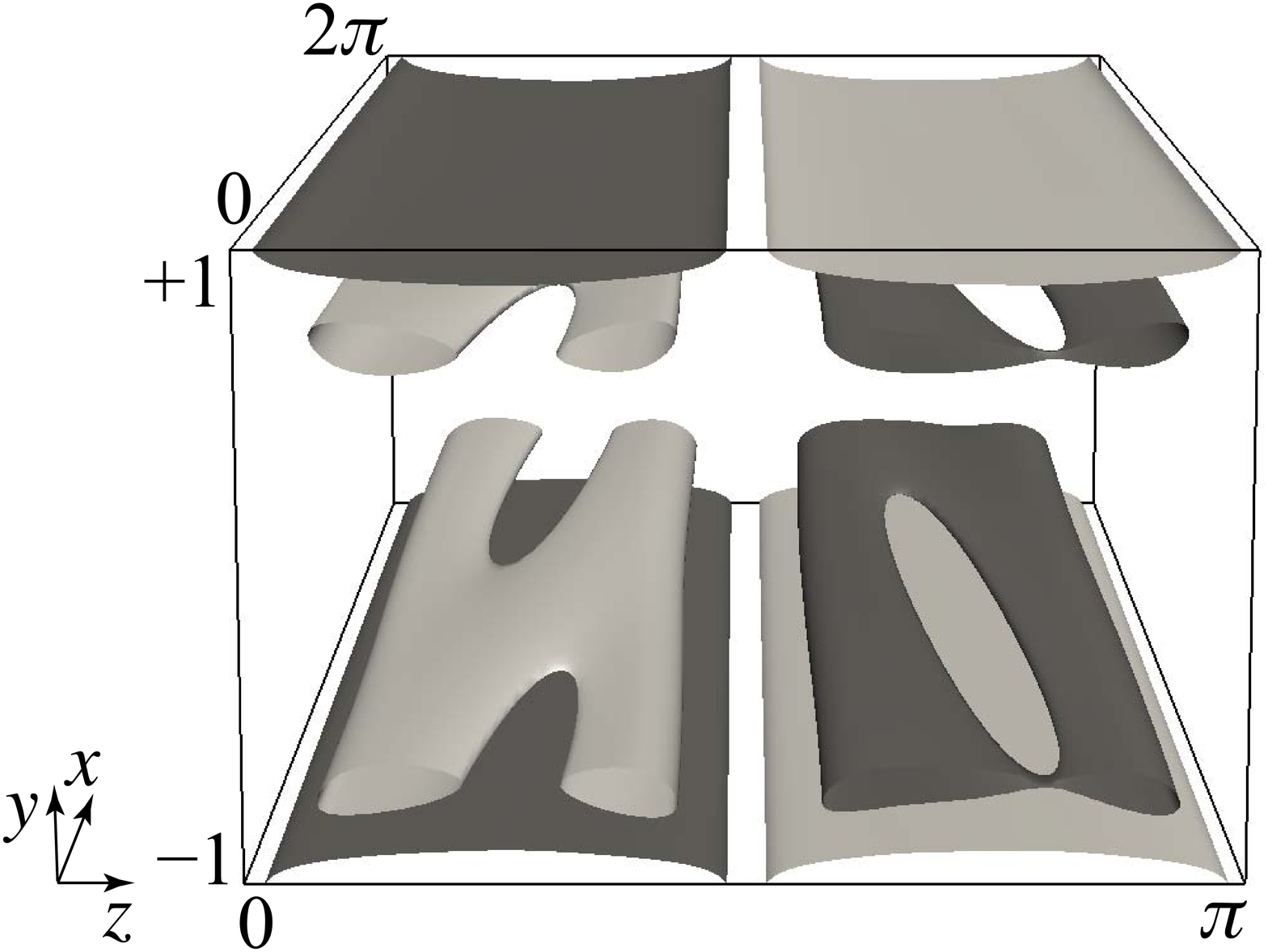}
	\end{minipage}
	\hspace{0.1em}
	\begin{minipage}{.32\linewidth}
	(\textit{c})\\
	\includegraphics[clip,width=\linewidth]{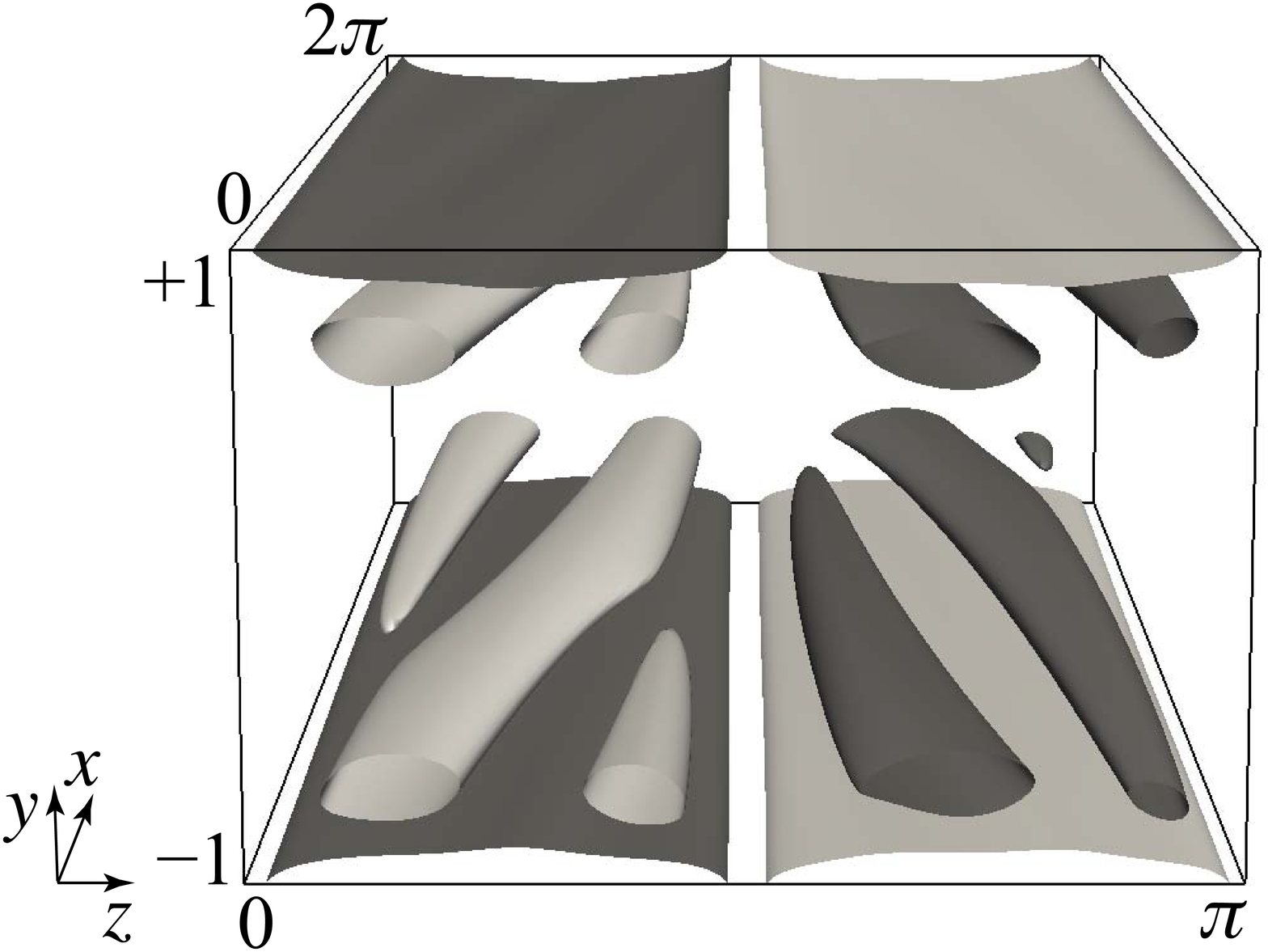}
	\end{minipage}

\caption{The optimal flow fields for $\lambda=0.1$. (\textit{a})
$Re=247.4$,  (\textit{b})
$Re=247.5$, (\textit{c})
$Re=249$.
Light \GKK{(or dark)} \GKK{grey} \GGK{objects represent}
\GGK{isosurfaces} of $\omega_{x}=+0.4$ \GKK{(or $-0.4$)}.
\label{fig14}}
\end{figure}

\begin{figure} 
\centering
	\begin{minipage}{.48\linewidth}
	(\textit{a})\\
	\includegraphics[clip,width=\linewidth]{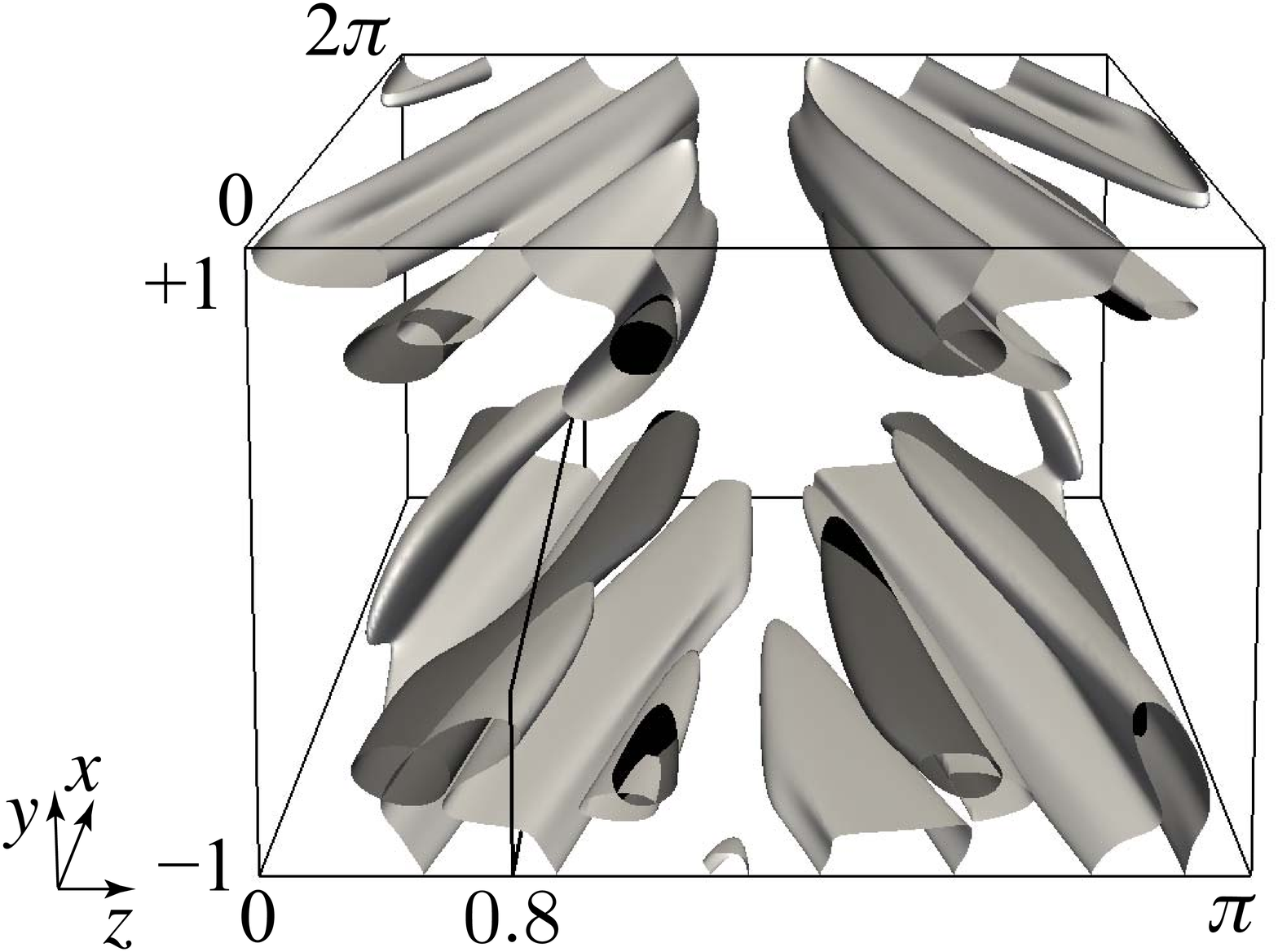}
	\end{minipage}
	\hspace{1em}
	\begin{minipage}{.48\linewidth}
	(\textit{b})\\
	\includegraphics[clip,width=\linewidth]{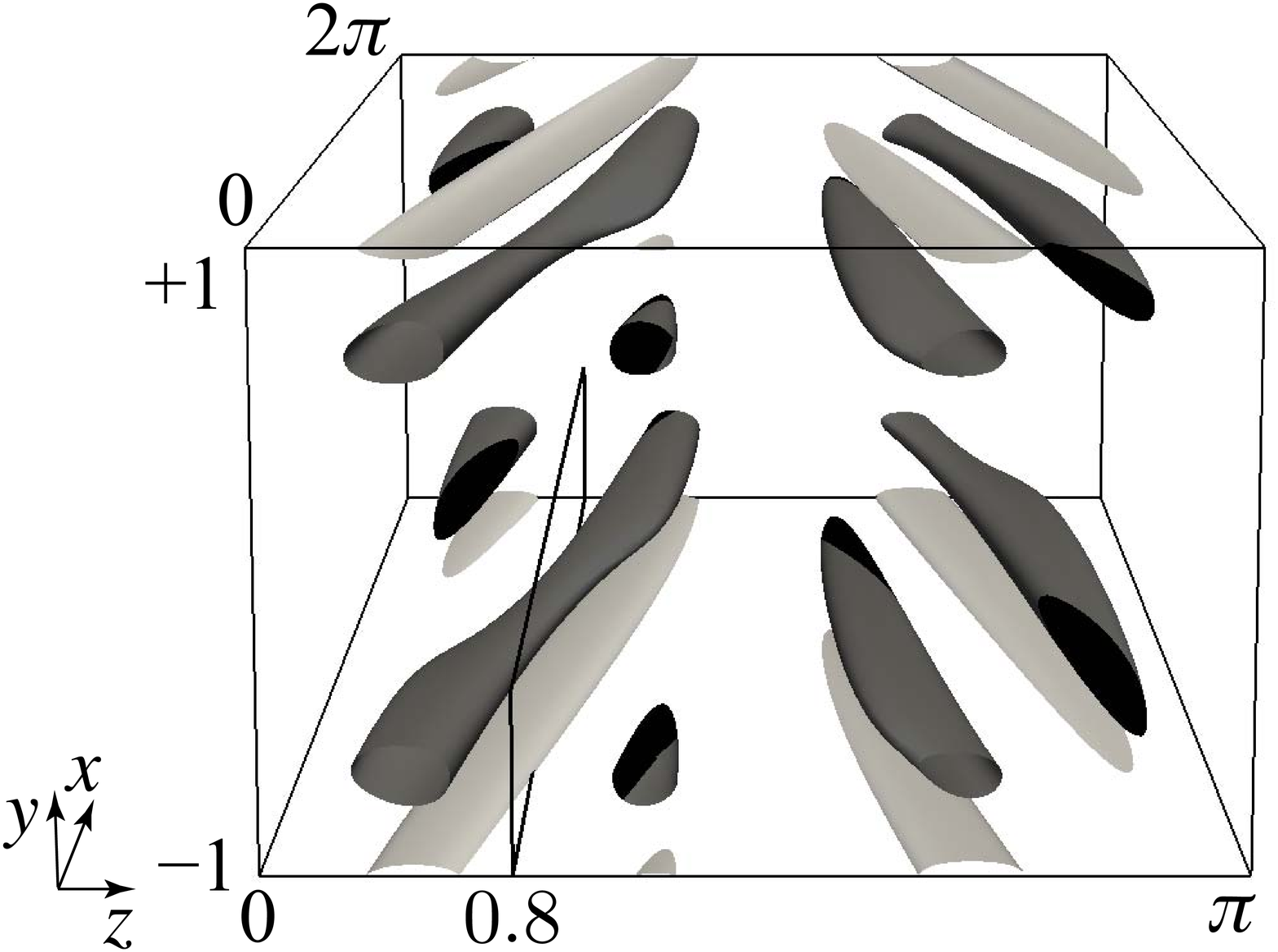}
	\end{minipage}

\caption{
\GGK{Vortical structures} and
significant increases
\GGK{in}
the local scalar and energy dissipation
\GGK{for the three-dimensional optimal velocity field
against the two-dimensional optimal velocity field}
at $Re=249$ for $\lambda=0.1$. 
The
\GGK{dark \GKK{grey} \SMM{(or black)} objects
represent
isosurfaces, $Q=+0.03$,} of the second invariant
\GGK{of velocity gradient tensor,
on which 
the positive (or negative) spanwise vorticity fluctuation $\omega_{z}'$
are shown.}
Light \GKK{grey} \GGK{objects}
\GGK{indicate isosurfaces} of (\textit{a}) \GGK{an increase in}
the local scalar dissipation,
$D_{S}-D_{S}^{\rm 2D}=+0.001$, (\textit{b})
\GGK{an increase in} the local energy dissipation,
$D_{E}-D_{E}^{\rm 2D}=+0.001$.
\SM{The frame in the figure\SMM{s} indicates the plane $z=0.8$.}
\label{fig15}}
\end{figure}

Figure \ref{fig13} shows the scalar dissipation
\GK{in} the optimal state
\GK{as a function of $Re$} for $\lambda=0.1$
\GK{to present} a bifurcation diagram of the solution to
the Euler--Lagrange equation.
The two-dimensional solution emerges from supercritical pitchfork
bifurcation \GK{on the laminar solution} at $Re=35$.
The two-dimensional solution exhibits a streamwise-independent
velocity field representing a pair of streamwise convection rolls
\GK{shown in \SMM{figure}~\ref{fig5}(\textit{a,b})},
and satisfying two spatial symmetries
\begin{eqnarray}
\label{eq6-1}
\displaystyle
[u,v,w](x,y,z)&=&[u,v,-w](x,y,-z), \\
\label{eq6-2}
\displaystyle
[u,v,w](x,y,z)&=&[-u,v,-w](x,-y,z).
\end{eqnarray}
As $Re$ increases further,
\GK{the \SM{secondary}} pitchfork bifurcation occurs on the two-dimensional
solution branch at $Re=247.5$ to create a three-dimensional solution
with a shift-and-reflect\SMM{ion} symmetry
\begin{eqnarray}
\label{eq6-3}
\displaystyle
[u,v,w](x,y,z)&=&[u,v,-w](x+L_{x}/2,y,-z),
\end{eqnarray}
and the symmetry (\ref{eq6-2}),
\GK{which brings about the significantly higher scalar dissipation.}
When $\lambda$ is changed,
\GK{the corresponding} bifurcation
arises at different values of $Re$, but the qualitative characteristics of the bifurcation diagram is not changed
(the pitchfork bifurcation occurs at larger $Re$ as $\lambda$ increases).
Figure \ref{fig14} shows the isosurface of the streamwise vorticity $\omega_{x}$ at \GK{the Reynolds \GKK{numbers} close to the onset of
the three-dimensional solution},
$Re=247.4,\,247.5$ and $249$.
As shown in \SMM{figure} \ref{fig6}, the optimal state has the streamwise large-scale circulation rolls.
They generate sheet-like distribution of 
\GK{strong} vorticity on the walls due to the no-slip
condition, \GK{
and just above the sheets the opposite-signed
tube-like (or ribbon-like) streamwise vortices appear.}
At $Re=247.5$, the
\GK{near-wall tubular} structures
\GK{exhibit streamwise dependence, that is three-dimensionality.}
\GK{At
$Re=249$
tubular vortices bear spanwise inclination
as already seen in \SMM{figure}~\ref{fig5}(\GKK{\textit{c,d}}) at
higher $Re$.}
It is important to
\GK{recall}
that \GK{the} scalar dissipation is enhanced with \GK{the}
appearance of the three-dimensional structures as shown in \SMM{figure} \ref{fig13}.
Figure \ref{fig15} \GK{visualises} isosurfaces of a positive value of
the second invariant of the velocity gradient tensor
\begin{eqnarray}
\label{eq6-4}
Q=-\frac{\partial u_{i}}{\partial x_{j}}\frac{\partial u_{j}}{\partial x_{i}}
\end{eqnarray}
for the optimal state at $Re=249$.
The spanwise vorticity fluctuation $\omega_{z}'=\omega_{z}-{\left< \omega_{z} \right>}_{xz}$ is also shown on the $Q$ isosurfaces by greyscale.
It can be seen that
most of all the extracted tube-like
structures exhibits $\omega_{z}'>0$.
That is, the vortex tubes are tilted in the spanwise direction so that they
\GK{may} produce the spanwise vorticity
\GK{antiparallel} to the vorticity of the background shear flow.
In the figures, the regions with significant increase of the local dissipation from the two-dimensional solution \GK{at the same value of $Re$}
are also shown.
The local scalar and energy dissipation are given \GK{respectively} by
\begin{eqnarray}
\label{eq6-5}
D_{S}=\frac{1}{RePr}{\left( \frac{\partial T}{\partial x_{j}} \right)}^{2}
,\hspace{1em}
D_{E}=\frac{1}{Re}{\left( \frac{\partial u_{i}}{\partial x_{j}} \right)}^{2},
\end{eqnarray}
\GK{and thus it follows}
that \GK{${\left< D_{S} \right>}_{xyz}=\varepsilon_{S}$} and
\GK{${\left< D_{E} \right>}_{xyz}=\varepsilon_{E}$}.
\SMM{In} figure \ref{fig15}, the regions of $D_{E}-D_{E}^{\rm 2D}>0$ \SMM{are found to} appear
just below the vortex tubes which are flanked with the regions of
$D_{S}-D_{S}^{\rm 2D}>0$,
\GK{where the superscript 2D means the two-dimensional solution}.

\begin{figure} 
\centering
	\begin{minipage}{.3\linewidth}
	(\textit{a})\\
	\includegraphics[clip,width=\linewidth]{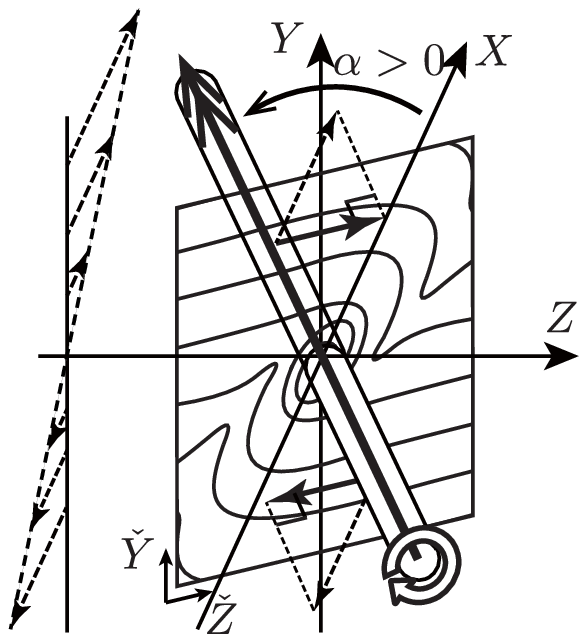}
	\end{minipage}
	\hspace{1em}
	\begin{minipage}{.3\linewidth}
	(\textit{b})\\
	\includegraphics[clip,width=\linewidth]{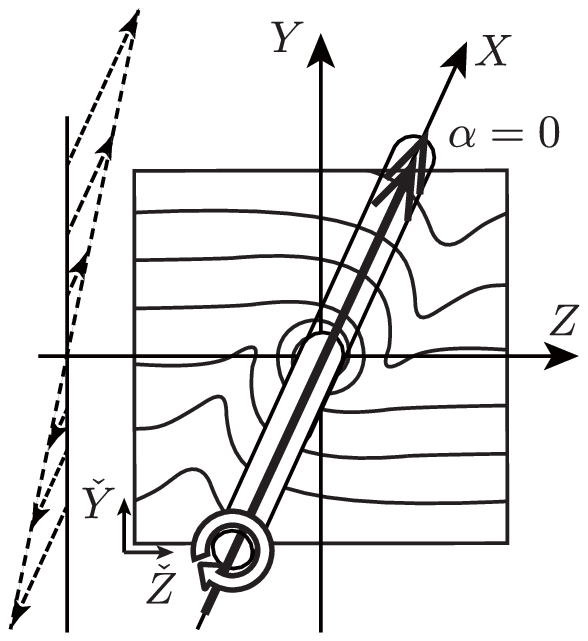}
	\end{minipage}
	\hspace{1em}
	\begin{minipage}{.3\linewidth}
	(\textit{c})\\
	\includegraphics[clip,width=\linewidth]{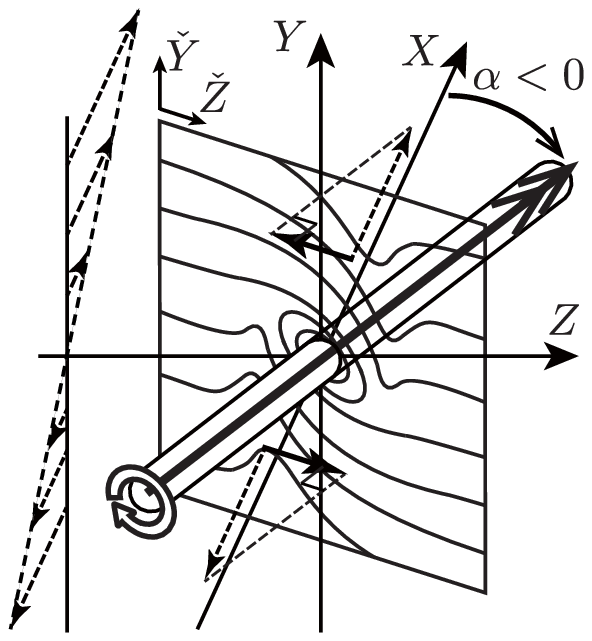}
	\end{minipage}

	\vspace{1em}

	\begin{minipage}{.3\linewidth}
	(\textit{d})\\
	\includegraphics[clip,width=\linewidth]{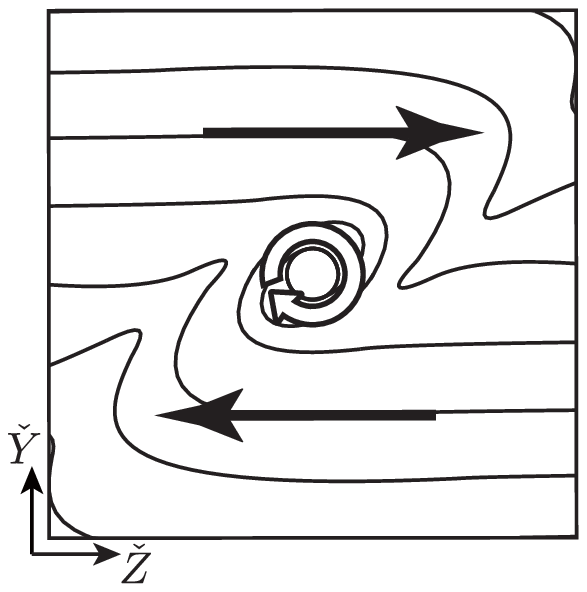}
	\end{minipage}
	\hspace{1em}
	\begin{minipage}{.3\linewidth}
	(\textit{e})\\
	\includegraphics[clip,width=\linewidth]{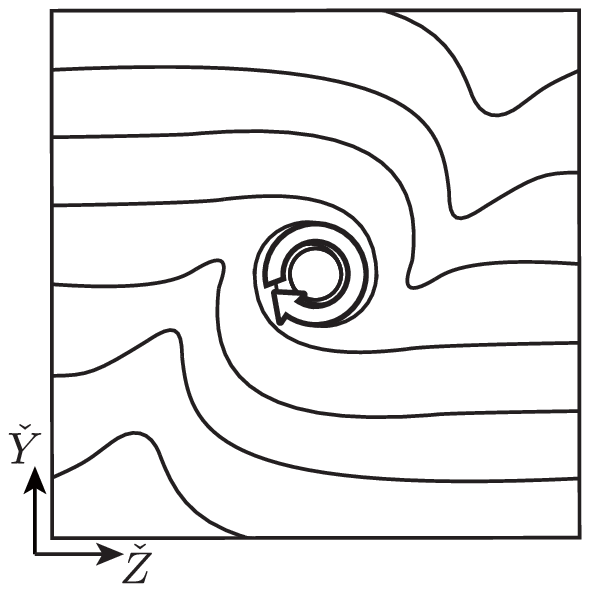}
	\end{minipage}
	\hspace{1em}
	\begin{minipage}{.3\linewidth}
	(\textit{f})\\
	\includegraphics[clip,width=\linewidth]{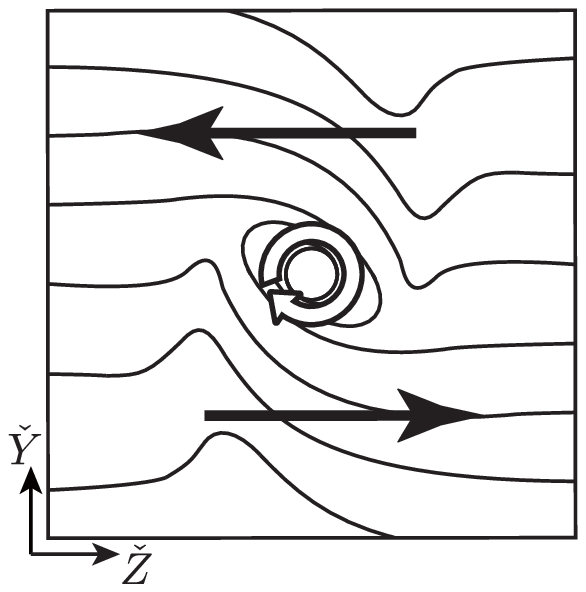}
	\end{minipage}

\caption{\GGK{Illustration} of scalar \GGK{fields} around
\GGK{cyclonic, neutral and ani-cyclonic}
vortex \GGK{tubes} in shear flow.
(\textit{a,b,c}) Configuration of the vortex tube
\GGK{in} the shear flow. 
The vortex tube is inclined at an angle $\alpha$ from the $X$-axis on the plane $Y=0$.
(\textit{d,e,f})  Isocontours of the scalar in the ($\check{Z}$,$\check{Y}$)-plane
normal to the central axis of the vortex \GGK{tube}.
(\textit{a,d})\ \GGK{the cyclonic case ($\alpha>0$)},
(\textit{b,e})\ \GGK{the neutral case ($\alpha=0$)},
(\textit{c,f})\ \GGK{the anticyclonic case ($\alpha<0$).}
\label{fig16}}
\end{figure}

\begin{figure} 
\centering
	\begin{minipage}{.8\linewidth}
	(\textit{a})\\
	\includegraphics[clip,width=\linewidth]{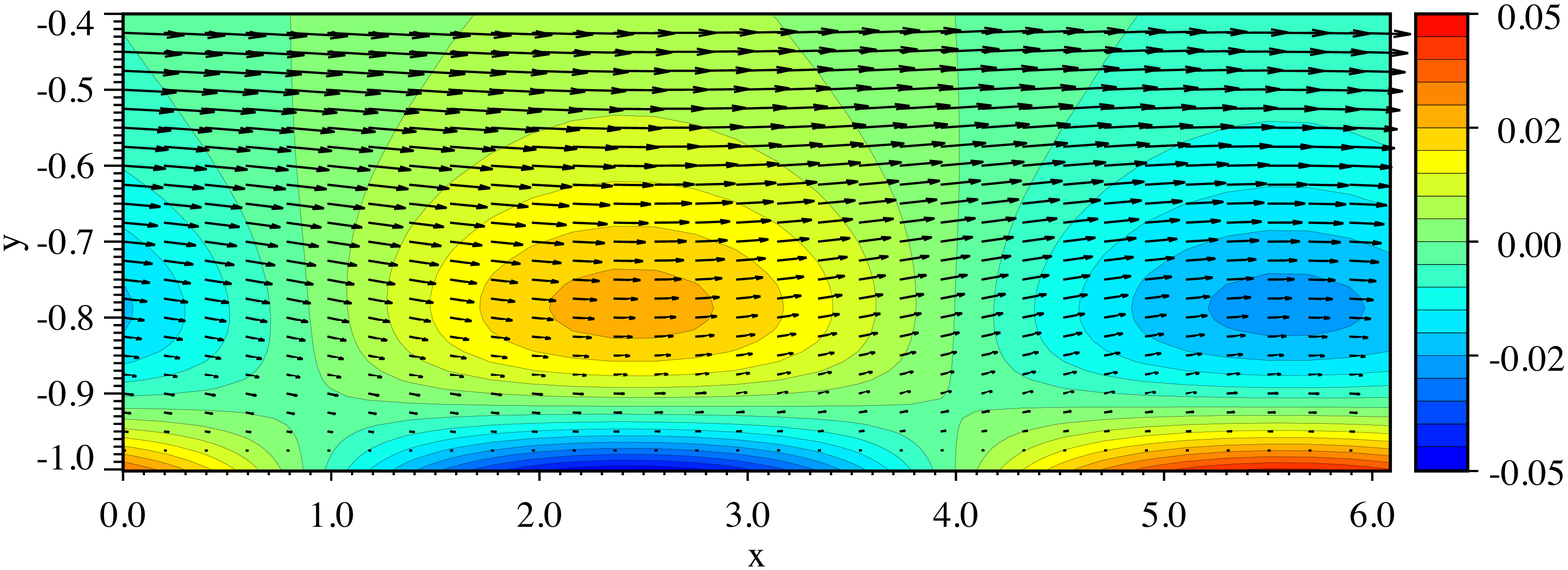}
	\end{minipage}
	
	\vspace{1em}
	
	\begin{minipage}{.8\linewidth}
	(\textit{b})\\
	\includegraphics[clip,width=\linewidth]{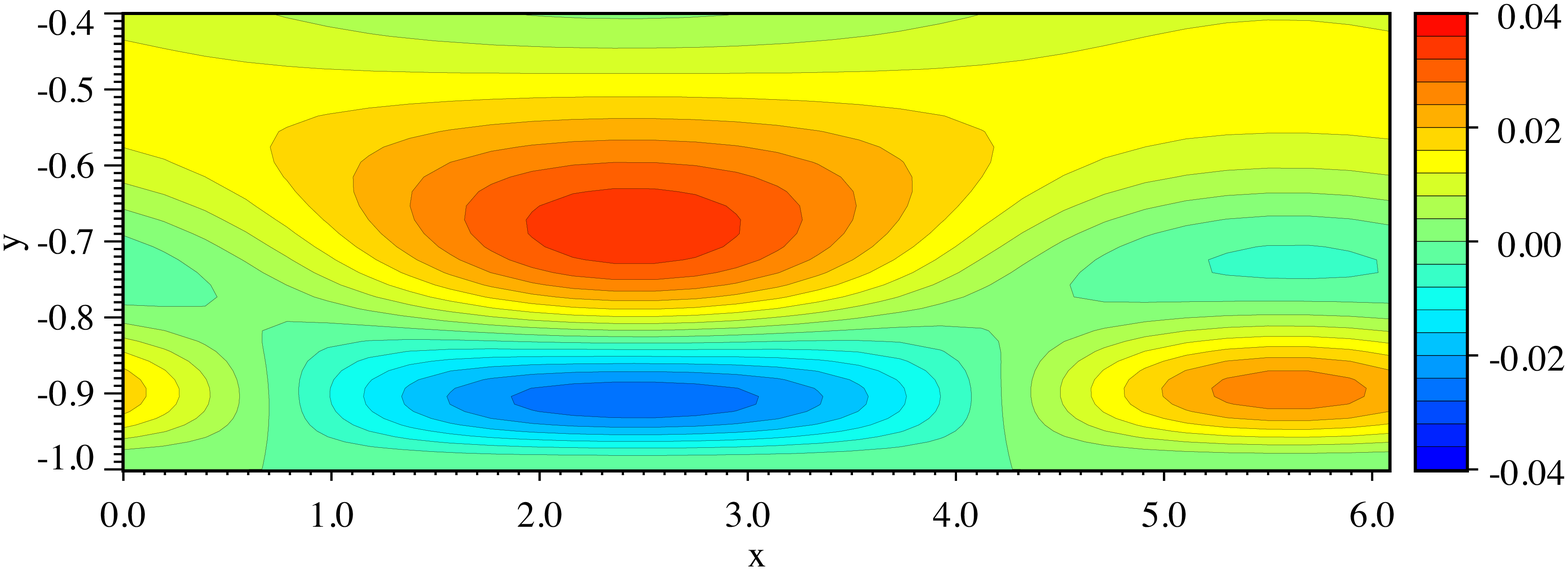}
	\end{minipage}
	
	\vspace{1em}
	
	\begin{minipage}{.8\linewidth}
	(\textit{c})\\
	\includegraphics[clip,width=\linewidth]{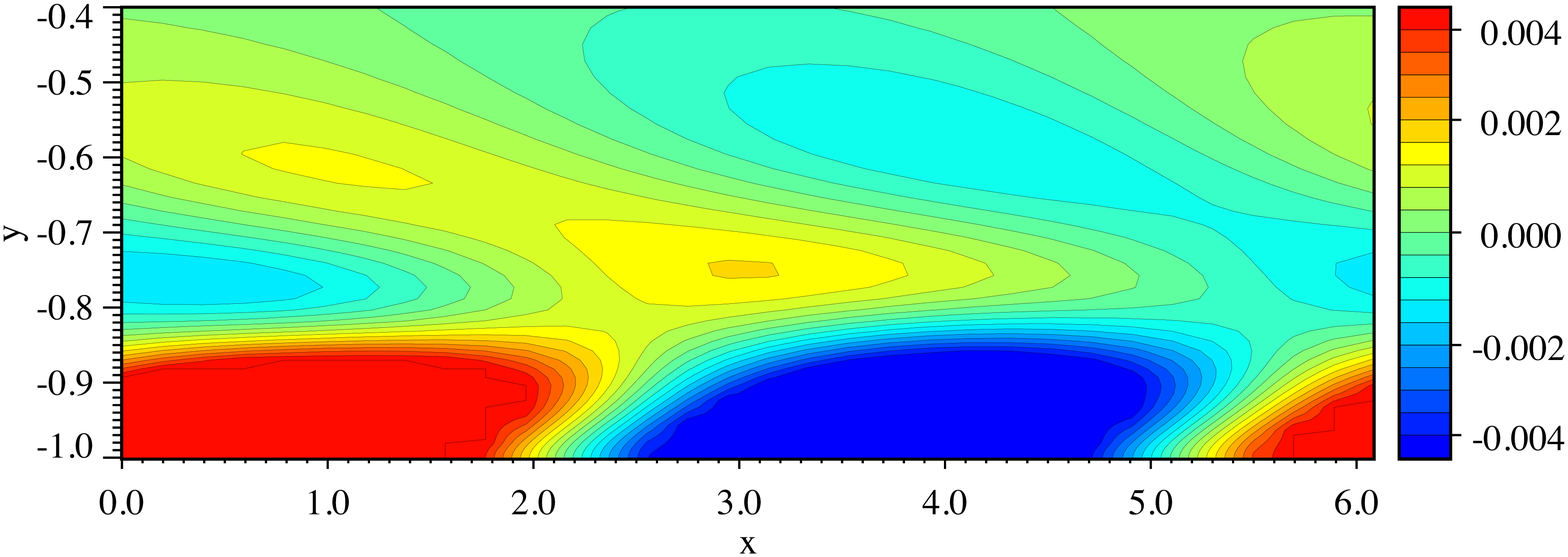}
	\end{minipage}
	
	\vspace{1em}
	
	\begin{minipage}{.8\linewidth}
	(\textit{d})\\
	\includegraphics[clip,width=\linewidth]{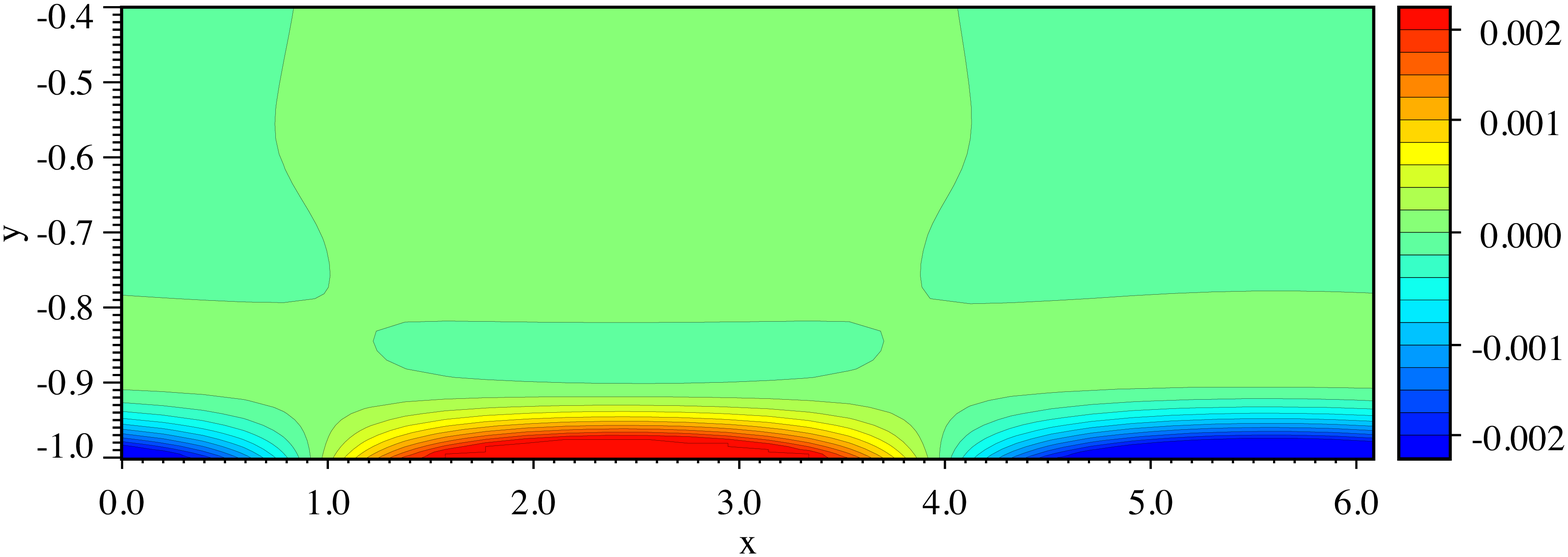}
	\end{minipage}

\caption{(Colour online)
\GGK{Two-dimensional cuts of the}
vorticity field, \GGK{the velocity field,} and the \GGK{local scalar and energy} dissipation
\GGK{fields}
at \SM{$Re=249$} for $\lambda=0.1$ \SMM{at $z=0.8$}.
\SM{(\textit{a}) Isocontours of the spanwise vorticity fluctuation $\omega_{z}'$ and the velocity vectors $(u+1,v)$ relative to the lower moving wall in the plane $z=0.8$ shown in \SMM{figure} \ref{fig15}.
(\textit{b}) Isocontours of the second invariant of velocity gradient tensor $Q$, (\textit{c}) a difference in the local scalar dissipation $D_{S}-D_{S}^{2D}$, (\textit{d}) a difference in the local energy dissipation $D_{E}-D_{E}^{2D}$ in the plane $z=0.8$}.
\label{fig17}}
\end{figure}

\begin{figure} 
\centering
	\begin{minipage}{.48\linewidth}
	(\textit{a})\\
	\includegraphics[clip,width=\linewidth]{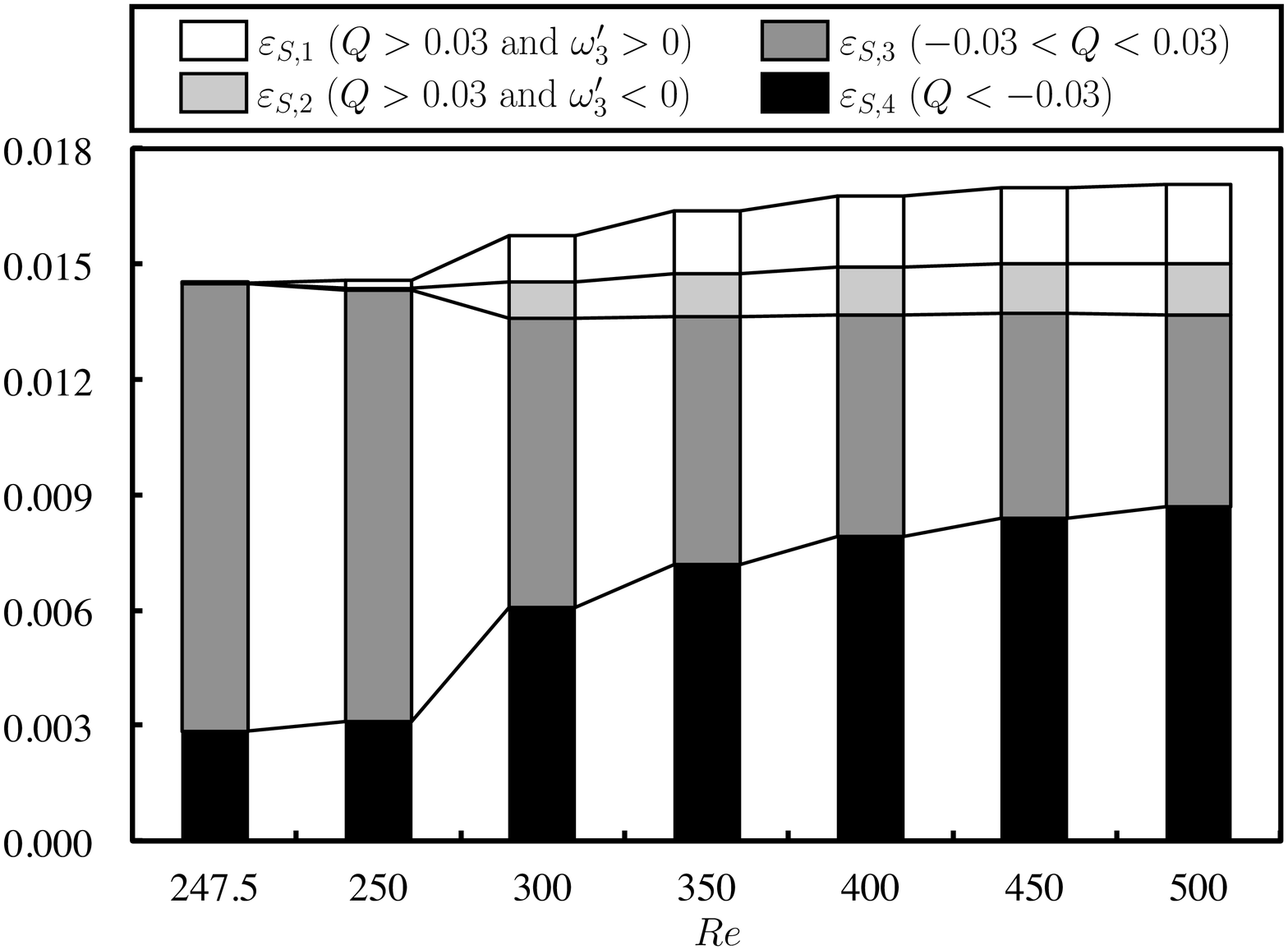}
	\end{minipage}
	\hspace{1em}
	\begin{minipage}{.48\linewidth}
	(\textit{b})\\
	\includegraphics[clip,width=\linewidth]{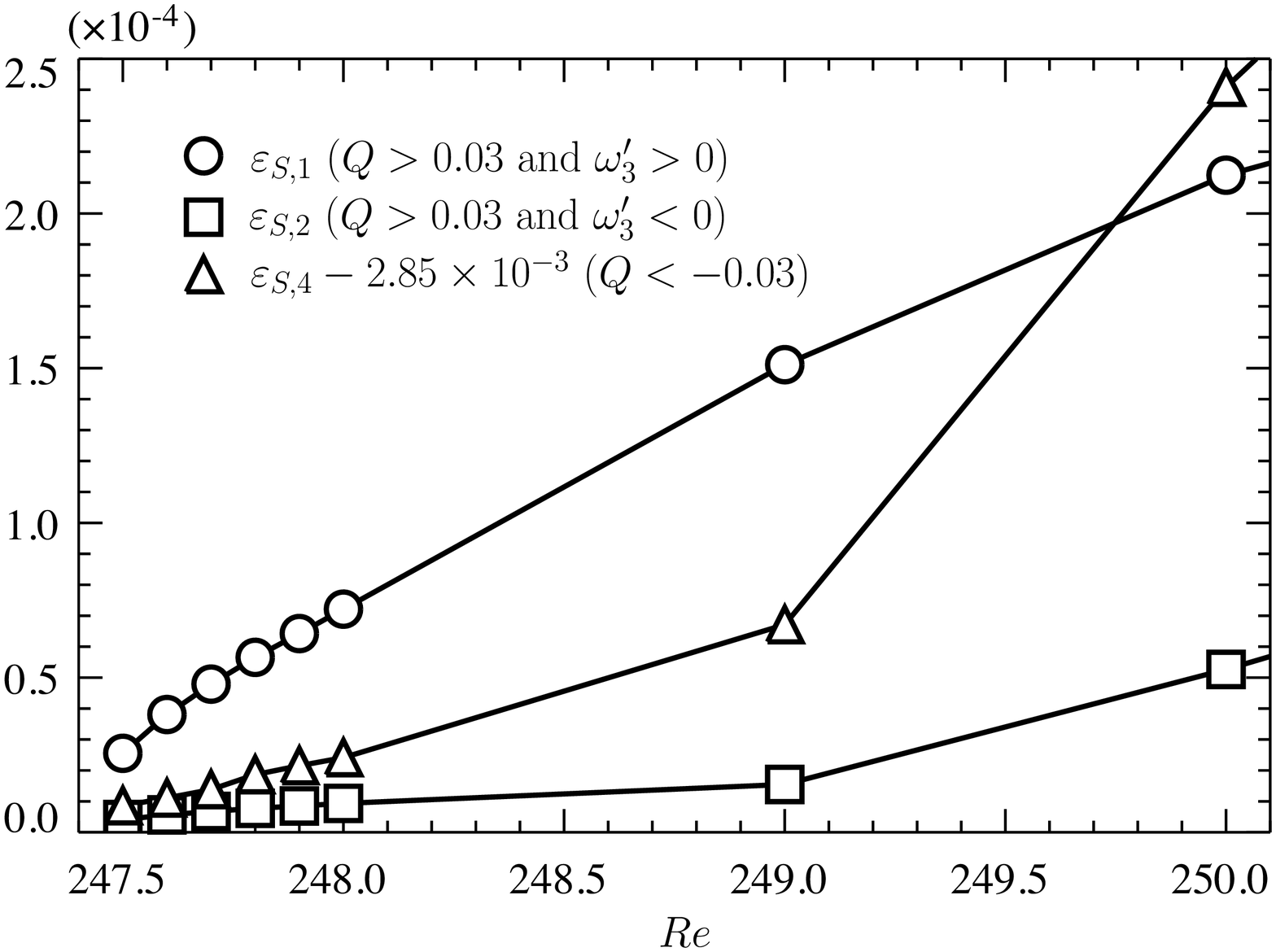}
	\end{minipage}

\caption{Contributions
\GGK{from the} four regions distinguished by the second invariant
\GGK{of the velocity gradient tensor} $Q$ and the spanwise vorticity fluctuation $\omega_{z}'$ to \GGK{the} total scalar dissipation for $\lambda=0.1$.
\GGK{Note that in}
(\textit{b}), \GGK{$\varepsilon_{S,4}-2.85\times 10^{-3}$
is shown instead of $\varepsilon_{S,4}$}.
\label{fig18}}
\end{figure}

The mechanism responsible for the increase in the local scalar dissipation in the \GK{flanks}
of the tilted \GK{quasi-streamwise}
vortex tube can be explained in terms of \GK{the following} simple model.
Figure \ref{fig16} shows illustrations of the scalar field around a vortex tube in shear flow.
Suppose that the vortex tube is inclined at an angle $\alpha$ from the streamwise (\GK{$X$}-) direction on the plane $Y=0$.
If $\alpha>0$ (or $\alpha<0$), the vortex has the spanwise (\GK{$Z$}-) vorticity component with the same (or opposite) sign as that of the shear flow.
Hereafter,
\GK{the tilted vortex tubes of the same (or opposite) signed
spanwise vorticity as that of the shear flow
is referred to as a cyclonic (or anticyclonic) vortex.}
\GKK{Figures \ref{fig16}(\textit{d,e,f})} show the isocontours of the scalar in the
\GK{$(\check Y,\check Z)$}-plane normal to the central
axis of the vortex tube.
As illustrated in these figures, the vortex tube spirally wraps
\GK{the isosurfaces} of
\GK{the} scalar around itself
\GK{through the convective transfer induced by swirling flow}.
For $\alpha=0$ \GK{(\SMM{figure}~\ref{fig16}\textit{b},\textit{e})},
the shear flow has no effect on the scalar field
\GK{because it is tangent to the scalar isosurfaces everywhere}.
\GK{In the case of $\alpha\neq 0$, however, the shear flow
has the normal velocity component to the isosurfaces, and so
it will play a distinct role in the convective scalar transfer
depending on the sign of the inclination angle $\alpha$.}
When $\alpha>0$
\GK{(i.e., the cyclonic inclination)},
the spacing of the wrapped \GK{isosurfaces}
of the scalar is widened by the cross-axial
\GK{velocity component of the} shear flow
as shown in \GK{\SMM{figure}~\ref{fig16}(\textit{a},\textit{d})}.
As a result, the local scalar gradient is decreased around the vortex tube, and thus the scalar dissipation is reduced.
On the other hand, when \SM{$\alpha<0$}
\GK{(i.e., the anticyclonic
inclination)}, the spacing of the wrapped
\GK{isosurfaces} is tightened as shown in 
\SMM{figure~\ref{fig16}(\textit{c},\textit{f})}.
\GK{It turns out that}
the scalar dissipation is locally intensified
\GGK{in the flanks of the anticyclonic vortex tubes, which have been observed
in the optimal states at high Reynolds numbers.}
\GGK{Note that although
the scalar dissipation is enhanced in the lower left
and upper right of the anticyclone in the above simple model
(\SMM{figure}~\ref{fig16}\textit{d}), in reality
the higher scalar dissipation is observed in the lower position
of the anticyclone, where the mean scalar gradient is much
larger than in the upper position.}

Appendix \ref{appB} analytically
\GK{demonstrates}
this local effect of the
\GK{anticyclonic}
inclination on the total scalar dissipation
\GK{(or equivalently the wall heat flux in plane Couette flow)}
for a tilted diffusing vortex tube with
circulation $\Gamma(>0)$
in
uniform shear flow of a shear rate $S(>0)$.
\GK{In this flow system}
the total scalar dissipation $\varepsilon_{S}^{*}$ at
\GK{early time} $St|\alpha|\ll1$
\GK{and high vortex \SM{P\'eclet} number}
$\Gamma/\kappa\gg1$ is evaluated
\GK{analytically} as
\begin{eqnarray}
\label{eq6-6}
\displaystyle
\varepsilon_{S}^{*}\approx{\left.\varepsilon_{S}^{*}\right|}_{\alpha=0}-\alpha\left[ \frac{1}{4}\kappa S^{3}\Gamma\ln{\left( \frac{\Gamma}{2\pi\kappa} \right)}t^{2} \right],
\end{eqnarray}
where the
leading\GK{-order} term
in the right-hand side represents the total scalar dissipation for
\GK{a purely aligned vortex with the shear flow ($\alpha=0$)}.
For the cyclonic case $\alpha>0$ (or the anticyclonic case $\alpha<0$), the total scalar dissipation is reduced (or enhanced) in comparison with that of the neutral case of $\alpha=0$.
This result
\GK{has been} obtained for
\GK{self-similar} diffusing vortex
\GK{free from the presence of a wall};
however, the \GK{essential} physical mechanism
\GK{proposed} in \SMM{figure}~\ref{fig16}
is thought to be
\GK{valid even} for the
\GK{time-independent} optimal state \GK{in a plane channel}.

Besides, the near-wall vortex tube 
\GGK{of the anticyclonic inclination}
has another role in heat transfer enhancement\GGK{, which will be more significant at higher Reynolds numbers}.
Figure \ref{fig17}\GGK{(\textit{a}--\textit{d})} 
respectively show the isocontours of
$\omega_{z}'$, \GKK{$Q$, $D_{S}-D_{S}^{\rm 2D}$ and $D_{E}-D_{E}^{\rm 2D}$} at $Re=\GKK{249}$.
\GGK{In \SMM{figure} \ref{fig17}\GKK{(\textit{a})}
\SMM{are shown} the velocity vectors}
relative to the lower wall \GGK{in} the ($x,y$)-plane.
The anticyclonic vortex
\GGK{of $\omega_{z}'>0$} is 
\GGK{found 
around \GKK{$(x,y)=(2.4,-0.7)$}}.
\SMM{In} \SMM{figure} \ref{fig17}(\textit{c}), the  
\GGK{higher scalar dissipation} \GGK{can be}
observed in the 
\GKK{downstream (or
\GGK{right-hand-side}) flank} of the vortex.
This local enhancement
\GGK{of $D_{S}$} is attributed to the above-described tightening of wrapped isotherms by the cross-axial shear flow just downstream (larger $x$) of the anticyclonic vortex.
On the other hand, the anticyclonic vortex induces the downward flow 
\GGK{towards} the lower wall upstream of
\GGK{(or in the left-hand side of)} the vortex.
The symmetric counterpart of this near-wall vortex 
\GKK{exists} on the upper wall \GGK{as well}.
The high- (or low-) temperature fluid is pushed on the lower cold (or upper hot) wall by the near-wall anticyclonic vortex, leading to a large temperature gradient near the walls.
As a result, the local scalar dissipation is significantly increased near the 
wall upstream of the \GGK{anticyclonic vortices}.
\GGK{The distribution of $D_{E}$ is much differ from $D_{S}$ \SMM{(cf. \GKK{figure} \ref{fig17}\textit{c,d})}.
The local energy dissipation is significantly increased just below the anticyclonic vortex \SMM{due to} the no-slip boundary condition.}

Figure \ref{fig18} shows the fractional contribution to the total scalar dissipation.
Each contribution represents the 
\GGK{integration} over \GGK{one of}
the separated \GGK{four regions} by using $Q$ and \GGK{$\omega_{z}'$},
defined as
\begin{eqnarray}
\label{eq6-7}
\displaystyle
\varepsilon_{S,n}=\frac{1}{RePrV}\int_{\Omega_{n}}{\left( \frac{\partial \theta}{\partial x_{j}} \right)}^{2}{\rm d}V \hspace{2em}(n=1,2,3,4),
\end{eqnarray}
where $\Omega_{1}$, $\Omega_{2}$, $\Omega_{3}$ and $\Omega_{4}$ denote the \GGK{anticyclonic} region of
$Q>0.03$ and \GGK{$\omega_{z}'>0$}, the \GGK{cyclonic} region of $Q>0.03$ and \GGK{$\omega_{z}'<0$}, the shear region with $|Q|<0.03$ (i.e. comparable vorticity and strain), and the strain region
\GGK{of $Q<-0.03$}, respectively.
Note that $\varepsilon_{S}=\varepsilon_{S,1}+\varepsilon_{S,2}+\varepsilon_{S,3}+\varepsilon_{S,4}+1/(RePr)$.
$\varepsilon_{S,3}$ is dominant at $Re=247.5$ and $Re=250$ \SMM{(figure18{\textit a})}.
The main contributor to $\varepsilon_{S,3}$ is the near-wall `impinging jet' in between the large-scale circulation rolls, which \GGK{significantly} increases the heat flux on the walls \GGK{as in the case of thermal convection}.
This contribution is\GGK{, however,} decreased as $Re$ increases.
Immediately after the bifurcation 
\GGK{of} the three-dimensional optimal field
\GGK{from the two-dimensional branch}, $\varepsilon_{S,1}$
\GGK{and $\varepsilon_{S,4}$ are} 
\GGK{remarkably} increased as shown in \SMM{figure} \ref{fig18}(\textit{b}).
\GGK{These are 
consequences} of the
\GGK{anticyclonic inclination of the streamwise vortex tubes} 
\GGK{leading to the} local enhancement of the temperature gradient
\GGK{by the cross-axial mean shear around the anticyclonic tube as well as by
the tube-induced flow
towards the wall, in the vicinity of which the strong
strain
arises}.
At large \GGK{Reynolds numbers}
$\varepsilon_{S,4}$ \GGK{becomes dominant.}

\section{Summary and discussion}\label{sec7} 
We have
\GGK{explored} optimal heat transfer enhancement in plane Couette flow.
Optimal
time-independent velocity \GGK{fields}
\GGK{have been found}
by \GGK{maximisation} of an objective functional defined as the excess of
\GGK{a wall heat flux (total scalar dissipation) over
total energy dissipation
under the constraint of incompressibility of the velocity
and an advection-diffusion equation for temperature}.
At quite \GGK{a
low} Reynolds number \GGK{$Re\sim \GKK{10^0}$},
\GGK{an} optimal state is 
laminar flow, but non-trivial solutions appear at \SMM{a} higher 
\GGK{Reynolds number $Re\sim \GKK{10^1}$}.
We have
obtained a streamwise-independent two-dimensional field as an optimal state.
The velocity field consists of large-scale circulation rolls that play a role in heat transfer enhancement with respect to
laminar \GGK{conductive heat transfer as in thermal convection}.
A three-dimensional optimal state has been observed to arise from pitchfork bifurcation on the two-dimensional branch
\GGK{at $Re\sim \GKK{10^2}$}.
The three-dimensional velocity field possesses similar large-scale rolls to those in the two-dimensional solutions; \GGK{however,} small-scale \GGK{quasi-}streamwise vortex tubes appear near the walls to be tilted in the spanwise direction from the streamwise direction.
The velocity fields
\GGK{exhibit} extremely high performance for heat transfer enhancement even at \GGK{much higher Reynolds numbers
$Re\sim \GKK{10^3}$}.
In the optimal states, higher scalar dissipation (i.e. \SMM{a} higher wall heat flux) is achieved with less energy dissipation than the time-averaged dissipation
\GGK{in} a \GGK{corresponding} turbulent state.

In the three-dimensional optimal state at
\GGK{high} $Re$, the \GGK{spanwise} spacing of the \GGK{quasi-streamwise} vortex tubes near the wall is proportionate to \GKK{their} distance from the wall.
The spacing of the largest \GGK{vortices, i.e. the large-scale circulation rolls,} scales with the outer length (i.e.
channel \GGK{half} width $h$), and it is independent of $Re$.
On the other hand, the spacing of the smallest vortex scales with the inner length which is given by the temperature difference and the temperature gradient on the wall.
We have observed quasi-streamwise vortical structures with hierarchical self similarity
\GGK{up to} the distance from the wall, \GGK{$0.2h$}.
Interestingly, the hierarchical structure leads to \SMM{a} logarithmic-like \SMM{mean} temperature profile in the region in which
\GGK{a convective} heat flux is dominant.

We have
\GGK{presented} the \GGK{two} mechanisms for heat transfer enhancement in the
\GGK{three-dimensional} optimal state
\GGK{in comparison to the streamwise-independent two-dimensional state}.
\GGK{The quasi-}streamwise vortex \GGK{tubes incline} in the opposite direction \GKK{to}
the vorticity of the
\GGK{mean} shear flow.
\GGK{This anticyclonic inclination of the vortex tubes brings about a remarkable increase in the local and thus total scalar dissipation.} 
\GGK{
The isotherms wrapped around the
anticyclonic vortices undergo the cross-axial mean shear, so that the
spacing of the wrapped isotherms is narrower and so the temperature gradient is steeper than those around a purely streamwise (two-dimensional) vortex tube, intensifying scalar
dissipation and so a wall heat flux.}
The \GGK{high-\SM{P\'eclet}-number and eartly-time} asymptotic analysis of the simple \GGK{analytical} model for a tilted straight vortex tube in \GGK{uniform} shear flow
\GGK{demonstrates} that 
\GGK{the}
\GGK{intensification of the local scalar dissipation around the vortex tube leads to an increase in total scalar dissipation.}
\GGK{Moreover, the tilted anticyclonic vortices
highly induce the flow towards the wall to push low- (or high-) temperature fluids on the hot (or cold) wall, enhancing the wall heat flux.}
\GGK{In} wall turbulence and
uniformly sheared turbulence, \GGK{however,} streamwise vortices
\GGK{have often been observed to incline} 
in the cyclonic direction
rather than
in the anticyclonic direction.
In
\GGK{real heat transfer enhancement of turbulent shear flows, therefore,}
\GGK{the introduction of an anticyclonic inclination of quasi-streamwise vortices 
could be new passive control strategy for heat transfer enhancement.}
\GGK{Our preliminary study on the introduction of an anticyclonic quasi-streamwise vortex in laminar \SMM{wall-bounded} shear flows has confirmed \SMM{significant} heat transfer enhancement in comparison to that of a cyclonic or purely streamwise vortex.}

\cite{Yamamoto2013} \GGK{have}
\GGK{observed}
a
\GGK{roughly spanwise-independent travelling-wave-like velocity field in their controlled turbulent channel flow
by optimisation of the dissimilarity
between heat and momentum transfer}.
In \GGK{such a maximisation} problem of the dissimilarity
the
\GGK{Navier--Stokes equation
should be taken into consideration as a constraint on the maximisation.}
The
optimal \GGK{states found in this work are} the
\GGK{outcomes} of pursuing \GGK{an incompressible}
velocity field which achieves
\GGK{more wall heat flux and less energy dissipation rather than
less skin friction,
and therefore
they are distinct from the roughly spanwise-independent dissimilar state
observed by \cite{Yamamoto2013}.}

\begin{figure} 
\centering
	\begin{minipage}{.8\linewidth}
	\includegraphics[clip,width=\linewidth]{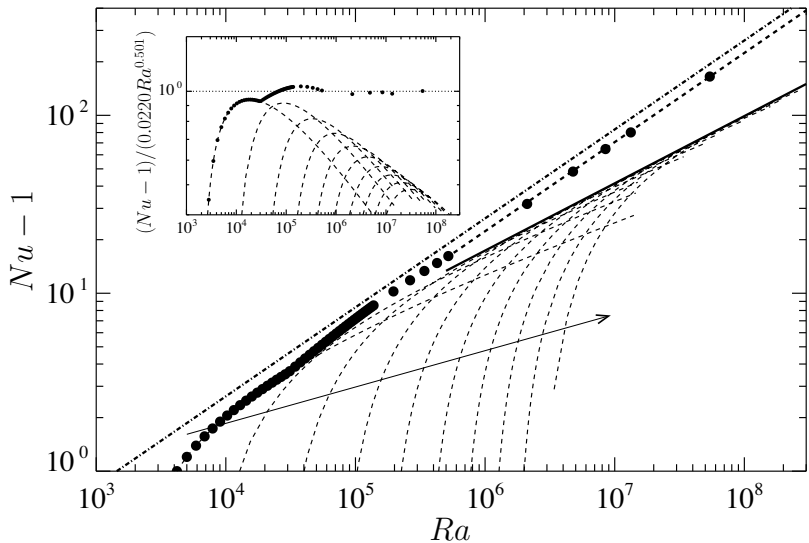}
	\end{minipage}

\caption{Nusselt number $Nu$ as a function of Rayleigh number $Ra$.
\SMM{Symbolds} denote the optimal state for $\lambda=0.1$.
The thick dashed line shows the power fit $Nu-1=0.0220Ra^{0.501}$ determined in the range $5\times10^5<Ra<6\times10^7$.
The dashed curves stand for
\GGK{the values optimised only within a streamwise-independent
two-dimensional velocity field of}
the aspect ratio $L_{z}/L_{y}=\pi/2,\pi/4,\pi/6,\pi/8,\pi/10,\pi/12,\pi/14,\pi/16,\pi/18,\pi/20$ along the arrow.
The thick solid line indicates the envelope fit $Nu-1=0.0937Ra^{0.378}$ determined in the range $10^7<Ra<10^8$.
\SMM{The dash-dotted line represents the upper bound in the Rayleigh--B\'enard convection with the no-slip boundary condition, $Nu-1=0.02634Ra^{1/2}$ \citep{Plasting2003}.}
\GGK{The} inset shows the compensated $Nu$.
\label{fig19}}
\end{figure}

\cite{Hassanzadeh2014} have \GGK{recently} reported \GGK{time-independent} optimal states for two-dimensional convection between \GGK{two parallel impermeable but free-slip plates of a constant \SM{temperature} difference}
by \GGK{maximising a} heat flux under the three constraints: divergence-free velocity field, temperature field satisfying \GGK{an} advection-diffusion equation, and fixed enstrophy.
They related \GGK{their} optimal states to the Rayleigh--B\'{e}nard convection problem, and found the scaling of
\GGK{the Nusselt number $Nu$ with
the Rayleigh number $Ra$}
as $Nu\sim Ra^{5/12}$, which agrees with the upper bounds
 in the case of \GGK{the} free-slip
\GGK{\GKK{boundary} condition,}
derived by the background method \citep{Lerley2006,Whitehead2012}.
Let us
\GGK{make} a similar
\GGK{estimate} for the
\GGK{optimal states obtained in the present study}.
For the combined natural
\GGK{and} forced convection in plane Couette flow,
\GGK{the} total energy budget equation is given by
\begin{eqnarray}
\label{eq7-1}
\displaystyle
8Re^{3}Pr^{2}(2\varepsilon_{E}-C_{f})=Ra(Nu-1).
\end{eqnarray}
By estimating $Ra$ from $\varepsilon_{E}$, $C_{f}$ and $Nu=(RePr/2)St$
\GGK{in the present optimal states}, we 
\GGK{have} found the scaling \GGK{of $Nu$ with $Ra$} as $Nu-1=0.0220Ra^{0.501}$ for the \GGK{three-dimensional}
optimal state at $Re=10^{3}-10^{4}$,
\GGK{or equivalently $Ra=5\times10^5-6\times10^7$
(see \SMM{figure} \ref{fig19}).}
The \GGK{observed} exponent $1/2$ \GGK{of $Ra$}
\GGK{is consistent with} the scaling of the upper bound in the Rayleigh--B\'{e}nard convection with
\GGK{the} no-slip boundary condition \citep{Doering1996,Plasting2003}, and \GGK{also}
accords with Kolmogorov's scaling \SMM{law} of the turbulent energy dissipation, that is $\varepsilon_{E}=\rm const.$ as $Re\rightarrow\infty$.
As shown in \SMM{figure} \ref{fig8}\GGK{(\textit{b})}, the optimal state at large $Re$
\GGK{achieves nearly constant} scalar and energy dissipation,
and it
\GGK{exhibits a} three-dimensional velocity field
\GGK{accommodating} hierarchical
\GGK{vortical structures} as
\GGK{discussed} in \S \ref{sec5}.
These results suggest that the three-dimensional velocity field with hierarchical self similarity might
\GGK{also} lead to optimal heat transfer between two parallel no-slip
\GGK{plates relatively at rest} under
\GGK{fixed} energy dissipation.

\section*{Acknowledgements}\label{sec:Acknowledgements} 
We are grateful to Professor C. R. Doering at
University of Michigan 
for
\GGK{useful} discussions.
This work was partially supported by a Grant-in-Aid Scientific Research (Grant Nos. 25249014, 26630055) from the Japanese Society for Promotion of Science (JSPS).
S.M. is supported by JSPS Grant-in-Aid for JSPS Fellows Grant Number 16J00685.
This research partly used computational resources under Collaborative Research Program for Young Scientists provided by Academic Center for Computing and Media Studies, Kyoto University.

\appendix

\begin{figure} 
\centering
	\begin{minipage}{.7\linewidth}
	\includegraphics[clip,width=\linewidth]{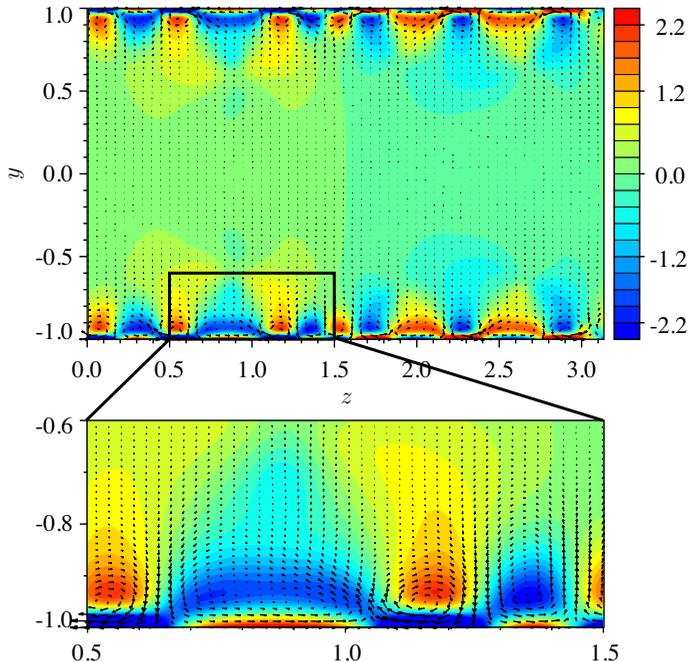}
	\end{minipage}
\caption{(Colour online) \SM{The \SMM{cross-stream} external force vectors $(f_{z},f_{y})$ and the streamwise vorticity $\omega_{x}$ in the plane $x=0$ for the corresponding optimal state at $Re=1000$ for $\lambda=0.1$.}
\label{fig20}}
\end{figure}

\begin{figure} 
\centering
	\begin{minipage}{.85\linewidth}
	(\textit{a})\\
	\includegraphics[clip,width=\linewidth]{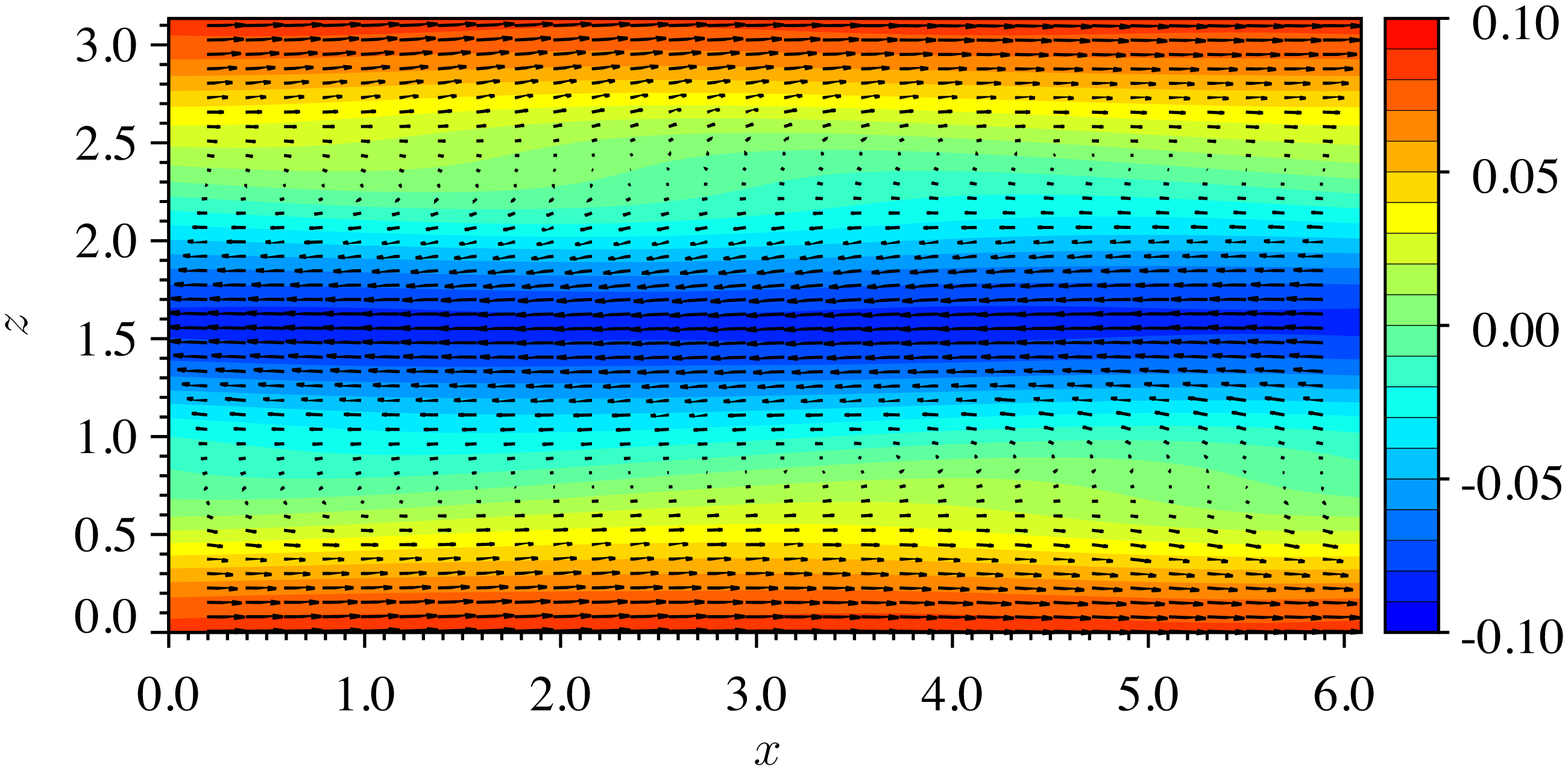}
	\end{minipage}

	\vspace{1em}

	\begin{minipage}{.85\linewidth}
	(\textit{b})\\
	\includegraphics[clip,width=\linewidth]{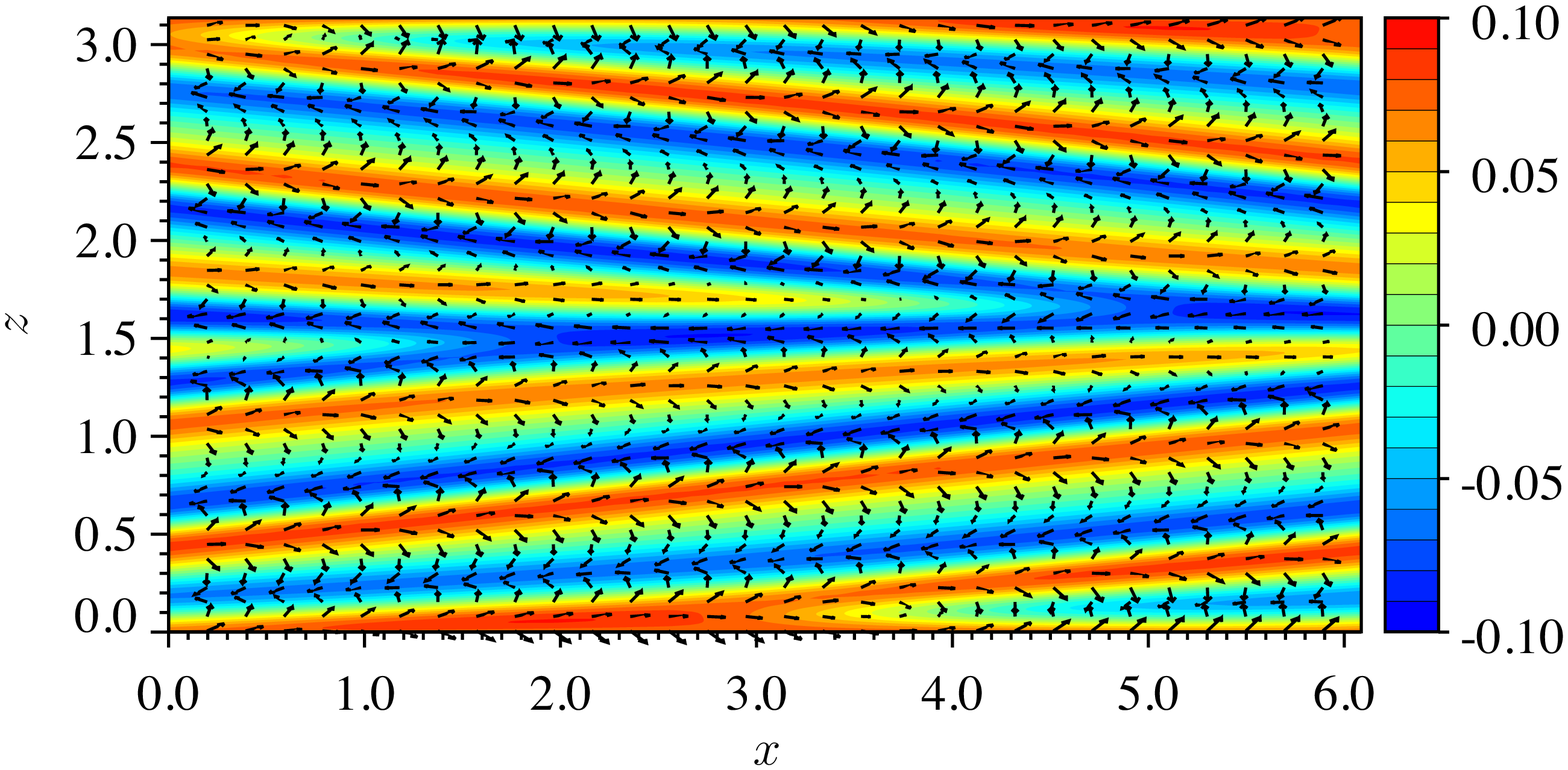}
	\end{minipage}

\caption{(Colour online) Spatial distribution of the 
\GGK{wall-parallel} external force and the wall-normal velocity
\GGK{for the corresponding} optimal state at $Re=1000$ for $\lambda=0.1$.
The \SMM{wall-parallel} external force vectors $(f_{x},f_{z})$ and the wall-normal velocity $v$ in the plane \SMM{(\textit{a})} $y=0$, (\textit{b}) $y=-0.9$. 
\label{fig21}}
\end{figure}

\begin{figure} 
\centering
	\begin{minipage}{.48\linewidth}
	(\textit{a})\\
	\includegraphics[clip,width=\linewidth]{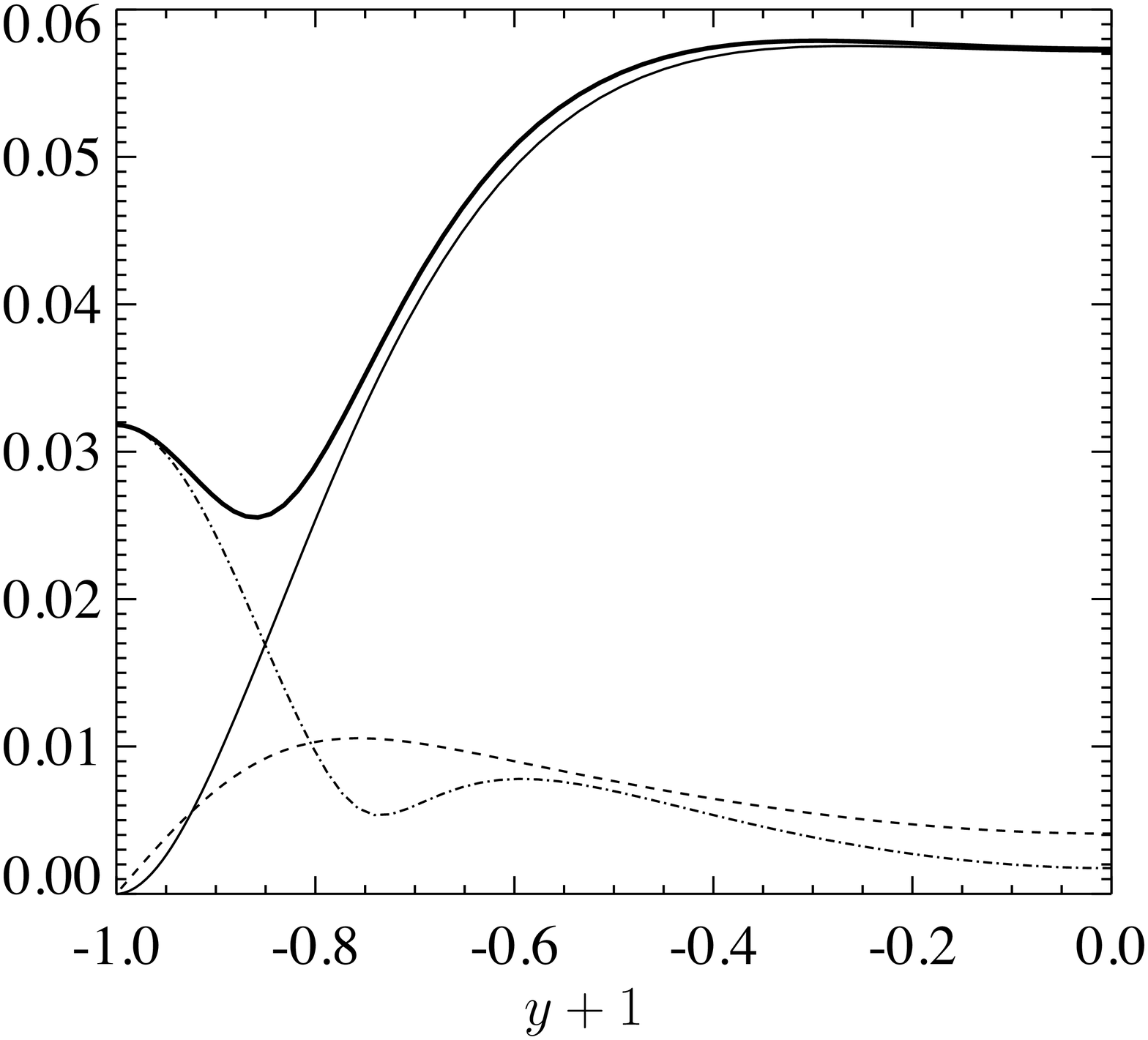}
	\end{minipage}
	\hspace{1em}
	\begin{minipage}{.48\linewidth}
	(\textit{b})\\
	\includegraphics[clip,width=\linewidth]{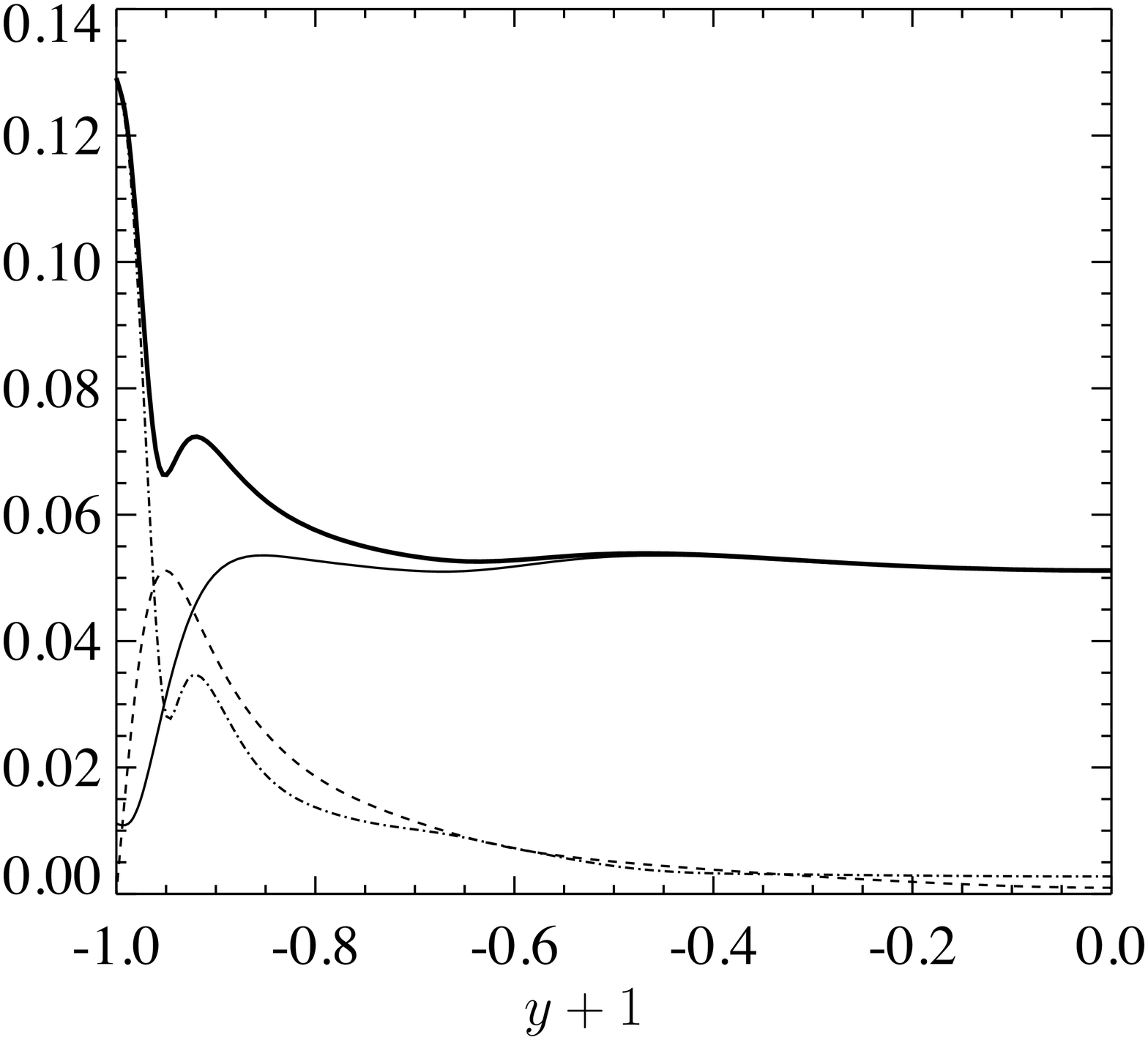}
	\end{minipage}

\caption[]{Root mean square of the external force
\SMM{for the corresponding optimal state}
\GGK{as a function of the distance to the lower wall $y+1$}
for $\lambda=0.1$ at (\textit{a}) $Re=200$, (\textit{b}) $Re=1000$.
\bsolid, ${\left< {|\mbox{\boldmath$f$}|}^{2} \right>}_{xz}^{1/2}$; \solid, ${\left< f_{x}^{2} \right>}_{xz}^{1/2}$; \dashed, ${\left< f_{y}^{2} \right>}_{xz}^{1/2}$; \chndot, ${\left< f_{z}^{2} \right>}_{xz}^{1/2}$.
\label{fig22}}
\end{figure}

\begin{figure} 
\centering
	\begin{minipage}{.48\linewidth}
	(\textit{a})\\
	\includegraphics[clip,width=\linewidth]{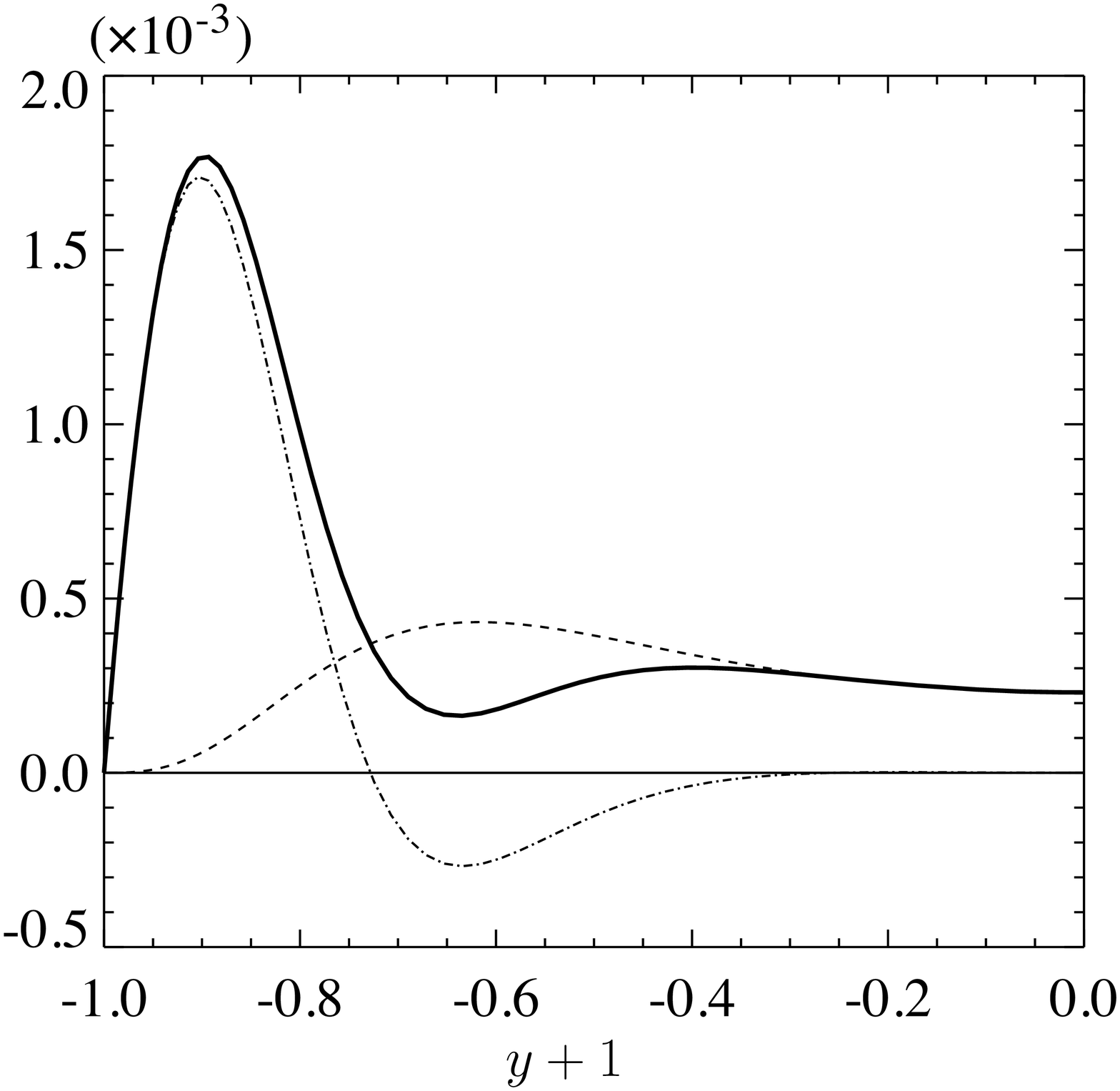}
	\end{minipage}
	\hspace{1em}
	\begin{minipage}{.48\linewidth}
	(\textit{b})\\
	\includegraphics[clip,width=\linewidth]{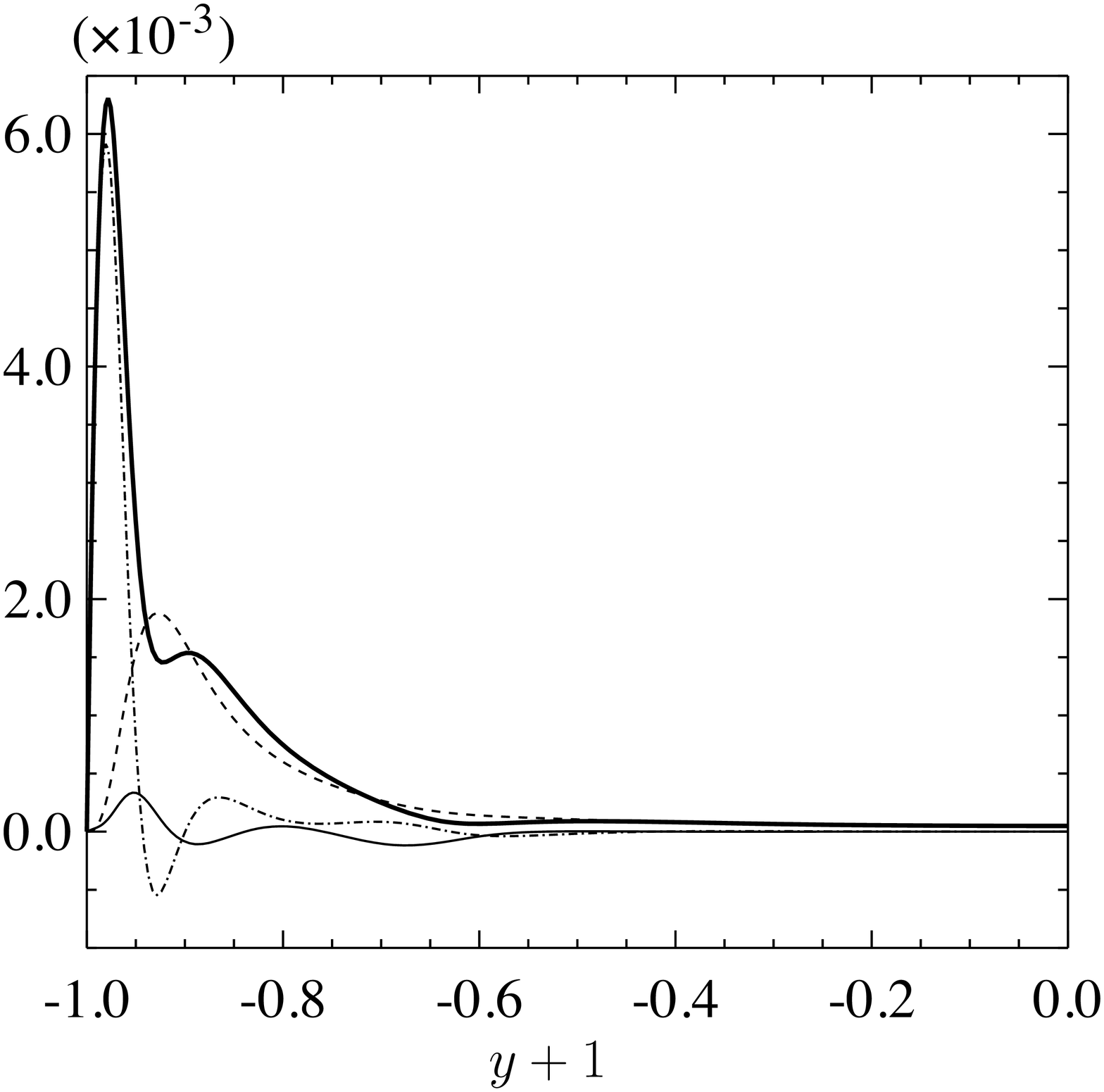}
	\end{minipage}

\caption[]{Mean energy input by the external force
\SMM{for the corresponding optimal state}
for $\lambda=0.1$ at (\textit{a}) $Re=200$, (\textit{b}) $Re=1000$.
\bsolid, ${\left< u_{i}f_{i} \right>}_{xz}$; \solid, ${\left< uf_{x} \right>}_{xz}$; \dashed, ${\left< vf_{y} \right>}_{xz}$; \chndot, ${\left< wf_{z} \right>}_{xz}$.
\label{fig23}}
\end{figure}

\section{External body force}\label{appA} 
\GGK{As a consequence of the optimisation} of the objective functional $J$, we have \GGK{obtained} incompressible time-independent velocity fields.
The 
\GGK{resulting} velocity fields do not 
\GGK{satisfy} the Navier--Stokes equation.
Thus, in order to \GGK{realise} the optimal state as a steady solution to the Navier--Stokes equation, we need external force.
In this \SMM{appendix}, we estimate the external body force to discuss its characteristics.

From the steady Navier--Stokes equation, the
\GGK{dimensionless} external body force
\GGK{to be added} can be
\GGK{expressed as}
\begin{eqnarray}
\label{a1}
f_{i}=\frac{\partial p}{\partial x_{i}}+u_{j}\frac{\partial u_{i}}{\partial x_{j}}-\frac{1}{Re}\frac{\partial^{2}u_{i}}{\partial x_{j}^{2}}
\end{eqnarray}
\GGK{in terms of the optimised velocity field $\mbox{\boldmath$u$}$.}
We
\GGK{now search for} a \GGK{solenoidal} force field,
\GGK{the wall-normal component of which is null on the walls,
that is $f_{y}(y=\pm1)=0$.}
By taking the divergence of
equation (\ref{a1}), we 
\GGK{have} the Poisson equation for the pressure $p$,
\begin{eqnarray}
\label{a3}
\frac{\partial^{2} p}{\partial x_{i}^{2}}=\frac{\partial}{\partial x_{i}}\left( u_{j}\frac{\partial u_{i}}{\partial x_{j}} \right).
\end{eqnarray}
The wall-normal component of
equation (\ref{a1}) and the 
\GGK{no-slip and impermeable
condition on the walls (i.e., $\mbox{\boldmath$u$}(y=\pm1)=0$)} 
\GGK{yield} the \GGK{usual} boundary condition for $p$
\GGK{on the walls $y=\pm 1$,}
\begin{eqnarray}
\label{a4}
\frac{\partial p}{\partial y}=
\frac{1}{Re}\GGK{\frac{\partial^{2}v}{\partial y^{2}}.}
\end{eqnarray}
We obtain $p$
\GGK{as a}
solution to equation \GGK{(\ref{a3})
supplemented by (\ref{a4}).}
\GGK{It turns out that} the external force field $\mbox{\boldmath$f$}$
\GGK{can be}
determined
\GGK{by substituting $\mbox{\boldmath$u$}$ and $p$ in} equation (\ref{a1}).

Figure \ref{fig20} and \ref{fig21} show the 
\GGK{cross-stream components of the} external force and the
\GGK{streamwise vorticity for the corresponding optimal field} 
at $Re=1000$ for $\lambda=0.1$.
\SMM{The} external force
\GGK{appears}
so that it may generate small-scale streamwise vortical motion near the \GGK{walls}. 
\GGK{At the same time}
the external force has a role in driving the large-scale circulation rolls.
\SMM{The} streamwise external force $f_{x}$ plays another important role in maintaining the steady state. 
\GGK{It can be seen}
in \SMM{figure} \ref{fig21}(\textit{a})
\GGK{that} $f_{x}$ is dominant in the central region of the channel, and the sign of $f_{x}$ corresponds to that of $v$, 
\GGK{implying }
that the external force works to suppress the advection of the streamwise velocity $u$ caused by the large-scale circulation rolls.
In other words, the generation of
large-scale \GGK{streaks} is inhibited by the external force.
Near the \GGK{walls}, the force 
\SMM{acts}
each streamwise vortex
\SMM{(figure \ref{fig21}\textit{b})}.
If the velocity field were to obey the Navier--Stokes equation
\GGK{without forcing},
the vortex tube would wrap isocontours of the streamwise velocity around itself \GGK{as in the wrapping of isotherms}
to form near-wall streaks.
Thus, the vortex tubes 
\GGK{of} the anticyclonic
\GGK{inclination}
would \GGK{also}
increase the energy dissipation 
\GGK{in addition to the scalar dissipation}
by the 
\GGK{same} mechanism described in \S\ref{sec6} \citep[see also][]{Kawahara2005}.
The external force eliminates this wrapping effect of the streamwise velocity, and thus only the scalar dissipation is enhanced.

Figure \ref{fig22} shows the root mean square of the external force 
\GGK{as a function of the distance to the lower wall} at $Re=200$ and $1000$.
At $Re=200$, the optimal state is 
\GGK{\SM{given} by} a streamwise-independent two-dimensional
\GGK{velocity field.}
\GGK{Note that the}
\GGK{longitudinal axis of the figure}
\GGK{denotes} the ratio of the magnitude of the external force to the inertial force \GGK{$U^2/h$}.
In both cases, the spanwise component ${\left< f_{z}^{2} \right>}_{xz}^{1/2}$ is dominant near the wall.
The spanwise force creates the large-scale circulation rolls and the small-scale streamwise vortices.
The streamwise component is rapidly increased with \GGK{the}
distance from the wall, and
\GGK{cross-streamwise force is} significantly small in the central region of the channel.

\GGK{The} net energy input by the streamwise force
\GGK{can be seen to be} significantly small
in \SMM{figure} \ref{fig23},
\GGK{which is a result of cancelling out of}
the \GGK{energy} input \GGK{($uf_{x}>0$)} and sink
\GGK{($uf_{x}<0$)}.
In
the two-dimensional \GGK{optimal}
state \GGK{at $Re=200$}, \SMM{strictly} ${\left< uf_{x} \right>}_{xz}=0$ at any height due to the \GGK{flow} symmetry.

\section{\GGK{Effects} of
\SMM{the inclination of a straight}
vortex tube on
scalar dissipation in
uniform shear flow}\label{appB} 
\begin{figure} 
\centering
	\begin{minipage}{.35\linewidth}
	\includegraphics[clip,width=\linewidth]{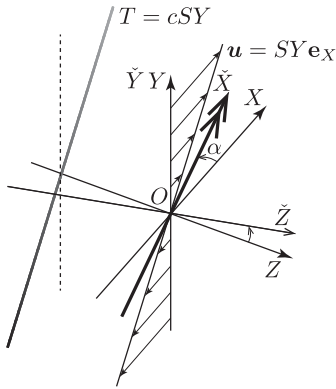}
	\end{minipage}

\caption{Flow configuration \GGK{of the analytical model.
The central axis of the vortex tube (the $\check{X}$-axis)
is represented by the double arrow.}\label{fig24}}
\end{figure}

\cite{Kawahara2005} has obtained 
\GGK{high-Reynolds-number and early-time}
asymptotic solutions to the incompressible Navier--Stokes equation, which represent spiral vortex layers around a straight vortex tube in uniform shear flow.
\GGK{His solutions have} shown that the inclination of the vortex
\GGK{from} the streamwise direction
\GGK{in the spanwise direction} increases or decreases total energy dissipation.
In this
\GGK{appendix}, we 
\GGK{perform} a similar analysis to
\GGK{the} problem of heat transfer around a \GGK{straight} vortex tube
in \GGK{uniform} shear flow.

\subsection{Basic equations} 
Let us consider
\GGK{temporal evolution of a passive scalar starting from
an initial temperature field $cSY$}
around an infinitely straight vortex tube in uniform shear flow $SY\mbox{\boldmath$\rm e$}_{X}$, where $S$ ($>0$) is a constant shear rate,
\GGK{$cS$ is an initial uniform temperature gradient
($c$ being a constant whose dimension is temperature/velocity),}
and $\mbox{\boldmath$\rm e$}_{X}$ denotes a unit vector in the $X$-direction (see \SMM{figure} \ref{fig24}).
The central axis of the vortex tube is on the plane $Y=0$ and inclined at an angle $\alpha$ from the $X$-axis.
We introduce the other coordinate system $O\check{X}\check{Y}\check{Z}$ given by rotating the original coordinate system $OXYZ$ at the angle $\alpha$ around the $Y$-axis.
The $\check{X}$-axis corresponds to the central axis of the vortex tube.

If we suppose that the velocity,
the pressure \GGK{and so the temperature} fields
are uniform along the central axis of the tube, i.e., the $\check{X}$-axis, the velocity field $\mbox{\boldmath$u$}$ can be expressed using the 
\GGK{streamfunction} $\psi(\check{Y},\check{Z},t)$ as
\begin{eqnarray}
\label{b1}
\mbox{\boldmath$u$}=u\mbox{\boldmath$\rm e$}_{\check{X}}+\frac{\partial\psi}{\partial \check{Z}}\mbox{\boldmath$\rm e$}_{\check{Y}}+\left( -\frac{\partial\psi}{\partial \check{Y}}+S\check{Y}\sin\alpha \right)\mbox{\boldmath$\rm e$}_{\check{Z}}.
\end{eqnarray}
\GGK{The evolution of the temperature field $T(\check{Y},\check{Z},t)$ is}
governed by the advection-diffusion equation
\begin{eqnarray}
\label{b2}
\left\{ \frac{\partial}{\partial t}-\kappa\left( \frac{\partial}{\partial \check{Y}^{2}}+\frac{\partial}{\partial \check{Z}^{2}} \right) \right\}T-\frac{\partial(\psi,T)}{\partial(\check{Y},\check{Z})}=-S\check{Y}\sin\alpha\frac{\partial T}{\partial \check{Z}}.
\end{eqnarray}

By introducing the plane polar coordinates ($r,\varphi$) with $\check{Y}=r\cos\varphi$ and $\check{Z}=r\sin\varphi$, 
equation (\ref{b2}) is rewritten as
\begin{eqnarray}
\label{b4}
\left\{ \frac{\partial}{\partial t}-\kappa\left( \frac{\partial^{2}}{\partial r^{2}}+\frac{1}{r}\frac{\partial}{\partial r}+\frac{1}{r^{2}}\frac{\partial^{2}}{\partial \varphi^{2}} \right) \right\}T-\frac{1}{r}\frac{\partial(\psi,T)}{\partial(r,\varphi)}&{\ }& \nonumber \\
&{\ }&\hspace{-10em}=-\frac{1}{2}S\sin\alpha\left( r\sin 2\varphi\frac{\partial}{\partial r}+\cos 2\varphi\frac{\partial}{\partial\varphi}+\frac{\partial}{\partial\varphi} \right)T.
\end{eqnarray}
\GGK{The temperature at $r\rightarrow\infty$, far from the vortex tube,
is given by}
\begin{eqnarray}
\label{b5}
T=cS\check{Y},
\end{eqnarray}
\GGK{while} $T$ is assumed to be regular at $r=0$.

In the coordinate $(r,\varphi)$, the local scalar dissipation is given by
\begin{eqnarray}
\label{b6}
D_{S}^{*}(r,\varphi,t)=\kappa\left[ {\left( \frac{\partial T}{\partial r} \right)}^{2}+\frac{1}{r^{2}}{\left( \frac{\partial T}{\partial \varphi} \right)}^{2} \right].
\end{eqnarray}

\subsection{
\GGK{High-\SM{P\'eclet}-number asymptotic solution}
} 
Let us 
\GGK{compute}
the scalar dissipation around the diffusing vortex tube that is 
\GGK{axisymmetric} with respect to the $\check{X}$-axis as
\begin{eqnarray}
\label{b7}
\psi(r,t)=-\frac{\Gamma}{2\pi}\int_{0}^{\eta}\frac{1-{\rm e}^{-\xi^{2}}}{\xi}{\rm d}\xi,
\end{eqnarray}
where $\Gamma$ ($>0$) is the circulation, and 
\begin{eqnarray}
\label{b8}
\eta=\frac{1}{2}r(\kappa t)^{-1/2}
\end{eqnarray}
is a similarity variable.
\SMM{Here we have supposed that $Pr=1$ (i.e. $\nu=\kappa$)}. A self-similar solution
\GGK{to} equation (\ref{b4})
\GGK{for} $\alpha=0$ can be expressed
\GGK{as}
\begin{eqnarray}
\label{b9}
T(r,\varphi,t)=cSr{\rm Re}[g(\eta)\GGK{\mbox{e}}^{-{\rm i}\varphi}].
\end{eqnarray}
Substitution of (\ref{b7}) and (\ref{b9}) into (\ref{b4}) yields
\begin{eqnarray}
\label{b10}
\frac{{\rm d}^{2}g}{{\rm d}\eta^{2}}+\left( 2\eta+\frac{3}{\eta} \right)\frac{{\rm d}g}{{\rm d}\eta}+{\rm i}\frac{\Gamma}{2\pi\kappa}\frac{1-{\rm e}^{-\eta^{2}}}{\eta^{2}}g=0.
\end{eqnarray}
Boundary conditions to be imposed are that $rg(\eta)$ is regular at $\eta=0$ and $g(\infty)=1$.
The asymptotic solution to (\ref{b10}) at 
\GGK{high \SM{P\'eclet} number}
$\Gamma/\kappa\gg 1$ has been obtained by \cite{Moore1985} and \cite{Kawahara1997}.
In the far region $\eta\gg (\Gamma/\kappa)^{1/4}$, the solution can be written as
\begin{eqnarray}
\label{b11}
g\approx{\rm exp}\left(\GGK{\mbox{i}}\frac{P_{\Gamma}}{4\eta^{2}}-\frac{P_{\Gamma}^{2}}{48\eta^{6}} \right),
\end{eqnarray}
where
\begin{eqnarray}
\label{b12}
P_{\Gamma}=\frac{\Gamma}{2\pi\kappa}
\end{eqnarray}
\GGK{is} the \GGK{vortex} \SM{P\'eclet} number.
Substituting (\ref{b9}) with the asymptotic expression (\ref{b11}) into (\ref{b6}), we have
\begin{eqnarray}
\label{b13}
\hspace{-2em}D_{S}^{*}=\kappa c^{2}S^{2}\left[ \frac{1}{2}{\left| \eta \frac{{\rm d}g}{{\rm d}\eta} \right|}^{2}+\frac{1}{2}\eta\frac{{\rm d}{|g|}^{2}}{{\rm d}\eta}+{|g|}^{2} \right]+\kappa c^{2}S^{2}{\rm Re}\left[ \eta \frac{{\rm d}g}{{\rm d}\eta}\left( g+\frac{1}{2}\eta\frac{{\rm d}g}{{\rm d}\eta} \right)
\GGK{\mbox{e}}^{-2{\rm i}\varphi} \right].
\end{eqnarray}
Using (\ref{b13}), the total dissipation per unit axial length is evaluated as
\begin{eqnarray}
\label{b14}
\int_{0}^{\infty}\int_{0}^{2\pi}D_{S}^{*} r{\rm d}r{\rm d}\varphi=8\pi\kappa^{2}c^{2}S^{2}I_{0}t,
\end{eqnarray}
where
\begin{eqnarray}
\label{b15}
I_{0}=\int_{0}^{\infty}\left[ \frac{1}{2}{\left| \eta \frac{{\rm d}g}{{\rm d}\eta} \right|}^{2}+\frac{1}{2}\eta\frac{{\rm d}{|g|}^{2}}{{\rm d}\eta}+{|g|}^{2} \right] \eta{\rm d}\eta.
\end{eqnarray}

\subsection{
\GGK{Early-time asymptotic solution}} 
If we consider the limit of a small tilt of a vortex tube $|\alpha|\ll 1$,
a solution
to the advection-diffusion equation (\ref{b4})
\GGK{may} be written as
\begin{eqnarray}
\label{b16}
T(r,\varphi,t)=cSr[T_{0}(\eta,\varphi)+St\alpha T_{1}(\eta,\varphi)+\cdot\cdot\cdot]
\end{eqnarray}
at early time $St|\alpha|\ll 1$, where the leading-order temperature is given in a separation-of-variable form
\begin{eqnarray}
\label{b17}
T_{0}={\rm Re}[g(\eta){\rm e}^{-{\rm i}\varphi}],
\end{eqnarray}
and the higher-order \SM{temperature} is
\begin{eqnarray}
\label{b18}
T_{1}={\rm Re}\{ {\rm i}[g_{1}(\eta){\rm e}^{-{\rm i}\varphi}+g_{3}(\eta){\rm e}^{-3{\rm i}\varphi}] \}.
\end{eqnarray}
By substituting (\ref{b16}) with (\ref{b17}) and (\ref{b18}) to (\ref{b4}), we obtain, for the $\varphi$-dependent part,
\begin{eqnarray}
\label{b19}
\frac{{\rm d}^{2}g_{1}}{{\rm d}\eta^{2}}+\left( 2\eta+\frac{3}{\eta} \right)\frac{{\rm d}g_{1}}{{\rm d}\eta}-4g_{1}+{\rm i}\frac{\Gamma}{2\pi\kappa}\frac{1-{\rm e}^{-\eta^{2}}}{\eta^{2}}g_{1}=\zeta_{1}(\eta),
\end{eqnarray}
and obtain, for the $3\varphi$-dependent part,
\begin{eqnarray}
\label{b20}
\frac{{\rm d}^{2}g_{3}}{{\rm d}\eta^{2}}+\left( 2\eta+\frac{3}{\eta} \right)\frac{{\rm d}g_{3}}{{\rm d}\eta}-4\left( 1+\frac{2}{\eta^{2}} \right)g_{3}+3{\rm i}\frac{\Gamma}{2\pi\kappa}\frac{1-{\rm e}^{-\eta^{2}}}{\eta^{2}}g_{3}=\zeta_{3}(\eta).
\end{eqnarray}
The inhomogeneous terms of (\ref{b19}) and (\ref{b20}) are
\GGK{given by}
\begin{eqnarray}
\label{b21}
\displaystyle
\zeta_{1}&=&\eta\left( \frac{{\rm d}g}{{\rm d}\eta} \right)^{\dag}-4{\rm i}{\rm Im}(g), \\
\label{b22}
\displaystyle
\zeta_{3}&=&\eta\frac{{\rm d}g}{{\rm d}\eta},
\end{eqnarray}
where $\dag$ denotes \GGK{a complex conjugate}.
Boundary conditions to be imposed are that $rtg_{1}(\eta)$ and $rtg_{3}(\eta)$ are regular at $\eta=0$, and $rtg_{1}(\eta)\rightarrow 0$ and $rtg_{3}(\eta)\rightarrow 0$ as $\eta\rightarrow\infty$.
At  $\eta\gg(\Gamma/\kappa)^{1/4}$, by using the asymptotic expression (\ref{b11}), $\zeta_{1}$ and $\zeta_{3}$ can be written,
\GGK{respectively,} as
\begin{eqnarray}
\label{b23}
\displaystyle
\zeta_{1}&\approx& {\rm i}\frac{P_{\Gamma}}{2\eta^{2}}{\rm exp}\left( -{\rm i}\frac{P_{\Gamma}}{4\eta^{2}}-\frac{P_{\Gamma}^{2}}{48\eta^{6}} \right)
-4{\rm i}\sin \left( \frac{P_{\Gamma}}{4\eta^{2}} \right){\rm exp}\left( -\frac{P_{\Gamma}^{2}}{48\eta^{6}} \right), \\
\label{b24}
\displaystyle
\zeta_{3}&\approx& -{\rm i}\frac{P_{\Gamma}}{2\eta^{2}}{\rm exp}\left( {\rm i}\frac{P_{\Gamma}}{4\eta^{2}}-\frac{P_{\Gamma}^{2}}{48\eta^{6}} \right).
\end{eqnarray}
If we consider the far region $\eta\gg(\Gamma/\kappa)^{1/4}$, all the terms
\GGK{including}
${\rm e}^{-\eta^{2}}$ can be neglected, and thus we can rewrite (\ref{b19}) and (\ref{b20}) as
\begin{eqnarray}
\label{b25}
&\displaystyle
\frac{{\rm d}^{2}g_{1}}{{\rm d}\eta^{2}}+\left( 2\eta+\frac{3}{\eta} \right)\frac{{\rm d}g_{1}}{{\rm d}\eta}-\left(4-{\rm i}\frac{\Gamma}{2\pi\kappa}\frac{1}{\eta^{2}} \right)g_{1}=\zeta_{1},& \\
\label{b26}
&\displaystyle
\frac{{\rm d}^{2}g_{3}}{{\rm d}\eta^{2}}+\left( 2\eta+\frac{3}{\eta} \right)\frac{{\rm d}g_{3}}{{\rm d}\eta}-\left[ 4+\left(8-3{\rm i}\frac{\Gamma}{2\pi\kappa} \right)\frac{1}{\eta^{2}} \right]g_{3}=\zeta_{3}.&
\end{eqnarray}
At large \SM{P\'eclet} number
\GGK{$\Gamma/(2\pi\kappa)\gg1$}, the asymptotic solutions to (\ref{b25}) and (\ref{b26}) are
\GGK{given, respectively, by}
\begin{eqnarray}
\label{b27}
\displaystyle
{\rm Re}(g_{1})&=&\frac{3}{4}\left[ \cos\left( \frac{P_{\Gamma}}{4\eta^{2}} \right)-\frac{4\eta^{2}}{P_{\Gamma}}\sin\left( \frac{P_{\Gamma}}{4\eta^{2}} \right) \right]{\rm exp}\left( -\frac{P_{\Gamma}^{2}}{48\eta^{6}} \right), \\
\nonumber \\
\label{b28}
\displaystyle
{\rm Im}(g_{1})&=&\frac{1}{4}\sin\left( \frac{P_{\Gamma}}{4\eta^{2}} \right){\rm exp}\left( -\frac{P_{\Gamma}^{2}}{48\eta^{6}} \right),
\end{eqnarray}
\begin{eqnarray}
\label{b29}
\displaystyle
g_{3}=-\left( 1-\frac{{\rm i}\eta^{2}}{2P_{\Gamma}} \right){\rm exp}\left( {\rm i}\frac{P_{\Gamma}}{4\eta^{2}}-\frac{P_{\Gamma}^{2}}{48\eta^{6}} \right)
-\frac{{\rm i}\eta^{2}}{2P_{\Gamma}}{\rm exp}\left( {\rm i}\frac{3P_{\Gamma}}{4\eta^{2}}-\frac{3P_{\Gamma}^{2}}{16\eta^{6}} \right).
\end{eqnarray}
Substituting the expansion (\ref{b16}) up to the order $St|\alpha|$ with (\ref{b17}) and (\ref{b18}) into the local scalar dissipation (\ref{b6}), we have
\begin{eqnarray}
\label{b30}
D_{S}^{*}&=&\kappa c^{2}S^{2}\left[ \frac{1}{2}{\left| \eta\frac{{\rm d}g}{{\rm d}\eta} \right|}^{2}+\frac{1}{2}\eta\frac{{\rm d}{|g|}^{2}}{{\rm d}\eta}+{|g|}^{2} \right]+\kappa c^{2}S^{2}{\rm Re}\left[ \eta\frac{{\rm d}g}{{\rm d}\eta}\left( g+\frac{1}{2}\eta\frac{{\rm d}g}{{\rm d}\eta} \right){\rm e}^{-2{\rm i}\varphi} \right] \nonumber \\
&-&2\kappa c^{2}S^{3}t\alpha{\rm Re}\left\{ {\rm i}\left[ \frac{1}{2}\eta^{2}\frac{{\rm d}g}{{\rm d}\eta}\left( \frac{{\rm d}g_{1}}{{\rm d}\eta} \right)^{\dag}+\frac{1}{2}\eta\frac{{\rm d}}{{\rm d}\eta}(gg_{1}^{\dag})+gg_{1}^{\dag} \right] \right\} \nonumber \\
&+&2\kappa c^{2}S^{3}t\alpha{\rm Re}\left\{ {\rm i}\left[ \frac{1}{2}\eta^{2}\left( \left( \frac{{\rm d}g}{{\rm d}\eta} \right)^{\dag}\frac{{\rm d}g_{3}}{{\rm d}\eta}+\frac{{\rm d}g}{{\rm d}\eta}\frac{{\rm d}g_{1}}{{\rm d}\eta} \right)+\frac{1}{2}\eta\frac{{\rm d}}{{\rm d}\eta}(g^{\dag}g_{3}+gg_{1})+2{g}^{\dag}g_{3} \right]{\rm e}^{-2{\rm i}\varphi} \right\} \nonumber \\
&+&2\kappa c^{2}S^{3}t\alpha{\rm Re}\left\{ {\rm i}\left[ \frac{1}{2}\eta^{2} \frac{{\rm d}g}{{\rm d}\eta} \frac{{\rm d}g_{3}}{{\rm d}\eta}+\frac{1}{2}\eta \frac{{\rm d}}{{\rm d}\eta}(gg_{3})-gg_{3} \right]{\rm e}^{-4{\rm i}\varphi} \right\}.
\end{eqnarray}
From (\ref{b30}), the total 
\GGK{scalar} dissipation
\GGK{for} $|\alpha|\ll 1$ is evaluated as
\begin{eqnarray}
\label{b31}
\int_{0}^{\infty}\int_{0}^{2\pi}D_{S}^{*} r{\rm d}r{\rm d}\varphi=8\pi\kappa^{2}c^{2}S^{2}I_{0}t-16\pi\kappa^{2}c^{2}S^{3}\alpha I_{1}t^{2},
\end{eqnarray}
where
\begin{eqnarray}
\label{b32}
I_{1}=\int_{0}^{\infty}{\rm Re} \left\{ {\rm i}\left[ \frac{1}{2}\eta^{2} \frac{{\rm d}g}{{\rm d}\eta}{\left(  \frac{{\rm d}g_{1}}{{\rm d}\eta} \right)}^{\dag}+\frac{1}{2}\eta \frac{{\rm d}}{{\rm d}\eta}(gg_{1}^{\dag})+gg_{1}^{\dag} \right] \right\} \eta{\rm d}\eta.
\end{eqnarray}
At
\GGK{$P_{\Gamma}
\gg 1$}, using the expressions of $g$ and $g_{1}$ given by (\ref{b11}), (\ref{b27}) and (\ref{b28}), $I_{1}$ can be written as
\begin{eqnarray}
\label{b33}
I_{1}\approx\frac{P_{\Gamma}}{32}I_{2}
\end{eqnarray}
where
\begin{eqnarray}
\label{b34}
I_{2}=3\int_{0}^{\infty}\left[ -\frac{2}{\xi^{2}}\sin\xi+\frac{1}{\xi}(1+\cos\xi) \right]{\rm exp}\left( -\frac{\xi^{3}}{3P_{\Gamma}} \right){\rm d}\xi.
\end{eqnarray}
At $P_{\Gamma}\gg 1$, \SMM{we can estimate $I_{2}$ as
\begin{eqnarray}
I_{2}\approx\ln P_{\Gamma},
\end{eqnarray}
which tells us that}
\begin{eqnarray}
\label{b38}
I_{1}\approx\frac{P_{\Gamma}}{32}\ln P_{\Gamma}.
\end{eqnarray}
Therefore, at $\Gamma/\kappa\gg 1$, we obtain the second term in the right-hand side of (\ref{b31}) as
\begin{eqnarray}
\label{b39}
-16\pi\kappa^{2}c^{2}S^{3}\alpha I_{1}t^{2}\approx -\frac{1}{4}\alpha\kappa c^{2}S^{3}\Gamma\ln\left( \frac{\Gamma}{2\pi\kappa} \right) t^{2},
\end{eqnarray}
which represents 
\GGK{the dependence of the total scalar dissipation
on
the inclination of the vortex tube}.
\GGK{When
$\alpha <0$,
that is anticyclonic inclination
(or
$\alpha >0$, that is cyclonic inclination)}, the scalar dissipation is enhanced (or reduced) in comparison with that of the neutral case of $\alpha=0$.

\bibliography{ref}

\begin{thebibliography}{38}
\expandafter\ifx\csname natexlab\endcsname\relax\def\natexlab#1{#1}\fi
\def\au#1{#1} \def\ed#1{#1} \def\yr#1{#1}\def\at#1{#1}\def\jt#1{\textit{#1}}
  \def\bt#1{#1}\def\bvol#1{\textbf{#1}} \def\vol#1{#1} \def\pg#1{#1}
  \def\publ#1{#1}\def\arxiv#1{#1}\def\org#1{#1}\def\st#1{\textit{#1}}

\bibitem[Ahlers {\em et~al.\/}(2012)Ahlers, Bodenschatz, Funfschilling,
  Grossmann, He, Lohse, R.Stevens \& Verzicco]{Ahlers2012}
{\sc \au{Ahlers, G.}, \au{Bodenschatz, E.}, \au{Funfschilling, D.},
  \au{Grossmann, S.}, \au{He, X.}, \au{Lohse, D.}, \au{R.Stevens} \&
  \au{Verzicco, R.}} \yr{2012}  \at{Logarithmic temperature profiles in
  turbulent {R}ayleigh--{B}\'enard convection}.  \jt{Phys. Rev. Lett.}
  \bvol{109}~(114501).

\bibitem[Ahlers {\em et~al.\/}(2014)Ahlers, Bodenschatz \& He]{Ahlers2014}
{\sc \au{Ahlers, G.}, \au{Bodenschatz, E.} \& \au{He, X.}} \yr{2014}
  \at{Logarithmic temperature profiles of turbulent {R}ayleigh--{B}\'enard
  convection in the classical and ultimate state for a {P}randtl number of
  0.8}.  \jt{J. Fluid Mech.}  \bvol{758},  \pg{436--467}.

\bibitem[Bewley {\em et~al.\/}(2001)Bewley, Moin \& Temam]{Bewley2001}
{\sc \au{Bewley, T.~R.}, \au{Moin, P.} \& \au{Temam, R.}} \yr{2001}
  \at{{DNS}-based predictive control of turbulence: an optimal benchmark for
  feedback algorithms}.  \jt{J. Fluid Mech.}  \bvol{447},  \pg{179--225}.

\bibitem[Busse(1969)]{Busse1969}
{\sc \au{Busse, F.~H.}} \yr{1969}  \at{On {H}oward's upper bound for heat
  transport by turbulent convection}.  \jt{J. Fluid Mech.}  \bvol{37},
  \pg{457--477}.

\bibitem[Busse(1970)]{Busse1970}
{\sc \au{Busse, F.~H.}} \yr{1970}  \at{Bounds for turbulent shear flow}.
  \jt{J. Fluid Mech.}  \bvol{41},  \pg{219--240}.

\bibitem[Chilton \& Colburn(1934)]{Chilton1934}
{\sc \au{Chilton, T.~H.} \& \au{Colburn, A.~P.}} \yr{1934}  \at{Mass transfer
  (absorption) coefficients prediction from data on heat transfer and fluid
  friction}.  \jt{Ind. Eng. Chem.}  \bvol{26},  \pg{1183--1187}.

\bibitem[Dipprey \& Sabersky(1963)]{Dipprey1962}
{\sc \au{Dipprey, D.~F.} \& \au{Sabersky, R.~H.}} \yr{1963}  \at{Heat and
  momentum transfer in smooth and rough tubes at various {P}randtl numbers}.
  \jt{J. Heat and Mass Transfer}  \bvol{6},  \pg{329--353}.

\bibitem[Doering \& Constantin(1992)]{Doering1992}
{\sc \au{Doering, C.~R.} \& \au{Constantin, P.}} \yr{1992}  \at{Energy
  dissipation in shear driven turbulence}.  \jt{Phys. Rev. Lett.}  \bvol{69},
  \pg{1648--1651}.

\bibitem[Doering \& Constantin(1994)]{Doering1994}
{\sc \au{Doering, C.~R.} \& \au{Constantin, P.}} \yr{1994}  \at{Variational
  bounds on energy dissipation in incompressible flows. {I}. {S}hear flow}.
  \jt{Phys. Rev. {\rm E}}  \bvol{49}~(5),  \pg{4087--4099}.

\bibitem[Doering \& Constantin(1996)]{Doering1996}
{\sc \au{Doering, C.~R.} \& \au{Constantin, P.}} \yr{1996}  \at{Variational
  bounds on energy dissipation in incompressible flows.
  {I\hspace{-.1em}I\hspace{-.1em}I}. {C}onvection}.  \jt{Phys. Rev. {\rm E}}
  \bvol{53}~(6),  \pg{5957--5981}.

\bibitem[Doering {\em et~al.\/}(2006)Doering, Otto \& Reznikoff]{Doering2006}
{\sc \au{Doering, C.~R.}, \au{Otto, F.} \& \au{Reznikoff, M.~G.}} \yr{2006}
  \at{Bounds on vertical heat transport for infinite-{P}randtl-number
  {R}ayleigh-{B}\'enard convection}.  \jt{J. Fluid Mech.}  \bvol{560},
  \pg{229--241}.

\bibitem[Hasegawa \& Kasagi(2011)]{Hasegawa2011}
{\sc \au{Hasegawa, Y.} \& \au{Kasagi, N.}} \yr{2011}  \at{Dissimilar control of
  momentum and heat transfer in a fully developed turbulent channel flow}.
  \jt{J. Fluid Mech.}  \bvol{683},  \pg{57--93}.

\bibitem[Hassanzadeh {\em et~al.\/}(2014)Hassanzadeh, Chini \&
  Doering]{Hassanzadeh2014}
{\sc \au{Hassanzadeh, P.}, \au{Chini, G.~P.} \& \au{Doering, C.~R.}} \yr{2014}
  \at{Wall to wall optimal transport}.  \jt{J. Fluid Mech.}  \bvol{751},
  \pg{627--662}.

\bibitem[Howard(1963)]{Howard1963}
{\sc \au{Howard, L.~N.}} \yr{1963}  \at{Heat transport by turbulent
  convection}.  \jt{J. Fluid Mech.}  \bvol{17},  \pg{405--432}.

\bibitem[Kasagi {\em et~al.\/}(2012)Kasagi, Hasegawa, Fukagata \&
  Iwamoto]{Kasagi2012}
{\sc \au{Kasagi, N.}, \au{Hasegawa, Y.}, \au{Fukagata, K.} \& \au{Iwamoto, K.}}
  \yr{2012}  \at{Control of turbulent transport less friction and more heat
  transfer}.  \jt{ASME J. Heat Transfer}  \bvol{134}~(031009).

\bibitem[Kasagi {\em et~al.\/}(2009)Kasagi, Suzuki \& Fukagata]{Kasagi2009}
{\sc \au{Kasagi, N.}, \au{Suzuki, Y.} \& \au{Fukagata, K.}} \yr{2009}
  \at{Microelectromechanical systems-based feedback control of turbulence for
  skin friction reduction}.  \jt{Annu. Rev. Fluid Mech.}  \bvol{41},
  \pg{231--251}.

\bibitem[Kawahara(2005)]{Kawahara2005}
{\sc \au{Kawahara, G.}} \yr{2005}  \at{Energy dissipation in spiral vortex
  layers wrapped around a straight vortex tube}.  \jt{Phys. Fluids}
  \bvol{17}~(055111).

\bibitem[Kawahara {\em et~al.\/}(1997)Kawahara, Kida, Tanaka \&
  Yanase]{Kawahara1997}
{\sc \au{Kawahara, G.}, \au{Kida, S.}, \au{Tanaka, M.} \& \au{Yanase, S.}}
  \yr{1997}  \at{Wrap, tilt and stretch of vorticity lines around a strong thin
  straight vortex tube in a simple shear flow}.  \jt{J. Fluid Mech.}
  \bvol{353},  \pg{115--162}.

\bibitem[Kerswell(2001)]{Kerswell2001}
{\sc \au{Kerswell, R.~R.}} \yr{2001}  \at{New results in the variational
  approach to turbulent {B}oussinesq convection}.  \jt{Phys. Fluids}
  \bvol{13},  \pg{192--209}.

\bibitem[Lerley {\em et~al.\/}(2006)Lerley, Kerswell \& Plasting]{Lerley2006}
{\sc \au{Lerley, G.~R.}, \au{Kerswell, R.~R.} \& \au{Plasting, S.~C.}}
  \yr{2006}  \at{Infinite-{P}randtl-number convection. {P}art 2. {A} singular
  limit of upper bound theory}.  \jt{J. Fluid Mech.}  \bvol{560},
  \pg{159--228}.

\bibitem[Malkus(1954)]{Malkus1954}
{\sc \au{Malkus, W. V.~R.}} \yr{1954}  \at{The heat transport and spectrum of
  thermal turbulence}.  \jt{Proc. R. Soc. Lond. {\rm A}}  \bvol{225},
  \pg{196--212}.

\bibitem[Moore(1985)]{Moore1985}
{\sc \au{Moore, D.W.}} \yr{1985}  \at{The interaction of a diffusing line
  vortex and an aligned shear flow}.  \jt{Proc. R. Soc. London}  \bvol{Ser. A
  399},  \pg{367--375}.

\bibitem[Nicodemus {\em et~al.\/}(1997)Nicodemus, Grossmann \&
  Holthaus]{Nicodemus1997}
{\sc \au{Nicodemus, R.}, \au{Grossmann, S.} \& \au{Holthaus, M.}} \yr{1997}
  \at{Improved variational principle for bounds on energy dissipation in
  turbulent shear flow}.  \jt{Physica {\rm D}}  \bvol{101},  \pg{178--190}.

\bibitem[Nicodemus {\em et~al.\/}(1998{\natexlab{{\em a\/}}})Nicodemus,
  Grossmann \& Holthaus]{Nicodemus1998a}
{\sc \au{Nicodemus, R.}, \au{Grossmann, S.} \& \au{Holthaus, M.}}
  \yr{1998{\natexlab{{\em a\/}}}}  \at{The background flow method. {P}art 1.
  {C}onstructive approach to bounds on energy dissipation}.  \jt{J. Fluid
  Mech.}  \bvol{363},  \pg{281--300}.

\bibitem[Nicodemus {\em et~al.\/}(1998{\natexlab{{\em b\/}}})Nicodemus,
  Grossmann \& Holthaus]{Nicodemus1998b}
{\sc \au{Nicodemus, R.}, \au{Grossmann, S.} \& \au{Holthaus, M.}}
  \yr{1998{\natexlab{{\em b\/}}}}  \at{The background flow method. {P}art 2.
  {A}symptotic theory of dissipation bounds}.  \jt{J. Fluid Mech.}  \bvol{363},
   \pg{301--323}.

\bibitem[Otero {\em et~al.\/}(2002)Otero, Wittenberg, Worthing \&
  Doering]{Otero2002}
{\sc \au{Otero, J.}, \au{Wittenberg, R.~W.}, \au{Worthing, R.~A.} \&
  \au{Doering, C.~R.}} \yr{2002}  \at{Bounds on {R}ayleigh--{B}\'enard
  convection with an imposed heat flux}.  \jt{J. Fluid Mech.}  \bvol{473},
  \pg{191--199}.

\bibitem[Plasting \& Kerswell(2003)]{Plasting2003}
{\sc \au{Plasting, S.~C.} \& \au{Kerswell, R.~R.}} \yr{2003}  \at{Improved
  upper bound on the energy dissipation rate in plane {C}ouette flow: the full
  solution to {B}usse's problem and the {C}onstantin--{D}oering--{Hopf} problem
  with one-dimensional background field}.  \jt{J. Fluid Mech.}  \bvol{477},
  \pg{363--379}.

\bibitem[Reynolds(1874)]{Reynolds1874}
{\sc \au{Reynolds, O.}} \yr{1874}  \at{On the extent and action of the heating
  surface of steam boilers}.  \jt{Proc. Lit. Phil. Soc. Manchester}  \bvol{14},
   \pg{7--12}.

\bibitem[Robertson \& Johnson(1970)]{Robertson1970}
{\sc \au{Robertson, J.~M.} \& \au{Johnson, H.~F.}} \yr{1970}  \at{Turbulence
  structure in plane {C}ouette flow}.  \jt{ASCE J. Eng. Mech. Div. Proc.}
  \bvol{96},  \pg{1171--1182}.

\bibitem[Saad \& Schultz(1986)]{Saad1986}
{\sc \au{Saad, Y.} \& \au{Schultz, M.~H.}} \yr{1986}  \at{{GMRES}: {A}
  generalized minimal residual algorithm for solving nonsymmetric linear
  systems}.  \jt{SIAM J. Sci. and Stat. Comput.}  \bvol{7}~(3),  \pg{856--869}.

\bibitem[Sasamori {\em et~al.\/}(2014)Sasamori, Mamori, Iwamoto \&
  Murata]{Sasamori2014}
{\sc \au{Sasamori, M.}, \au{Mamori, H.}, \au{Iwamoto, K.} \& \au{Murata, A.}}
  \yr{2014}  \at{Experimental study on drag-reduction effect due to sinusoidal
  riblets in turbulent channel flow}.  \jt{Exp. Fluids}  \bvol{55}~(1828).

\bibitem[Sondak {\em et~al.\/}(2015)Sondak, Smith \& Waleffe]{Sondak2015}
{\sc \au{Sondak, D.}, \au{Smith, L.~M.} \& \au{Waleffe, F.}} \yr{2015}
  \at{Optimal heat transport solutions for {R}ayleigh--{B}\'enard convection}.
  \jt{J. Fluid Mech.}  \bvol{784},  \pg{565--595}.

\bibitem[Suga {\em et~al.\/}(2011)Suga, Mori \& Kaneda]{Suga2011}
{\sc \au{Suga, K.}, \au{Mori, M.} \& \au{Kaneda, M.}} \yr{2011}  \at{Vortex
  structure of turbulence over permeable walls}.  \jt{Int. J. Heat and Fluid
  Flow}  \bvol{32},  \pg{586--595}.

\bibitem[Townsend(1976)]{Townsend1976}
{\sc \au{Townsend, A.~A.}} \yr{1976} {\em The structure of turbulent shear
  flow\/}.  \publ{Cambridge University Press}.

\bibitem[Viswanath(2007)]{Viswanath2007}
{\sc \au{Viswanath, D.}} \yr{2007}  \at{Recurrent motions within plane
  {C}ouette turbulence}.  \jt{J. Fluid Mech.}  \bvol{580},  \pg{339--358}.

\bibitem[Viswanath(2009)]{Viswanath2009}
{\sc \au{Viswanath, D.}} \yr{2009}  \at{The critical layer in pipe flow at high
  {R}eynolds number}.  \jt{Philos. Trans. R. Soc. A}  \bvol{367},
  \pg{561--576}.

\bibitem[Whitehead \& Doering(2012)]{Whitehead2012}
{\sc \au{Whitehead, J.~P.} \& \au{Doering, C.~R.}} \yr{2012}  \at{Rigid bounds
  on heat transport by a fluid between slippery boundaries}.  \jt{J. Fluid
  Mech.}  \bvol{707},  \pg{241--259}.

\bibitem[Yamamoto {\em et~al.\/}(2013)Yamamoto, Hasegawa \&
  Kasagi]{Yamamoto2013}
{\sc \au{Yamamoto, A.}, \au{Hasegawa, Y.} \& \au{Kasagi, N.}} \yr{2013}
  \at{Optimal control of dissimilar heat and momentum transfer in a fully
  developed turbulent channel flow}.  \jt{J. Fluid Mech.}  \bvol{733},
  \pg{189--220}.

\end{thebibliography}
\bibliographystyle{jfm}

\end{document}